\documentclass[a4paper,11pt]{article}
\pdfoutput=1 

\usepackage{jheppub} 
\usepackage{amsmath,amssymb,amsthm,amscd,graphicx}

\input epsf.sty

\makeatletter
\newsavebox{\@brx}
\newcommand{\llangle}[1][]{\savebox{\@brx}{\(\m@th{#1\langle}\)}%
  \mathopen{\copy\@brx\kern-0.5\wd\@brx\usebox{\@brx}}}
\newcommand{\rrangle}[1][]{\savebox{\@brx}{\(\m@th{#1\rangle}\)}%
  \mathclose{\copy\@brx\kern-0.5\wd\@brx\usebox{\@brx}}}
\makeatother

\theoremstyle{definition}

\newcommand{\CH}{{\cal H}}

\newcommand{\CK}{{\cal K}}

\newcommand{\CO}{{\cal O}}
\newcommand{\CP}{{\cal P}}
\newcommand{\CQ}{{\cal Q}}

\newcommand{\CS}{{\cal S}}
\newcommand{\CT}{{\cal T}}

\def\IZ{{\mathbb Z}}
\def\IR{{\mathbb R}}
\def\IC{{\mathbb C}}
\def\IP{{\mathbb P}}

\def\IF{{\mathbb F}}
\newcommand{\tr}{{\rm Tr}}
\newcommand{\re}{{\rm e}}
\newcommand{\ri}{{\rm i}}
\newcommand{\rd}{{\rm d}}

\newcommand{\mx}{\mathsf{x}}
\newcommand{\my}{\mathsf{y}}
\newcommand{\mq}{\mathsf{q}}
\newcommand{\mm}{\mathsf{p}}
\newcommand{\im}{\mathsf{i}}

\newcommand{\mb}{\mathsf{b}}

\newcommand{\mO}{\mathsf{O}}

\newcommand{\mR}{\mathsf{R}}
\newcommand{\mJ}{\mathsf{J}}
\newcommand{\fad}{\operatorname{\Phi}_{\mathsf{b}}}

\newcommand{\be}{\begin{equation}}
\newcommand{\ee}{\end{equation}}
\newcommand{\ba}{\begin{aligned}}
\newcommand{\ea}{\end{aligned}}
\newcommand{\ben}{\begin{eqnarray}\displaystyle}
\newcommand{\een}{\end{eqnarray}}

\newcommand{\sectiono}[1]{\section{#1}\setcounter{equation}{0}}

\newdimen\tableauside\tableauside=1.0ex
\newdimen\tableaurule\tableaurule=0.4pt
\newdimen\tableaustep
\def\phantomhrule#1{\hbox{\vbox to0pt{\hrule height\tableaurule width#1\vss}}}
\def\phantomvrule#1{\vbox{\hbox to0pt{\vrule width\tableaurule height#1\hss}}}
\def\sqr{\vbox{%
  \phantomhrule\tableaustep
  \hbox{\phantomvrule\tableaustep\kern\tableaustep\phantomvrule\tableaustep}%
  \hbox{\vbox{\phantomhrule\tableauside}\kern-\tableaurule}}}
\def\squares#1{\hbox{\count0=#1\noindent\loop\sqr
  \advance\count0 by-1 \ifnum\count0>0\repeat}}
\def\tableau#1{\vcenter{\offinterlineskip
  \tableaustep=\tableauside\advance\tableaustep by-\tableaurule
  \kern\normallineskip\hbox
    {\kern\normallineskip\vbox
      {\gettableau#1 0 }%
     \kern\normallineskip\kern\tableaurule}%
  \kern\normallineskip\kern\tableaurule}}
\def\gettableau#1{\ifnum#1=0\let\next=\null\else
\squares{#1}\let\next=\gettableau\fi\next}

\newcommand{\figref}[1]{Fig.~\protect\ref{#1}}

\title{\boldmath Exact eigenfunctions and the open topological string}

\author{Marcos Mari\~no and Szabolcs Zakany}

\affiliation{D\'epartement de Physique Th\'eorique et Section de Math\'ematiques\\
Universit\'e de Gen\`eve, Gen\`eve, CH-1211 Switzerland}

\emailAdd{marcos.marino@unige.ch, szabolcs.zakany@unige.ch} 

\abstract{Mirror curves to toric Calabi--Yau threefolds can be quantized and lead to trace class operators on the real line. The eigenvalues of these operators are encoded in 
the BPS invariants of the underlying threefold, but much less is known about their eigenfunctions. In this paper we first develop methods in 
spectral theory to compute these eigenfunctions. We also provide a matrix integral representation which allows to study them in a 't Hooft limit, where 
they are described by standard topological open string amplitudes. Based on these results, we propose a conjecture for the exact eigenfunctions which involves 
both the WKB wavefunction and the standard topological string wavefunction. This conjecture 
can be made completely explicit in the maximally supersymmetric, or self-dual case, which we work out in detail for local $\IP^1 \times \IP^1$. 
In this case, our conjectural eigenfunctions turn out to be closely related to Baker--Akhiezer functions on the mirror curve, 
and they are in full agreement with first-principle calculations in spectral theory.}

\begin{document}

\maketitle
\flushbottom

\sectiono{Introduction}

It has been suspected for a long time that topological strings on toric Calabi--Yau (CY) manifolds can be encoded in simple quantum mechanical systems. 
To identify these systems, it was suggested in \cite{adkmv} that the quantization of the mirror curve to the CY leads naturally to a 
quantum mechanical operator, and that the eigenfunctions 
of this operator should be closely 
related to the open topological string partition function. Recently, some of these ideas have been developed into a precise 
correspondence between the spectral theory of 
quantum mirror curves, and topological string theory \cite{ghm,cgm} (see \cite{mmrev} for a review). According to this correspondence, 
given a toric CY manifold, with a mirror curve of genus $g_\Sigma$, one can construct $g_\Sigma$ trace class operators on $L^2(\IR)$. It has been conjectured that the generalized Fredholm determinant of these operators, 
which is an entire function of the $g_\Sigma$ moduli of the curve, 
can be computed {\it exactly} from the BPS invariants of the underlying CY. This leads in particular 
to exact quantization conditions for these operators. Moreover, the standard topological string 
emerges in a 't Hooft-type large $N$ limit from the spectral traces of the operators. This limit is a strongly coupled limit of the 
quantum-mechanical problem. At weak coupling, one recovers 
the perturbative WKB analysis of \cite{mirmor,acdkv}, which involves the refined topological string in the so-called 
Nekrasov--Shatashvili (NS) limit \cite{ns}.

The correspondence of \cite{ghm,cgm} builds upon previous results for the spectral problem \cite{km,hw}, 
and is largely inspired by developments in the Fermi gas approach 
to Chern--Simons--matter theories \cite{mp,hmo,hmo2,calvo-m, hmo3,hmmo,cgm8}. Various aspects of the correspondence 
have been extensively studied in the last two years. We have by now 
a precise characterization of the operators associated to the mirror curves \cite{kasm,kmz,lst,cgum} 
and first principles calculations of their spectral traces \cite{hmo,py,oz}. 
The conjectures on Fredholm determinants and quantization conditions has been tested with flying colors in 
many geometries \cite{ghm,gkmr, cgm,cgum}, both numerically and analytically, and proved in one important example in \cite{bgt}. 
There is a matrix model formulation of the spectral traces \cite{mz,kmz} where one can study their 't Hooft limit with standard large $N$ techniques, providing 
in this way further, analytic tests of the proposals of \cite{ghm,cgm}. The conjecture on the Fredholm determinant leads to a concrete realization of the 
``non-perturbative" partition function proposed in \cite{eynard,em}, as noted in \cite{cgm8,ghm} and further studied in \cite{grassi}. In addition, the reformulation of the 
quantization conditions of \cite{ghm} in \cite{wzh} has led to an explicit proposal for the exact spectrum of 
relavitistic Toda lattices \cite{hm} and cluster integrable systems \cite{fhm}. There are also connections with spectral problems 
arising in condensed matter physics \cite{butterfly}. 

In order to have a complete solution of the spectral problem, however, one needs as well explicit descriptions of the eigenfunctions, 
and not only of the eigenvalues. In this paper, we present a preliminary analysis of the problem, based on the approach of \cite{ghm,mz}. 
Our strategy is based on three steps. 

First of all, we give a detailed description of the eigenfunctions from the point of view of spectral theory. Since many of the operators we are 
interested in have explicit integral kernels, we can 
study them by using classical results in Fredholm theory, as well as the more recent results of \cite{tw,oz}. In this framework, 
one can introduce wavefunctions which formally satisfy the difference equation obtained from the mirror curve, 
but are not actual eigenfunctions. This is similar to what happens in conventional Quantum Mechanics, where one can study functions which 
solve the Schr\"odinger equation but do not have the required decay properties at infinity. By an abuse of language, we will call these 
wavefunctions ``off-shell". They are generalizations of the Fredholm determinant, and in particular they can be 
computed by {\it convergent} expansions in the exponentiated 
energy or fugacity $\kappa$. This is in contrast with the WKB method, where one obtains formal, asymptotic expansions which have to be 
resummed with Borel techniques. Moreover, the coefficients of this convergent expansion can be computed analytically for selected values of $\hbar$, 
and they can be regarded as generalizations of the fermionic spectral traces. 

Second, we extend the results in \cite{mz,kmz} and provide matrix model representations for these wavefunctions. This makes it possible to 
study their 't Hoof limit, and we find that it is governed by the standard, open topological string partition function. More precisely, one obtains the 
so-called topological string wavefunction, which is 
a formal asymptotic expansion involving integrated versions of the open topological string amplitudes. This vindicates the philosophy 
of \cite{adkmv}, but with two twists. The first twist is that the standard topological string appears in the strong coupling limit of the spectral problem, and not 
in the weak coupling limit, as envisaged originally in \cite{adkmv}. This of course happened already in the closed string sector, as shown 
in \cite{ghm}. The other twist is more subtle. 
Open topological string amplitudes depend on a choice of frame for closed and open string moduli. However, 
there is a preferred frame for these moduli in which one makes contact with enumerative invariants. We call this frame the large radius frame. It turns out that, 
when we write down the 't Hooft limit of the wavefunction in this frame, we find {\it two} copies of the open topological string. These copies correspond to the two different 
sheets of the underlying Riemann surface defined by the mirror curve. 

Armed with these results, we can address the problem of finding the exact eigenfunctions for the spectral problem 
when $\hbar$ is real. In our third step, we propose an answer to this problem 
which generalizes the conjecture for the Fredholm determinant in \cite{ghm}. According to our proposal, the ``off-shell" eigenfunctions 
involve the all-orders resummation of the WKB partition function, which controls the weakly coupled regime, 
and the topological string wavefunction, which controls the strongly coupled regime. Our proposal can be made completely explicit in the so-called maximally 
supersymmetric $\hbar=2\pi$. In this case, as argued in \cite{cgm8,ghm}, we expect important simplifications, 
and indeed we can write down closed form formulae for the wavefunctions, in terms of Jacobi theta functions. The resulting expressions are very similar 
to Baker--Akhiezer functions, and make contact with the tau functions studied in \cite{em,be}. These wavefunctions can be compared with the explicit results in 
Fredholm theory and we find a remarkable agreement.

The resulting picture is very interesting, both physically and mathematically. Physically, the structure we find is similar to the one obtained in \cite{mmss}, in the context of 
non-critical strings. The building blocks of the wavefunction involve two different types of open topological string amplitudes, corresponding to the two 
different sheets of the mirror curve. They have branch cuts and singularities. However, when we sum these two contributions to obtain the final 
answer for the wavefunction, the branch cuts and singularities {\it disappear}, and we obtain an {\it entire} function on the complex plane. As noted in 
\cite{mmss}, this indicates that the mirror geometry, as probed by the D-brane 
wavefunction, disappears non-perturbatively, and is replaced by the complex plane. 
The singularities appearing in the building blocks are expected: they are the well-known turning point singularities of the standard WKB approximation. 
In our calculation, these singularities disappear in the end by including non-perturbative corrections (which come 
from the standard topological string), and by adding the contributions of the two Riemann sheets. 
 
 As Michael Berry once noted, ``only wimps specialize in the general case. Real scientists pursue examples" \cite{berry}. In line 
 with this philosophy, most of the results mentioned above have been obtained in one particular example, namely, the 
 operator for the toric CY known as local $\IF_0$. We also present some of the results in the case of local $\IP^2$, but the full 
 details for this and other cases will appear elsewhere. We believe that the general picture that we have obtained will apply to many other examples. 
 
 Before closing this introduction, we should mention the relationship between our approach and other perspectives 
 on the problem. In \cite{amir}, the problem of solving for the eigenfunctions has been addressed by using the exact WKB method. 
 A proposal for the exact eigenfunctions is made which mimics the quantization condition 
 in the form given in \cite{wzh}: one should consider the quotient of the resummed WKB wavefunction, and its S-dual transform. 
This proposal contains some of the ingredients of the answer (for example, for $\hbar=2 \pi$ it also leads to the function (\ref{sigx})), but it does not seem to reproduce the correct 
 eigenfunctions.
 
 A popular approach to obtaining eigenfunctions of this type of spectral problems is to use the 5d instanton partition function in 
 the presence of defects, and in the NS limit. The resulting object is a resummation of the WKB wavefunction. However, in the case of interest here, 
 namely when $\hbar$ is real, the 5d partition functions (with and without defects) are afflicted with a dense set of poles. These are only removed when 
 non-perturbative contributions in $\hbar$ are added. Therefore, the 5d instanton partition function with 
 defects gives just a formal wavefunction. In contrast, in this paper we are interested in the {\it actual} 
 eigenfunctions\footnote{A recent attempt to obtain the eigenfunctions in the framework of 5d instanton partition functions has been made in \cite{sciarappa}, 
 but at the end of the day its proposal is the same of \cite{amir}, therefore it shares the same problems.}.

A third approach to the problem is based on topological recursion. The topological 
string wavefunction can be defined for arbitrary spectral curves by using the topological recursion of \cite{eo}. It was noted in examples in \cite{gs} that the resulting 
wavefunction is equal to the perturbative WKB wavefunction associated to the formal quantization of the spectral curve, also called ``quantum curve" (see \cite{mulase,norbury} for 
reviews). Very recently it has been proved that this is a general property for curves of genus zero \cite{boucharde}. 
For mirror curves of higher genus, the wavefunction constructed in the context of topological 
 recursion is in principle different from the standard WKB wavefunction. Our proposal for the wavefunction includes both of them, and it is 
also closely related to the non-perturbative framework for topological recursion proposed in \cite{eynard, em,be}. One important difference 
however between our approach and the approach based on topological recursion is that our eigenfunctions, even off-shell, are well-defined beyond perturbation theory, while 
the literature on ``quantum curves" deals only with formal, asymptotic expansions.

This paper is organized as follows. In section 2 we develop tools to compute the eigenfunctions, by relying on classical Fredholm theory and on more recent 
results of Tracy--Widom. Our main example is the operator associated to local $\IF_0$, but we also generalize some of the results to other operators. We perform in particular 
a detailed calculation in the maximally supersymmetric case of local $\IF_0$ which will set the standards for our conjecture in section 4. In section 3, we provide a matrix model 
representation of the eigenfunctions, we study their 't Hooft limit, and we make contact with the topological 
string wavefunction. In section 4 we present a conjecture for the exact eigenfunctions which generalizes the conjecture for the spectral determinant in \cite{ghm}. In addition, we 
present completely explicit results for the eigenfunctions of local $\IF_0$ in the maximally supersymmetric case. They fully agree with our computations 
in section 2. Finally, section 5 contains conclusions and prospects for future work. The first Appendix contains technical details on Bergmann kernels, theta functions, and the 
Abel--Jacobi map of the elliptic curve for local $\IF_0$. The second Appendix explains how to calculate the eigenfunctions for local $\IF_0$ numerically, in order to test the results 
of Fredholm theory.

\sectiono{Eigenfunctions and spectral theory}

\label{sect-2}
\subsection{Operators from mirror curves}

The construction of \cite{ghm,cgm} associates a set of $g_\Sigma$ trace class operator on $\CH=L^2(\IR)$ to any toric CY 
threefold with a mirror curve $X$ of genus $g_\Sigma$. In this paper, for simplicity, we will restrict to the case $g_\Sigma=1$. In this case, 
the complex moduli of the curve include a ``true" geometric modulus, which we will denote by $\kappa$, and a set of ``mass" parameters $\xi_i$, $i=1, \cdots, s-1$, where $s$ 
depends on the geometry 
under consideration \cite{hkp,hkrs}. From the point of view of the Newton polytope associated to $X$, the modulus corresponds to the single inner point, while the mass parameters 
are points in the boundary. The mirror curve can be put in the form
\be
\label{ex-W}
W(\re^x, \re^y)= \CO(x,y)+ \kappa=0,  
\ee
where $\CO(x,y)$ has the form
\be
\label{coxp}
 \CO (x,y)=\sum_{i=1}^{s+2} \exp\left( \nu^{(i)}_1 x+  \nu^{(i)}_2 y + f_i(\boldsymbol{\xi}) \right), 
 \ee
and $f_i(\boldsymbol{\xi})$ are suitable functions of the parameters $\xi_j$. The vectors $\nu^{(i)}_{1,2}$ can be read from the fan defining the toric 
CY threefold. The operator associated to $X$ can be regarded as a quantization of the above mirror curve: we promote $x$, $y$ to self-adjoint Heisenberg 
operators $\mathsf{x}$, $\mathsf{y}$ satisfying the commutation relation 
\be
[\mathsf{x}, \mathsf{y}]=\im\hbar. 
\ee
By using Weyl's quantization prescription, 
\be
\re^{a x+  b y} \rightarrow \re^{a \mx + b \my}, 
\ee
the function $\CO(x,y)$ becomes a self-adjoint operator, which 
will be denoted by $\mathsf{O}$. If the mass parameters satisfy appropriate positivity conditions, the operator $\mO$ has a number of important 
properties. The operator 
\be
\rho=\mO^{-1}, 
\ee
acting on $L^2(\IR)$, turns out to be of trace class in all known examples \cite{kasm,lst}. Therefore, it has a discrete spectrum $\re^{-E_n}$, $n=0,1,2, \cdots$, 
and its Fredholm determinant 
\be
\label{fred-det}
\Xi(\kappa)={\rm det}(1+ \kappa \rho)
\ee
is well-defined and is an entire function of $\kappa$.

Our main example in this paper will be the CY known as local $\IF_0$, where the curve can be put in the form 
\be
\label{spec-curve}
\re^x + m_{\IF_0} \re^{-x}+ \re^y + \re^{-y}+ \kappa=0. 
\ee
This geometry has a single mass parameter $m_{\IF_0}$, and for simplicity we will choose most of the time $m_{\IF_0}=1$ (the generalization to arbitrary $m_{\IF_0}$ should be straightforward). The resulting 
operator is 
\be
\label{f0op}
\mO= \re^{\mx} + m_{\IF_0}\re^{-\mx}+ \re^{\my} + \re^{-\my}. 
\ee

\subsection{The eigenfunctions according to Fredholm}
\label{sectFredholm}

We are interested in solving the spectral problem for the operator $\rho$ (or equivalently, for the operator $\mO$), i.e. we want to solve 
for its eigenvalues and its eigenfunctions. A useful approach to this problem is Fredholm theory, which gives convergent expansions for the eigenfunctions and 
for the Fredholm determinant (see for example \cite{integral}). In this theory, the basic ingredient is the integral 
kernel of the operator $\rho$, which we will denote by $\rho(x_i, x_j)$. The 
fermionic spectral traces are defined by 
\be
Z(N)={1\over N!} \int_{\IR^N}  \rho \begin{pmatrix}  x_1 & \cdots & x_N \\
 x_1 & \cdots &x_N \end{pmatrix}\rd x_1 \cdots \rd x_N, 
 \ee
where
 \be
 \rho \begin{pmatrix}  x_1 & \cdots & x_N \\
 x_1 & \cdots &x_N \end{pmatrix}={\rm det}\left[ \rho(x_i, x_j) \right]_{i,j=1,\cdots, N}. 
 \ee
In terms of these traces, the Fredholm determinant is given by 
\be
\Xi(\kappa)=1+ \sum_{N\ge 1} \kappa^N Z(N), 
\ee
and the series in $\kappa$ converges for any $\kappa\in \IC$. The eigenvalues of $\mO$ are minus the zeroes of the Fredholm 
determinant, i.e. when $\kappa=-\re^{E_n}$ we have 
\be
\Xi\left( -\re^{E_n} \right)=0, \qquad n=0, 1, \cdots. 
\ee
In order to obtain the eigenfunctions of $\rho$, we consider the {\it resolvent}, 
\be
\mR={\rho \over 1+\kappa \rho}. 
\ee
Since $\mO$ has a discrete spectrum, the resolvent can be written as 
\be
\mR= {1\over \mO+ \kappa}= \sum_{n \ge 0} {|\varphi_n\rangle \langle \varphi_n |\over \re^{E_n} + \kappa} , 
\ee
so the residue of the simple pole at $\kappa=-\re^{E_n}$ gives the projector onto the $n$-th eigenfunction. In terms of the integral kernel of the resolvent, 
\be
R(x,t; \kappa) =\langle x| \mR | t\rangle, 
\ee
we have
\be
\label{res-res}
{\rm Res}_{\kappa= -\re^{E_n}} R(x,t; \kappa)= \varphi_n(x) \varphi_n^*(t). 
\ee
 One of the main results of Fredholm theory is that the resolvent can be written as 
\be
R(x,t;\kappa)= {D(x,t;\kappa) \over \Xi(\kappa)}, 
\ee
where
\be
D(x,t;\kappa)= \sum_{N\ge 0} \kappa^N B_N(x,t), 
\ee
and 
\be
B_N(x,t)= {1\over N!} \int_{\IR^N} \rho \begin{pmatrix} x & x_1 & \cdots & x_N \\
t& x_1 & \cdots &x_N \end{pmatrix}\rd x_1 \cdots \rd x_N.
\ee
In this equation, the integrand is given by 
\be
\rho \begin{pmatrix} x & x_1 & \cdots & x_N \\
t& x_1 & \cdots &x_N \end{pmatrix}={\rm det}\left[ \rho(\mu_i, \nu_j) \right]_{i,j=0, 1, \cdots, N}, 
\ee
where
\be
\label{coords}
\ba
\mu_0=x, \quad \mu_i=x_i, \quad i=1,\cdots, N, \\
\nu_0=t, \quad \nu_i=x_i, \quad i=1,\cdots, N. 
\ea
\ee
Since the Fredholm determinant can be written as an infinite product, 
\be
\Xi(\kappa)= \prod_{m \ge0} (1+ \kappa \re^{-E_m} ), 
\ee
the residue in (\ref{res-res}) can be calculated as follows:
\be
\label{prod-phi}
\varphi_n(x) \varphi_n^*(t)= \lim_{\kappa \to -\re^{E_n}} (\kappa+ \re^{E_n}){D(x,t; \kappa) \over \Xi(\kappa)}=
{1\over \Xi'(-\re^{E_n} ) } D(x, t; -\re^{E_n} ) . 
\ee
We assumed here that the eigenvalues are not degenerate. If $(x_0, t_0)$ are such that $D\left(x_0, t_0; -\re^{E_n}  \right) \not=0$, one finds in addition 
\be
{\varphi_n(x) \over \varphi_n(x_0)} = {D\left(x, t_0; -\re^{E_n}  \right) \over  D\left(x_0, t_0; -\re^{E_n}  \right)}. 
\ee
This is Fredholm's formula for the eigenfunctions of the operator $\rho$ in terms of the determinant $D(x,t; \kappa)$. It requires previous knowledge of the 
spectrum from Fredholm's determinant.

So far our considerations have been general. Further mileage can be obtained if we take into account the particular form of the 
integral kernels appearing in the theory of quantized mirror curves. Let us then assume that the integral kernel of the operator $\rho$ is of the following form, 
\be
\label{rhov}
\rho(x,t)= {{\sqrt{v(x)}}{\sqrt{v(t)}} \over 2 \cosh\left( {x-t \over 2 \xi}\right)}, 
\ee
where $\xi$ is a constant. Since $\rho$ is of trace class, the function $v(x)$ belongs to $L^1(\IR)$ and goes to zero at infinity. 
As shown in \cite{kmz}, and as we will recall shortly, the integral kernel of the inverse of (\ref{f0op}) has this structure, so our considerations 
here are sufficient to address the case of local $\IF_0$. Integral 
kernels of the form (\ref{rhov}) were studied by Zamolodchikov in \cite{zamo} and by Tracy and Widom in \cite{tw}. We will also write the kernel (\ref{rhov}) as 
\be
\rho(x,y)= \frac{\sqrt{E(x)}\sqrt{E(y)}}{M(x)+M(y)},
\ee
where
\be
\label{defEM}
M(x)=\re^{x/\xi}, \qquad \qquad E(x)=v(x)M(x).
\ee
As shown in \cite{kmz}, following previous developments in \cite{kwy,mp,mz}, one can use the Cauchy identity to obtain a matrix model-like expression for the fermionic 
spectral trace, 
\be
\label{znmm}
Z(N)= {1\over N!} \int_{\IR^N}  \rd ^N x\,  \prod_{i=1}^N v(x_i) {\prod_{i<j} \left[ 2 \sinh \left( {x_i - x_j \over 2 \xi } \right)\right]^2 
\over \prod_{i,j} 2 \cosh\left( {x_i - x_j \over 2  \xi } \right)}. 
\ee
Let us denote by
\be
\llangle f(x_1, \cdots, x_N) \rrangle
\ee
an {\it unnormalized} vacuum expectation value in the matrix model above, i.e.  
\be
\llangle f(x_1, \cdots, x_N) \rrangle= {1\over N!} \int_{\IR^N}  \rd ^N x\,  f(x_1, \cdots, x_N) \prod_{i=1}^N v(x_i) {\prod_{i<j} \left[ 2 \sinh \left( {x_i - x_j \over 2 \xi } \right)\right]^2 
\over \prod_{i,j} 2 \cosh\left( {x_i - x_j \over 2  \xi } \right)}. 
\ee
Then, applying again Cauchy's identity, one finds that the numerator of the resolvent is given by
\be 
\label{dxy}
D(x, t; \kappa) = \rho(x,t) \sum_{N\ge 0} \llangle[\Bigl] \prod_{i=1}^N \tanh\left( {x- x_i \over 2 \xi } \right) \tanh\left( {t- x_i \over 2 \xi } \right)  \rrangle[\Bigr] \kappa^N. 
\ee
Morally speaking, the resolvent contains information about the square of the eigenfunction. In order to extract the eigenfunction itself, we will consider 
the limiting behavior of the resolvent when one of its arguments go to infinity. Let us first define:
\be
\label{psiN}
\Psi_N(x)= \llangle[\biggl]  \prod_{i=1}^N \tanh\left( {x- x_i \over 2 \xi} \right)\rrangle[\biggr].
\ee
We will set $\Psi_0(x)= Z(N)$. Note that, if $x \in \IR$, 
this is the average of a bounded function in the convergent matrix integral (\ref{znmm}), therefore it is perfectly well defined for any $N \ge 0$ and 
any $x \in \IR$. Let us now study the limit $t \rightarrow \pm \infty$ of $D(x,t; \kappa)$. From (\ref{dxy}) we find
\be
\label{d+}
D(x,t) \approx {\sqrt{v(t) \over M(t)}} {\sqrt{E(x)}} \sum_{N=0}^\infty \Psi_N(x) \kappa^N, \qquad t \rightarrow \infty. 
\ee
On the other hand, we  have
\be
\label{d-}
D(x,t)  \approx {\sqrt{v(t) M(t)}} {\sqrt{E(x)} \over M(x)} \sum_{N=0}^\infty \Psi_N(x) (-\kappa)^N, \qquad t \rightarrow -\infty.
\ee
The dependence on $t$ as $t \rightarrow \pm  \infty$ is in both cases the same, i.e. 
\be
\re^{-{|t| \over 2\xi}} {\sqrt{v(t)}}. 
\ee
The above result suggests defining the following two functions, for arbitrary $\kappa$
\be
\label{josts}
\ba
\Xi_+(x; \kappa)&= {\sqrt{E(x)}} \sum_{N=0}^\infty \Psi_N(x) \kappa^N,\\
\Xi_-(x; \kappa)&= {\sqrt{E(x)} \over M(x)} \sum_{N=0}^\infty \Psi_N(x) (-\kappa)^N.
\ea
\ee
Since 
\be
\label{psinl}
\Psi_N(x) \rightarrow (\pm 1)^N Z(N), \qquad  x \rightarrow \pm \infty, 
\ee
they have the following asymptotic behaviour at infinity: 
\be
\label{asymxi}
\ba
 \Xi_+(x; \kappa) &\rightarrow \begin{cases}\Xi(\kappa) \re^{{x \over 2\xi}} {\sqrt{v(x)}} , & \text{ when $x\rightarrow \infty$,}\\
\Xi(-\kappa) \re^{x \over 2\xi} {\sqrt{v(x)}} , & \text{ when $x\rightarrow -\infty$},
 \end{cases}\\
\Xi_-(x; \kappa) &\rightarrow \begin{cases}\Xi(-\kappa) \re^{-{x \over 2\xi}} {\sqrt{v(x)}} , & \text{ when $x\rightarrow \infty$,}\\
\Xi(\kappa) \re^{-{x \over 2\xi}} {\sqrt{v(x)}} , & \text{ when $x\rightarrow -\infty$}.
 \end{cases}
 \ea
 \ee
 In other words, $\Xi_\pm (x; \kappa)$ decrease at $\mp \infty$, but not necessarily at $\pm \infty$. 
We recall that, in studying the Schr\"odinger equation for a one-dimensional confining potential, 
 one can introduce functions $\psi_{\pm}(x)$ which satisfy 
 \be
 \label{schrodinger}
 -\psi_{\pm} ''(x; E) + V(x) \psi_{\pm}(x;E)= E\psi_{\pm}(x; E), 
 \ee
 go to zero as $\mp \infty$, respectively, but are not necessarily square integrable, unless one 
 tunes the value of $E$ to be a true eigenvalue (see \cite{voros-zeta, voros-complex}). 
 The functions $\Xi_{\pm}(x; \kappa)$ are precisely the analogues of these functions for 
 our problem: as we will see in the next section, they are in the kernel of the operator $\mO+ \kappa$, they go to zero as $\pm \infty$, but 
 they are not actual eigenfunctions of $\rho$ for generic values of $\kappa$. When $\kappa=-\re^{E_n}$ is a zero of the Fredholm determinant, 
 one has
 \be
 \Xi_+(x; -\re^{E_n} ) = \Xi_-(x; -\re^{E_n} ), 
 \ee
 and they give the true, square-integrable eigenfunctions of the spectral problem, up to an overall normalization. 
 In this sense, $\Xi_{\pm}(x; \kappa)$ are also similar to the Jost functions of scattering 
 theory in one dimension, which become square integrable and proportional to each other when one considers 
 bound states (see for example \cite{leon}). Note that, when 
 $\kappa=-\re^{E_n}$, the leading terms of 
 $\Xi_\pm (x; \kappa)$ as $x\rightarrow \mp \infty$ vanish, since $\Xi(-\re^{E_n})=0$. 
 
 We would like to emphasize that the wavefunctions $\Xi_\pm(x; \kappa)$ contain a lot of spectral information about the operator $\rho$, and they 
are meaningful even ``off-shell," i.e. even when $\kappa$ does not correspond to an eigenvalue. In this paper we will in fact present conjectural 
results for these functions for arbitrary values of $\kappa$. 
 
 In order to obtain normalized eigenfunctions, we look at 
 the asymptotics of $\varphi_n(x)$. Let us assume that 
\be
\varphi_n(x) \approx c_n^{\pm} {\sqrt{v(x)}} \re^{-|x|/2}, \qquad x\rightarrow \pm \infty. 
\ee
Then, from (\ref{prod-phi}) and (\ref{d-}) we find 
\be
\varphi_n(x) \varphi^*_n(t) \approx {\Xi( \re^{E_n} ) \over \Xi'(-\re^{E_n} ) }  {\sqrt{v(x)}} \re^{-|x|/2}{\sqrt{v(t)}} \re^{-|t|/2},  \qquad x\rightarrow \infty, \, \, t \rightarrow -\infty. 
\ee
Therefore, we must have
\be
c_n^+ c_n^{-}={\Xi( \re^{E_n} ) \over \Xi'(-\re^{E_n} ) } . 
\ee
When the eigenfunctions have a definite parity, we have that $c_n^+=(-1)^n c_n^-=c_n$, and 
\be
c_n^2 = (-1)^n {\Xi( \re^{E_n} ) \over \Xi'(-\re^{E_n} ) }. 
\ee
Finally, by using (\ref{d-}) again, we obtain 
\be
\label{wavef}
\varphi_n(x) = {1\over {\sqrt{ (-1)^n \Xi(\re^{E_n}) \Xi'(-\re^{E_n})}}} \Xi_+(x; -\re^{E_n}).
\ee

\subsection{The eigenfunctions according to Tracy--Widom}

In \cite{tw}, Tracy and Widom studied the resolvent associated to kernels of the form (\ref{rhov}), and they provided tools that lead to 
an efficient computation of the fermionic spectral traces and of the functions $\Psi_N(x)$, as developed in \cite{hmo,py,oz}. 
Following \cite{tw}, let us define the functions
\be
\label{phieo}
\ba
\Phi_e(x)&={1\over {\sqrt{E(x)}}} \left[{1\over I-\kappa^2 \rho^2}  \sqrt{E}\right](x),\\
 \Phi_o(x)&={1\over {\sqrt{E(x)}}} \left[ {\kappa \rho \over I-\kappa^2 \rho^2} \sqrt{E}\right](x). 
 \ea
\ee
They can be written as power series in $\kappa$:
\be
\ba
\Phi_e(x)&= \sum_{n=0}^\infty \kappa^{2n} \phi_{2n}(x),\\
\Phi_o(x)&= \sum_{n=0}^\infty \kappa^{2n+1} \phi_{2n+1}(x),
\ea
\ee
with the normalization 
\be
\phi_0(x)=1, 
\ee
and the functions $\phi_j(x)$ satisfy the recursion relation 
\be
\label{phijdef}
|\phi_j \rangle= {1\over {\sqrt{E}}} \rho {\sqrt{E}} |\phi_{j-1} \rangle. 
\ee
Let us now decompose the resolvent $\mR$ as follows, 
\be
\mR={1\over \kappa} \left( \mR_o-\mR_e \right),
\ee
where
\be
\mR_e= {\kappa^2 \rho^2 \over 1- \kappa^2 \rho^2}, \qquad \mR_o= {\kappa \rho \over 1- \kappa^2 \rho^2}.
\ee
As shown in \cite{tw}, the integral kernels of the operators $\mR_{e,o}$ can be written in terms of the functions defined in (\ref{phieo}):
\be
\label{reo}
\ba
{1\over \kappa} R_e(x, t)&= {\sqrt{E(x)} \sqrt{E(t)} \over M(x)- M(t)} \left( \Phi_e(x) \Phi_o(t) - \Phi_o(x) \Phi_e(t) \right), \\
{1\over \kappa} R_o(x, t)&= {\sqrt{E(x)} \sqrt{E(t)} \over M(x)+M(t)} \left( \Phi_e(x) \Phi_e(t) - \Phi_o(x) \Phi_o(t) \right). 
\ea
\ee
It can be shown that, if the potential $v(x)$ decays exponentially at infinity, the functions $\phi_j(x)$ go to constants at $\pm \infty$. We will denote 
\be
\lim_{x\to \pm \infty} \Phi_{e,o}(x)= c_{e,o}^\pm (\kappa). 
\ee
In addition, the functions $\phi_j(x)$ with $j\not=0$ go to zero as $x \rightarrow \infty$, therefore we have
\be
\label{+limit}
c_e^+=1, \qquad c_o^+=0. 
\ee
We conclude from (\ref{reo}) that, as $t \rightarrow \infty$, 
\be
R(x,t; \kappa) \approx {\sqrt{v(t) \over M(t)}} {\sqrt{E(x)}} \left( \Phi_e(x)- \Phi_o(x) \right). 
\ee
By comparing with (\ref{d+}) we find that 
\be
\label{phi-psi1}
 \Phi_e(x)- \Phi_o(x) = {1\over \Xi(\kappa)} \sum_{N=0}^\infty \Psi_N(x) \kappa^N.
\ee
Taking the limit $x\rightarrow -\infty$ in (\ref{phi-psi1}) gives
\be
\label{quot}
c_e^- - c_o^-={\Xi(-\kappa) \over \Xi(\kappa)}. 
\ee
By changing the sign of $\kappa$ in (\ref{phi-psi1}) we obtain
\be
\label{plusc}
\Phi_e(x)+ \Phi_o(x) = {1\over \Xi(-\kappa)} \sum_{N=0}^\infty \Psi_N(x) (-\kappa)^N.
\ee
The results (\ref{phi-psi1}) and (\ref{plusc}) lead to the following expressions for the functions (\ref{josts}):
\be
\ba
\Xi_+(x; \kappa)&=\Xi( \kappa)  \re^{ {x\over 2 \xi}}  {\sqrt{v(x)}} \left( \Phi_e(x)-\Phi_o(x) \right),\\
\Xi_-(x; \kappa)&=\Xi(- \kappa)  \re^{- {x\over 2 \xi}} {\sqrt{v(x)}} \left( \Phi_e(x)+ \Phi_o(x) \right).
\ea
\ee
The recursive methods developed in \cite{hmo,py,oz} to compute the $\phi_j(x)$ make it possible to compute the 
functions (\ref{josts}) as a power series expansion in $\kappa$. 
General results of Fredholm theory imply that this series is convergent for any $\kappa$. In that sense, this method to calculate eigenfunctions is 
much better than the WKB method, which 
gives formal power series in $\hbar$. However, as we will see, the resummation of the WKB expansion provided by topological string theory will give us 
the functions (\ref{josts}), as exact functions of $\kappa$. 


We can now use the results of \cite{tw} to verify that the functions $\Xi_\pm (x; \kappa)$ formally satisfy 
\be
\label{or-eq}
\left(\mO+ \kappa \right) \Xi_\pm (x; \kappa)=0, 
\ee
for arbitrary $\kappa$, as mentioned above. To see this, let us write $\rho$ in operator form:
\be
\rho= {\sqrt{v(\mx)}} {2 \pi \xi \over 2 \cosh (\mm/2)}  {\sqrt{v(\mx)}}, 
\ee
where
\be
[\mx, \mm]= 2 \pi \ri \xi. 
\ee
In particular, 
\be
2 \pi \xi \mO={1\over  {\sqrt{v(\mx)}}} 2 \cosh (\mm/2){1\over  {\sqrt{v(\mx)}}}. 
\ee
As shown in \cite{tw} the function 
\be
\Phi_-(x)= \Phi_e(x)- \Phi_o(x) 
\ee
solves the difference equation 
\be
\Phi_-( x+ \pi \ri \xi) - \Phi_-( x-\pi \ri\xi)=2 \pi \ri \xi \kappa v(x) \Phi_+( x). 
\ee
Then, the function 
\be
\widetilde \Phi_-(x) = \re^{{x \over 2 \xi}} \Phi_-(x)
\ee
satisfies
\be
\widetilde\Phi_-( x+ \pi \ri\xi) +\widetilde \Phi_-( x-\pi \ri\xi)=-2 \pi \xi \kappa v(x) \widetilde \Phi_-( x), 
\ee
We conclude that the function 
\be
|\psi\rangle=|\widetilde \Phi_-\rangle
\ee
is a solution to the operator equation
\be
2 \cosh(\mm/2) |\psi\rangle=-2 \pi \xi \kappa v(\mx) |\psi\rangle, 
\ee
and 
\be
|\varphi\rangle = {\sqrt{v(\mx)}}|\psi\rangle
\ee
satisfies (\ref{or-eq}). But 
\be
\varphi(x)={1\over \Xi(\kappa)} \Xi_+(x; \kappa), 
\ee
so $\Xi_+(x; \kappa)$ satisfies (\ref{or-eq}), as we wanted to show. A similar argument can be made for $\Xi_-(x; \kappa)$, 
by using that 
\be
\Phi_+(x)= \Phi_e(x)+ \Phi_o(x) 
\ee
satisfies the difference equation
\be
\Phi_+( x+ \pi \ri \xi) - \Phi_+( x-\pi \ri\xi)=-2 \pi \ri \xi \kappa v(x) \Phi_+( x). 
\ee

Note that, even though the functions $\Xi_\pm (x; \kappa)$ are formally in the kernel of $\mO+\kappa$ for arbitrary $\kappa$, they 
are not eigenfunctions of $\rho$. The reason is that they are not in the domain of $\mO$, regarded as an unbounded 
operator on $L^2(\IR)$, unless $\kappa$ takes special values. In some cases, as we will see in this paper, the functions $\Xi_\pm (x; \kappa)$ 
are not even square integrable for general $\kappa$. The fact that difference equations like (\ref{or-eq}) have huge kernels has led to discussions in the literature on the criteria for selecting 
eigenfunctions. In \cite{ghm}, this problem was solved in a simple way by noting that, when $\hbar$ is real, the inverse operator 
$\rho$ is self-adjoint and of trace class on $L^2(\IR)$, so it leads to a well-defined and discrete spectrum. This is also the point of view that we will 
pursue in this paper. In particular, the problem of finding the eigenfunctions of (\ref{f0op}) is well posed and it has a unique solution. 

\subsection{An application: eigenfunctions for local $\IF_0$}

\begin{figure}[ht]
\begin{center}
\includegraphics[scale=0.55]{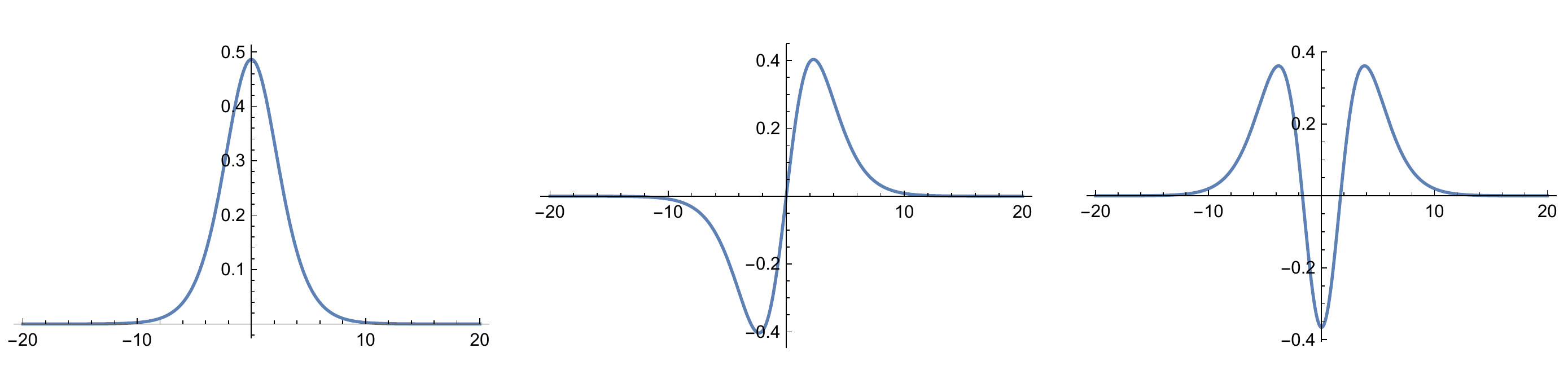}
\caption{The first three normalized wavefunctions $\varphi_n (q)$, $n=0,1,2$, for local $\IF_0$ with $\hbar=2\pi$ and $m_{\IF_0}=1$, 
as computed from (\ref{wavef}). We use the series in $\kappa$ of (\ref{josts}) up to order $\kappa^{19}$. The eigenvalues are computed 
from the approximation of the Fredholm determinant up to order $\kappa^{24}$. }
\label{figwf}
\end{center}
\end{figure}

As we mentioned above, our main example in this paper is the operator associated to the local $\IF_0$, or local ${ \IP}^1 \times \IP^1$ geometry, which we wrote 
down in (\ref{f0op}). We will often use the variables $\mb$ and $\mu$ such that
\be
\hbar=\pi \mb^2, \qquad \qquad m_{\IF_0}=\re^{2 \mb^2\mu}.
\ee
As explained in \cite{kmz}, in order to write down the integral kernel of $\rho= \mO^{-1}$, one has to perform the following change of variables, 
\be
\label{xyq}
\ba
\mx&={1\over {\sqrt{2}}} (\mm+\mq)+{3 \mb^2 \mu\over 2},\\
 \my&={1\over {\sqrt{2}}} (\mm-\mq)+{\mb^2 \mu\over 2}, 
 \ea
\ee
where the operators $\mq$, $\mm$ satisfy the canonical commutation relation 
\be
[\mq,\mm]=\ri \hbar. 
\ee
The kernel is given by \cite{kmz}
\be
\rho(q_1,q_2)=\re^{-\frac{\mb^2\mu}{2}}\frac{f(q_1) \overline{f(q_2)}}{2 \sqrt{2} \pi \mb^2 \cosh \left (\frac{q_1-q_2}{{\sqrt{2}} \mb^2} \right )}, 
\qquad  f(z)=\re^{{\sqrt{2}} z/4 } \frac{ \fad \left ({z\over {\sqrt{2}} \pi \mb}-\frac{\mb\mu}{2\pi}+\frac{\ri \mb}{4} \right )}{ \fad \left ({z\over {\sqrt{2}} \pi \mb}+\frac{\mb\mu}{2\pi}-\frac{\ri b}{4} \right ) }, 
\ee
where $\fad(z)$ is Faddeev's quantum dilogarithm (we follow the conventions of \cite{kasm}). This kernel is of the form (\ref{rhov}), with 
\be
\xi= {\mb^2 \over  {\sqrt{2}}}. 
\ee
A particularly simple case occurs when $m_{\IF_0}=1$ and $\hbar=2\pi$. This value of $\hbar$ corresponds to the so-called maximally 
supersymmetric or self-dual case, where we expect the theory to be simpler. For this value, $\mb= {\sqrt{2}}$, and the kernel reads
\be
\label{p1k}
\rho(q_1,q_2)=\frac{\sqrt{v(q_1)}\sqrt{v(q_2)}}{2 \cosh \left (\frac{ q_1-q_2}{2\sqrt{2}} \right )}, \qquad \qquad v(z)=\frac{1}{8\sqrt{2} \pi\cosh^2\left(\frac{z}{2\sqrt{2}}\right)}.
\ee
It is easy to compute the functions $\phi_j(q)$ recursively. Here are the first two:
\be
	\label{phiTW}
\ba
	\phi_1(q) &= \frac{\re^{q/{\sqrt{2}}} \left(q/\sqrt{2}-1\right)+1}{2\pi
   \left(\re^{ q/\sqrt{2}}-1\right)^2}, \\
   	\phi_2(q) &= \frac{-2 \re^{ \sqrt{2} q} \left( \left( q^2+\pi^2\right)-2 \sqrt{2}
   q+2\right)+\re^{  q/\sqrt{2} } \left(4 \sqrt{2} q+\pi
   ^2+4\right)+\left(\pi ^2-4\right) \re^{3   q/\sqrt{2}}+4}{32 \pi^2
   \left(\re^{ \sqrt{2}  q}-1\right)^2}.
\ea
\ee
From these functions, we can build the $\kappa$ expansion of $\Phi_{e,o}(x;\kappa)$, which gives a convergent expansion for the functions (\ref{josts}). 
We can then set $\kappa$ to be a zero of the Fredholm determinant, of the form $\kappa= -\re^{E_n}$, and in this way we obtain excellent 
approximations for the normalized eigenfunctions $\varphi_n(q)$, through (\ref{wavef}). 
The results for $n=0,1,2$ are shown Fig. \ref{figwf}. 
We can see that the eigenfunctions display the nodal structure typical of confining potentials, namely, the eigenfunction of the 
$n$-th energy level has $n$ zeroes. In addition, they have definite parity, since 
\be
\varphi_n (-q) =(-1)^n \varphi_n(q). 
\ee
This is expected from the form of the integral kernel. We checked the above results for the 
eigenfunctions by performing a numerical diagonalization of the operator $\mO$, and we explain this procedure in the Appendix \ref{app-deux}. 

\begin{figure}[ht]
\begin{center}
\includegraphics[scale=0.55]{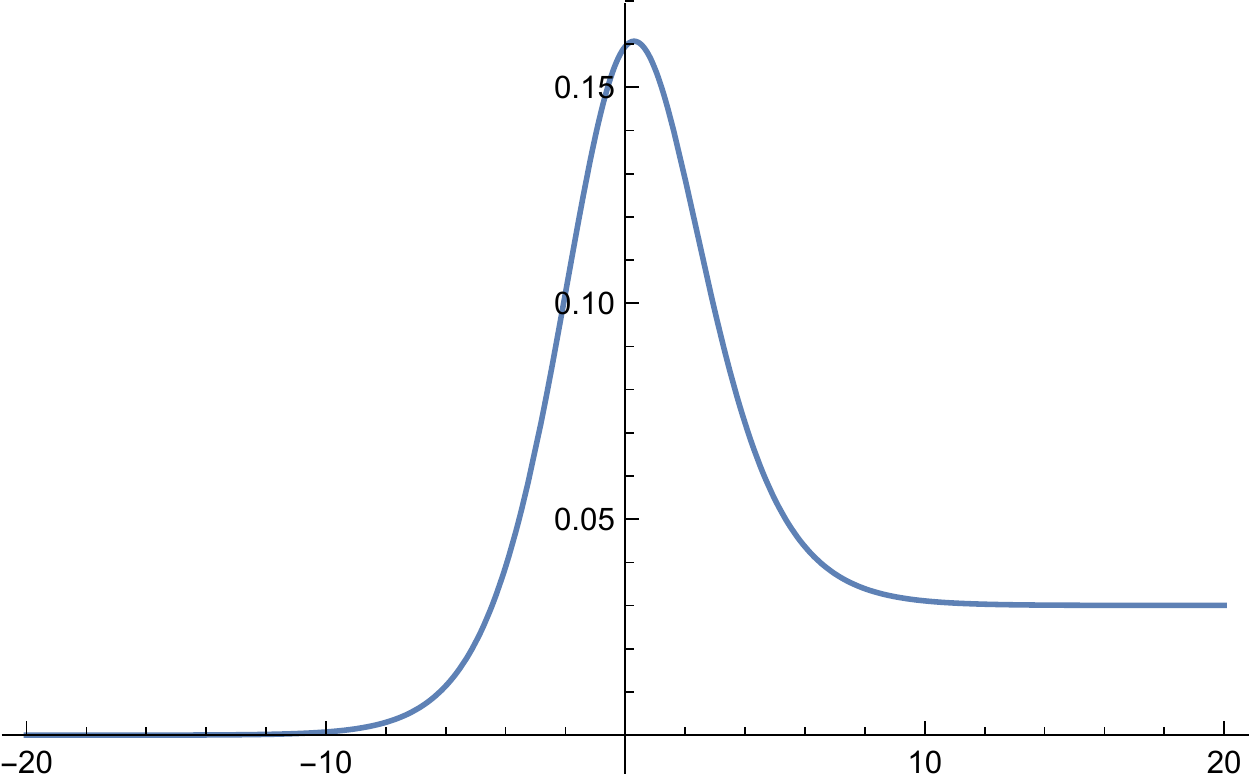}
\caption{The function $\Xi_+(q; \kappa)$ for $\kappa=-\re^{E_0-1/7}$, for local $\IF_0$ with $\hbar=2\pi$ and $m_{\IF_0}=1$. It decays as $q\rightarrow -\infty$ but goes to a 
non-zero constant when $q\rightarrow \infty$, therefore it is not square integrable.}
\label{modwf}
\end{center}
\end{figure}

When $\kappa$ is not a zero of the Fredholm determinant, we obtain functions which are formally in the kernel of the operator $\mO+\kappa$,  
but are not eigenfunctions of $\rho$. In the self-dual case $\hbar=2 \pi$, these functions are not square integrable. The function $\Xi_+(q; \kappa)$, 
for example, decays exponentially as $q\rightarrow -\infty$ for a generic $\kappa$, but will go to a non-zero constant as $q \rightarrow \infty$. An example of 
this behavior is shown in \figref{modwf}, which depicts the function $\Xi_+(q; \kappa)$ when $\kappa=-\re^{E_0-1/7}$. 

\subsection{Canonical transformation of the wavefunctions}

As we saw in the last section, the appropriate coordinates to write the integral kernel are not the same ones 
as the coordinates appearing in the original curve, but are related to them 
by a canonical transformation. The coordinates of the original mirror curve (\ref{spec-curve}) are somewhat special, 
since they lead to topological string amplitudes with a worldsheet instanton interpretation. Only in this set of coordinates we will be able to write 
closed form expressions for the wavefunctions. The rule for going from one set of coordinates to the other is 
particularly simple in the case of linear canonical transformations: it is just an integral transform of the wavefunction. We will now provide a brief review 
of this. Let $x,p$ canonically conjugate coordinates, and $\bar x$, $\bar p$ be the new coordinates 
obtained by a canonical transformation. Let $F(x,\bar x)$ be the generating functional of the transformation, characterized by 
\be
p = {\partial F\over \partial x}, \qquad \bar p= -{\partial F \over \partial \bar x}. 
\ee
Let us suppose that the canonical transformation is linear, 
\be
\begin{pmatrix} \bar x \\ \bar p\end{pmatrix} =\begin{pmatrix} a& b \\ c& d \end{pmatrix} \begin{pmatrix}  x \\  p\end{pmatrix}, \qquad \begin{pmatrix} a& b \\ c& d \end{pmatrix}  \in {\rm SL}(2, \IR), 
\ee
with inverse
\be
\begin{pmatrix}  x \\  p\end{pmatrix}= \begin{pmatrix} d& -b \\ -c& a \end{pmatrix} \begin{pmatrix} \bar x \\ \bar p\end{pmatrix}. 
\ee
The generating functional is then given by 
\be
F(x, \bar x)= -{1\over 2 b} \left( a x^2 + d \bar x^2 -2 x \bar x \right). 
\ee
The generating functional of the inverse transformation is
\be
F^{-1}(\bar x, x)=  {1\over 2 b} \left( a x^2 + d \bar x^2 -2 x \bar x \right)=-F(x, \bar x). 
\ee
Linear canonical transformations are implemented by unitary transformations acting on wavefunctions, 
and one finds (see for example \cite{moshinsky})
\be
\label{can-psi}
\xi(x)=\int_\IR U(x, \bar x) \bar \xi(\bar x)\rd \bar x, \qquad U(x, \bar x)= {1\over {\sqrt{2 \pi \hbar |b|}}} \re^{\ri F(x, \bar x)/\hbar}, 
\ee
where $\xi(x)$, $\bar \xi (\bar x)$ are the wavefunctions in the coordinates $x$, $\bar x$, respectively. 
We are interested in the canonical transformation of the wavefunction $\Xi_+(q; \kappa)$ to the mirror curve coordinates appearing in (\ref{spec-curve}). The coordinates are related by 
\be
	\begin{pmatrix}
	  q \\
	  p 
	 \end{pmatrix}
	  =
	\frac{1}{\sqrt{2}}
	   \begin{pmatrix}
	  1 & -1 \\
	  1 & 1 
	 \end{pmatrix}
	 \begin{pmatrix}
	  x \\
	  y 
	 \end{pmatrix}, 
\ee
and the wavefunction in the $x$-variable is given by 
 \be
 \label{psidef}
 \psi(x; \kappa) = \int_\IR U(x, q) \Xi_+(q; \kappa) \rd q, 
 \ee
 where 
\be
U(x,q)=\frac{1}{{\sqrt {2\pi \hbar}} \, 2^{-1/4}} \, {\rm exp}\left ( \frac{\ri}{2\hbar \, 2^{-1/2}}\left(2^{-1/2}q^2-2xq+2^{-1/2}x^2\right) \right ). 
\ee
We can write
\be
\label{psixn}
\psi(x; \kappa)= \sum_{N=0}^\infty \psi_N(x) \kappa^N, 
\ee
where
\be
\label{psin}
\psi_N(x) =\int_{-\infty}^{\infty}  U(x,q) \sqrt{E\left(q \right)} \Psi_N(q) \rd q.  
\ee
We will now give some concrete results for this canonical transformation in the case $\hbar=2 \pi$ and $m_{\IF_0}=1$. These results will be crucial in order 
to compare with the predictions of topological string theory. Note first that the integral transform (\ref{can-psi}) 
implements a unitary transformation which acts in principle on the Hilbert space $\CH= L^2(\IR)$. However, in (\ref{psin}) it acts on functions which 
are not square integrable (they decrease exponentially as $q\rightarrow -\infty$, but they go to a constant as $q \rightarrow \infty$). Fortunately, in this case, the integrals in the 
r.h.s. of (\ref{psin}) are well defined without further ado. Moreover, they can be computed analytically by using the results of \cite{gk}. The reason is that the integrands in the r.h.s. of (\ref{psin}) 
are given as sums
\be
	\sum_{\ell=0}^N f_\ell(q) q^\ell
\ee
where the $f_\ell(q)$ are quasi-periodic functions. Therefore, the integration in (\ref{psin}) can be done by using Lemma 2.1 of \cite{gk}, and it reduces to a residue computation. 
The terms $f_\ell(q) q^\ell$ with $\ell>0$ are not quasi-periodic because of the power $q^\ell$, but they can be generated from quasi periodic integrals by using that
\be
	q^\ell U(x,q)= \left [\frac{1}{\sqrt{2}} \left (2 \pi \ri \frac{\partial}{\partial x} + x \right ) \right ]^\ell U(x,q).
\ee
The result of the integration has the following structure, 
\be
\label{two-psix}
\psi_N(x)=\re^{-{\ri x^2\over 4 \pi}} \psi^{(-)}_N(x) + \re^{{\ri x^2\over 4 \pi}} \psi_N^{(+)}(x). 
\ee
The first term is due to residues at the $x$-dependent poles of the integrand in (\ref{psin}), while the second term is due to the $x$-independent poles. As we will see, 
the two contributions correspond to two different saddle-points of the integral transform in the 't Hooft limit, and they will be given by two different copies of the topological 
string wavefunction. One finds, for example, 
\be
\label{psix-ex}
\ba
\psi_0^{(-)}(x)&={\re^{\pi \ri /4}} {\re^x \left(\re^x-\ri\right) \over \sqrt{2\pi}(\re^{2 x}-1)}, \\
\psi_1^{(-)}(x)&={ \re^{\pi \ri /4} \over 4\sqrt{2} \pi^{3/2}  \left(\re^{2 x}-1\right)^3} \re^x \bigl(\re^{2 x} (-6 \pi +2 \ri (x-1))+\re^{5
   x}+\re^{4 x} (-2 \ri x-2 \pi +\ri)\\
   & \qquad+\re^x (2 x+4 \ri \pi +1)+\re^{3 x} (-2 (x+1)+4 \ri \pi
   )+\ri\bigr), \\
  \psi_0^{(+)}(x)&= -\frac{ \re^x}{\sqrt{\pi}(\re^{2 x}-1)},\\
  \psi_1^{(+)}(x)&=\frac{\re^x \left(\pi  \left(6 \re^x+10 \re^{3 x}+\ri \re^{4 x}-\ri\right)-4 \ri \re^x \left(\re^{2
   x}-1\right) x\right)}{8 \pi^{3/2}  \left(\re^{2 x}-1\right)^3},
   \ea
   \ee
and so on. Note that each $\psi_N^{(\pm)}(x)$ is singular at $x=0$, but $\psi_N(x)$ turns out to be an {\it entire} function on the full complex $x$-plane, for 
all $N \ge 0$.  The behavior at infinity of the function $\psi(x; \kappa)$ 
is similar to the one of $\Xi_+(q, \kappa)$: for generic, real $\kappa$, it decreases exponentially as $x \rightarrow -\infty$, but it is an oscillatory function as $x \rightarrow \infty$. 
When $\kappa$ is a zero of the Fredholm determinant, $\psi(x;\kappa)$ is square integrable, and an eigenfunction of the operator $\rho$, in the $x$ representation. 
Let us finally note that the canonical transformation of $\Xi_-(q, \kappa)$ can be easily deduced from the one of $\Xi_+(q, \kappa)$.

\subsection{Kernels of more general form}

Many steps of the previous sections relied on the fact the the integral kernel of $\rho$ takes the form (\ref{rhov}). In this section we will consider a generalization given by
\be
	\label{kernelgen}
	\rho(x,y)=\frac{\sqrt{v(x)}  \overline{\sqrt{v(y)}} }{2 \cosh \left ( \frac{x-y}{2\xi}+\ri \pi C \right )}=\frac{\sqrt{E(x)}  \overline{\sqrt{E(y)}} }{\alpha M(x)+\alpha^{-1}M(y)}, \qquad \qquad \alpha=\re^{\ri \pi C},
\ee
where $C$ is a rational number in the complement of $\mathbb Z+\frac{1}{2}$, and $E(x)$, $M(x)$ are defined as in (\ref{defEM}). Many kernels coming from the quantization of mirror curves (\ref{ex-W}) take such form \cite{kasm}. For example, the mirror curve of local $\mathbb P^2$ leads to the case where $C=\frac{1}{6}$, and the mirror curve of the canonical bundle over the weighted projective space $\mathbb P(m,n,1)$ leads to 
\be
C=\frac{m-n+1}{2(m+n+1)}.
\ee
These are the cases that give what are called ``three term operators" in \cite{kasm}, 
and they arise as degeneration limits of more complicated mirror curves. The case which was studied in the preceding sections is retrieved by 
setting $C=0$, and requiring the function $v(x)$ to be real. We will see now that many of the above considerations can be extended to this more general case.

The manipulations using the Cauchy determinant can be generalized easily. For the numerator of the resolvent, we obtain
\be 
D(x, t; \kappa) = \rho(x,t) \sum_{N\ge 0} \llangle[\Biggl] \prod_{i=1}^N \frac{\sinh \left( {x- x_i \over 2 \xi } \right)}{ \cosh\left( {\frac{x- x_i}{ 2\xi}+\ri \pi C  } \right) } \frac{\sinh\left( {t- x_i \over 2 \xi } \right)}{ \cosh\left( {\frac{t- x_i}{ 2\xi}-\ri \pi C } \right) } \rrangle[\Biggr] \kappa^N,
\ee
where the unnormalized expectation value is now given by
\be
\label{genkermm}
\llangle f(x_1, \cdots, x_N) \rrangle= {1\over N!} \int  \rd ^N x\,  f(x_1, \cdots, x_N) \prod_{i=1}^N |v(x_i)| {\prod_{i<j} \left[ 2 \sinh \left( {x_i - x_j \over 2 \xi } \right)\right]^2 
\over \prod_{i,j} 2 \cosh\left( { \frac{x_i - x_j }{ 2  \xi}+\ri \pi C } \right)},
\ee
Similarly as before, we also define
\be
	\label{phiNgen}
	\Psi_N(x)= \llangle[\biggl]  \prod_{i=1}^N \frac{\sinh \left( {x- x_i \over 2 \xi } \right)}{ \cosh\left( {\frac{x- x_i}{ 2\xi}+\ri \pi C  } \right) } \rrangle[\biggr].
\ee
and
\be
\ba
	\Xi_{+}(x;\kappa) &=   \sqrt{E(x)} \sum_{N=0}^{\infty}  (\alpha \kappa)^N \Psi_N(x), \\
	\Xi_{-}(x ;\kappa) &= \frac{\sqrt{E(x)}}{M(x)} \sum_{N=0}^{\infty} (-\alpha^{-1} \kappa)^N \Psi_N(x).
\ea
\ee
With these definitions, all the results at the end of section (\ref{sectFredholm}) are straightforwardly generalized. In particular, we have that
\be
\ba
	\label{Dlimitgen}
	D(x,t;\kappa) &\approx  \overline{\sqrt{v(t) \over M(t)}} \alpha \Xi_{+}(x), \qquad t \rightarrow \infty,\\
	D(x,t;\kappa) &\approx \overline{\sqrt{v(t) M(t)} } \alpha^{-1} \Xi_{-}(x), \qquad t \rightarrow \infty.
\ea
\ee
 The Tracy-Widom lemma can be partially generalized for these more general kernels. We can show for example (as done in \cite{oz}) that the $\ell^{\rm th}$ power of the kernel can be written as
\be
	\rho^{\ell}(x,y)=\alpha^{-1} \frac{ \sqrt{E(x)} \overline{\sqrt{E(y)}}}{ M(x)-\omega^\ell M(y)} \sum_{k=0}^{\ell-1} \omega^k \phi_{k}(x)\overline{\phi_{\ell-k-1}(y)},
\ee
where
\be
	\omega=-\alpha^{-2},
\ee
and where the functions $\phi_{k}(x)$ are given by the usual recursion
\be
	| \phi_{k} \rangle = \frac{1}{\sqrt{E}} \rho \sqrt{E} | \phi_{k-1} \rangle, \qquad \qquad \phi_0(x)=1.
\ee
As for the case in the previous section, it can be seen that the limiting behaviour of $\phi_{k}(x)$ for $k>0$ are given by
\be
\ba
	&\phi_k(x) \rightarrow 0, \qquad \qquad x \rightarrow \infty, \\
	&\phi_k(x) \rightarrow c_k, \qquad \qquad x \rightarrow -\infty.
\ea
\ee
Since the resolvent can be written in terms of powers of $\rho$, we have the alternative expression for $D(x,t;\kappa)$:
\be
	D(x,t;\kappa)=\Xi(\kappa)R(x,t,\kappa)=\Xi(\kappa) \sum_{\ell=0}^{\infty}(-\kappa)^{\ell} \rho^{\ell+1}(x,t).
\ee
%
Using the limiting behaviour of $\rho(x,t)$ when $t \rightarrow \pm \infty$ and comparing with (\ref{Dlimitgen}), we obtain
\be
\ba
\Xi_{+}(x;\kappa) &=   \Xi(\kappa) \sqrt{E(x)} \sum_{\ell=0}^{\infty}  (- \kappa)^\ell \phi_\ell(x), \\
	\Xi_{-}(x ;\kappa) &= \Xi(\kappa) \frac{\sqrt{E(x)}}{M(x)} \sum_{\ell=0}^{\infty} (-\kappa)^\ell \sum_{k=0}^{\ell} \omega^{k} \bar c_{\ell-k} \phi_k(x).
\ea
\ee
These functions satisfy the difference equation given by $\sf O$. This is easily checked for $\Xi_{+}(x;\kappa)$, since by construction we have
\be
	\rho \sqrt{E} | \phi_k \rangle = \sqrt{E}  | \phi_{k+1} \rangle.
\ee
It is then immediately verified that
\be
	{\sf O} \sqrt{E} | \phi_{k+1} \rangle = \sqrt{E}  | \phi_{k} \rangle, \qquad \qquad {\sf O} \sqrt{E} | \phi_{0} \rangle = 0,
\ee
which implies that $\Xi_+(x;\kappa)$ satisfies the difference equation (for any $\kappa$)
\be
	({\sf O}+\kappa) | \Xi_+ \rangle =0.
\ee
The functions $\Xi_\pm(x)$ play the same essential role as in the previous section: 
when the parameter $\kappa$ is chosen such that the Fredholm determinant $\Xi(\kappa)$ vanishes, 
then $\Xi_\pm(x;\kappa)$ become the (unnormalized) square integrable eigenfunctions of the operator $\rho$. In the next section, we will perform some 
computations for local $\mathbb P^2$ in the 't Hooft limit.

\sectiono{The 't Hooft limit of eigenfunctions and open topological strings}

\subsection{General considerations}

One of the main results of \cite{ghm} is that the conventional topological string emerges in a 't Hooft-like limit of the spectral problem, corresponding 
to the strongly coupled regime $\hbar \gg 1$. This is somewhat surprising, since in \cite{ns,mirmor,acdkv} the spectral problem has been related to the NS limit of the refined 
topological string. There is of course no contradiction between these two statements, since the NS string appears in the weakly coupled regime 
$\hbar \ll 1$. The contribution of the standard topological string is crucial in order to obtain exact results. 

The emergence of the conventional topological string is particularly clear when one considers the spectral traces $Z(N)$. To understand this in some detail, we will 
briefly review some results of \cite{ghm,mz}, by restricting ourselves to mirror curves of genus one. The spectral information of the operator $\rho$ is conjecturally encoded in the total 
grand potential associated to the CY. This is the sum of two pieces, 
\be
\label{jtotal}
\mathsf{J} (\mu,\hbar) = \mathsf{J}^{\rm WKB} (\mu, \hbar)+ \mathsf{J}^{\rm WS} (\mu,  \hbar), 
\ee
where we have introduced the chemical potential $\mu$, which is related to the modulus $\kappa$ by
\be
\kappa=\re^\mu. 
\ee
The first piece in (\ref{jtotal}) is the WKB grand potential, which can be obtained by resumming the WKB expansion. The results of 
\cite{acdkv,hkrs} show that this resummation can be performed 
in terms of BPS invariants of the topological string. More explicitly, let us consider the NS limit of the refined topological string free energy, which is given by 
\be
\label{NS-j}
\ba
F^{\rm NS}({\bf t}, \hbar) &={1\over 6 \hbar} \sum_{i,j,k=1}^s a_{ijk} t_i t_j t_k +\sum_{i=1}^s b^{\rm NS}_i t_i \hbar \\
& +\sum_{j_L, j_R} \sum_{w, {\bf d} } 
N^{{\bf d}}_{j_L, j_R}  \frac{\sin\frac{\hbar w}{2}(2j_L+1)\sin\frac{\hbar w}{2}(2j_R+1)}{2 w^2 \sin^3\frac{\hbar w}{2}} \re^{-w {\bf d}\cdot{\bf  t}}. 
\ea
\ee
In this equation, the coefficients $a_{ijk}$ are the standard cubic couplings of the classical prepotential of the CY, 
the coefficients $b_i^{\rm NS}$ are linear couplings at the next-to-leading order in $\hbar$, and  $N^{\bf d}_{j_L, j_R}$ are the refined BPS invariants (see 
for example \cite{mmrev} for more details). Then, the WKB grand potential is given by 
\be
\label{jm2}
\ba
\mathsf{J}^{\rm WKB}_S (\mu, \boldsymbol{\xi}, \hbar)&=\sum_{i=1}^s  {t_i(\hbar) \over 2 \pi}   {\partial F^{\rm NS}({\bf t}(\hbar), \hbar) \over \partial t_i} 
+{\hbar^2 \over 2 \pi} {\partial \over \partial \hbar} \left(  {F^{\rm NS}({\bf t}(\hbar), \hbar) \over \hbar} \right)\\
&+  \sum_{i=1}^s  {2 \pi \over \hbar} b_i t_i(\hbar) + A({\boldsymbol \xi}, \hbar), 
\ea
\ee
where ${\bf t} (\hbar)$ is the quantum mirror map of \cite{acdkv}, and $A({\boldsymbol \xi}, \hbar)$ is a constant. Since $\kappa$ is identified with a complex moduli of the mirror curve, 
the quantum mirror map contains 
the dependence on $\mu$. 
The second piece in (\ref{jtotal}), or ``worldsheet" grand potential, contains the information of the conventional topological string. 
It involves the resummation of worldsheet instantons provided by Gopakumar and Vafa \cite{gv}, which leads to the 
generating functional 
\be
\label{GVgf}
F^{\rm GV}\left({\bf t}, g_s\right)=\sum_{g\ge 0} \sum_{\bf d} \sum_{w=1}^\infty {1\over w} n_g^{ {\bf d}} \left(2 \sin { w g_s \over 2} \right)^{2g-2} \re^{-w {\bf d} \cdot {\bf t}}.  
\ee
In this equation, ${\bf t}$ is the vector of K\"ahler parameters of the CY, $g_s$ is the topological string coupling constant, 
and $n_g^{{\bf d}}$ are the Gopakumar--Vafa invariants of genus $g$ and with multi-degree ${\bf d}$. The ``worldsheet" grand potential is then given by 
\be
\label{jws}
\mathsf{J}^{\rm WS}(\mu,   \hbar)= F^{\rm GV}\left( {2 \pi \over \hbar}{\bf t}(\hbar)+ \pi \ri {\bf B}, {4 \pi^2 \over \hbar} \right).
\ee
The shift in the first argument involves ${\bf B}$, a constant vector (``B-field") which depends on the geometry under consideration (see \cite{ghm,cgm,mmrev} for more details). 
Note that the conventional string coupling constant is identified with the {\it dual} Planck constant,
\be
\label{hdual}
\hbar_{\rm D}={4 \pi^2 \over \hbar}. 
\ee
Therefore, the weakly coupled topological string corresponds to the strongly coupled quantum mechanical problem.
The conjecture of \cite{ghm} states that the Fredholm determinant of $\rho$, defined in (\ref{fred-det}), is given by 
\be
\label{spec-det}
\Xi (\kappa)= \sum_{n \in \IZ} \exp\left(  \mathsf{J} (\mu + 2 \pi \im n, \hbar) \right).  
\ee
This leads to the following exact expression for the fermionic spectral traces, 
\be
\label{int-Z}
Z(N) = {1\over 2 \pi \ri} \int_{\mathcal C} \re^{\mJ (\mu,  \hbar)- N \mu} \rd \mu, 
\ee
where ${\mathcal C}$ is a contour going from $\re^{-\im \pi/3} \infty$ to $\re^{\im \pi/3} \infty$ (this is the standard contour for 
the integral representation of the Airy function). 

As explained in detail in \cite{mz}, one can now consider the 't Hooft limit 
\be
\label{th1}
N \rightarrow \infty, \qquad \hbar \rightarrow \infty,
\ee
in which the 't Hooft parameter 
\be
\label{th2}
\lambda={N \over \hbar} 
\ee
remains fixed. In this limit, only the worldsheet grand potential contributes. It follows from \cite{abk} that the integral in (\ref{int-Z}) implements a change of frame, to the so-called 
conifold frame. We conclude that the fermionic spectral trace has the asymptotic expansion \cite{mz}
\be
\label{thooft}
\log Z(N) \sim \sum_{g=0}^\infty F_g(\lambda) \hbar^{2-2g}, 
\ee
where $F_g(\lambda)$ is the topological string free energy in the conifold frame, and $\lambda$ is proportional to the corresponding flat conifold 
coordinate (see \cite{mz,kmz} for details and examples). All the statements above can be generalized to higher genus mirror curves \cite{cgm}. The 't Hooft-like expansion 
(\ref{thooft}) is indeed very natural if one considers the matrix model representation of the spectral traces presented in \cite{mz,kmz} (for integral kernels of the form (\ref{rhov}), this is written down in (\ref{znmm})). 

In the case of eigenfunctions, the generalization of the fermionic spectral trace is the function (\ref{psiN}), which is the building block of the $\Xi_{\pm}(x; \kappa)$. Since the function 
$\Psi_N(x)$ is defined by a matrix integral representation, we can study its 't Hooft limit by using standard tools in large $N$ matrix models. Note that $\Psi_N(x)$ is the average value of 
a determinant-like expression. It plays, in the $O(2)$ matrix model of (\ref{znmm}), the same r\^ole that the average
\be
\label{detx}
\langle {\rm det}(x- M) \rangle
\ee
plays in conventional, Hermitian matrix models. It has been pointed out many times in the literature \cite{adkmv,mmss, cpw,em,be,gs, awata} 
that this average defines a (D-brane) wavefunction. Therefore, it is not 
surprising that the 't Hooft limit of $\Psi_N(x)$ is also related to the topological string wavefunction. Let us recall how this wavefunction is 
constructed, in the general setting of the topological recursion associated to a parametrized spectral curve $y=y(X)$. 
Given such a curve, one can construct an infinite sequence of meromorphic differentials \cite{eynard-co,eo}
\be
W_{g,h} (X_1, \cdots, X_h) \rd X_1 \cdots \rd X_h, \qquad g\ge 0, \quad h\ge 1. 
\ee
In the case $g=0$, $h=1$ (the ``disk" amplitude) one has 
\be
W_{0,1}(X) \rd X= y(X) \rd X, 
\ee
while the case $g=0$, $h=2$ (the ``annulus" amplitude) is essentially given by the Bergmann kernel of the curve $B(X_1, X_2)$, 
\be
W_{0,2}(X_1, X_2) \rd X_1 \rd X_2= B(X_1,X_2) \rd X_1 \rd X_2 - {\rd X_1 \rd X_2 \over (X_1-X_2)^2}. 
\ee
(Explicit expressions for the annulus amplitude will be presented below). When the spectral curve is the mirror curve of a toric CY, the above meromorphic differentials 
calculate open topological string amplitudes \cite{mmopen,bkmp} associated to the toric branes introduced in \cite{av,akv}. The topological string wavefunction, which we will denote by $\psi_{\rm top}(X, {\bf t}, g_s)$, is defined by the 
following asymptotic expansion around $g_s=0$:
\be
\label{psitop}
\psi_{\rm top}(X, {\bf t}, g_s)\sim \exp \left[ \sum_{g=0}^\infty \sum_{h =1}^\infty {(-\ri g_s)^{2g-2+h} \over h!} \int^X \cdots \int^X W_{g,h} (X_1, \cdots, X_h) \rd X_1 \cdots \rd X_h \right]. 
\ee
The ${\bf t}$ are flat coordinates for the moduli of the curve. In the case of convergent, Hermitian matrix models, (\ref{psitop}) gives the 't Hooft expansion of the vev (\ref{detx}). 
For general spectral curves, (\ref{psitop}) defines a formal asymptotic expansion. 


\subsection{A matrix model calculation}

We will now study the 't Hooft limit of $\Psi_N(q)$ in the case of local $\IP^1 \times \IP^1$. As we will see, 
the original coordinates $x,q$ appearing in (\ref{xyq}) are not the right scaling variables. 
Therefore, for the purposes of this section, we will relabel the original coordinates $q$, $p$ as $q_m$, $p_m$ (where the subscript refers to mirror curve), and we will introduce a new coordinate $q$ 
defined by
\be
	\label{tHooftscale}
q={2 \pi \over \hbar} q_m.  
\ee
 Let us denote by $\CQ$ the matrix with eigenvalues $q_1, \cdots, q_N$. We then have, 
\be
\label{cum}
\ba
{\Psi_N(q_m) \over Z(N)} &=
 \left\langle {\rm det} \left(\tanh\left( {q - \CQ \over 2 {\sqrt{2}} } \right)\right)\right \rangle= \left\langle \exp \left( \tr \log \left(\tanh\left( { q- \CQ \over 2{\sqrt{2}} }\right) \right) \right) \right \rangle\\
&= 
\exp \left[ \sum_{s=1}^\infty {1\over s!} \left\langle \left(\tr \log\left( \tanh \left( { q-\CQ\over 2{\sqrt{2}} }\right)\right) \right)^s \right\rangle^{(c)} \right]. 
\ea
\ee
In going to the last line, we have just used the definition of connected vevs. We now note that 
\be
\label{logt}
\log \left( \tanh\left( {q- q_i \over 2{\sqrt{2}}  } \right) \right) = \log \left( 1- Q^{-1} z_i \right) -  \log \left( 1+Q^{-1} z_i \right) , 
\ee
where
\be
z_i =\re^{q_i/{\sqrt{2}} }, \qquad Q= \re^{q/{\sqrt{2}}}. 
\ee
We will also introduce a canonically conjugate coordinate $p$, in such a way that the corresponding quantum operators satisfy the commutation 
relations
\be
[\mq, \mm]=\ri \hbar_{\rm D}.  
\ee
The dual Planck constant is given in (\ref{hdual}). We will also denote
\be
P=\re^{p/{\sqrt{2}}}.
\ee
We will also need the original coordinates for the mirror curve (\ref{spec-curve}). In this section, these coordinates will be denoted 
by $x_m$, $y_m$, and we will reserve the notation $x$, $y$ for the canonically conjugate coordinates related to $q$, $p$ by the linear canonical transformation,  
\be
\label{sym-trans}
\begin{pmatrix} x \\y \end{pmatrix}= {1\over {\sqrt{2}}} \begin{pmatrix} 1 & 1 \\ -1& 1 \end{pmatrix} \begin{pmatrix} q \\ p\end{pmatrix}. 
\ee
By recalling the relation (\ref{xyq}), we find that the original coordinates $x_m$, $y_m$ of the mirror curve are related to $x, y$ by  
\be
 x_m={\hbar \over 2 \pi}  x, \qquad y_m= {\hbar \over 2 \pi} y. 
  \ee
In taking the 't Hooft limit, we have to keep $q$, or equivalently, $Q$, fixed. We conclude that, in the 't Hooft limit, the open string moduli $x_m, y_m$ have to be taken to infinity, in such a way that
  \be
  {x_m \over \hbar}, \qquad {y_m \over \hbar}
  \ee
are kept fixed. 
  
We can now give a formula for (\ref{cum}) in terms of matrix model vevs. If we expand (\ref{logt}) in a series in $1/Q$, we find
\be
\tr \log \left( \tanh\left( {q- q_i \over 2 {\sqrt{2}} } \right) \right)= -2 \sum_{k=0}^\infty {1\over 2k+1} \tr M^{2k+1}Q^{-2k-1} , 
\ee
where $M$ is the diagonal matrix with entries $z_1, \cdots, z_N$. We then obtain the representation, 
\be
\label{psinm}
\ba
&\log \left({\Psi_N(q_m) \over Z(N)} \right) =\\
& \sum_{s=1}^\infty {(-2)^s \over  s!} \sum_{k_1, \cdots, k_s=0}^\infty {1\over (2k_1+1) \cdots (2k_s+1)}\langle  \tr M^{2k_1+1} \cdots  \tr M^{2k_s+1} \rangle^{(c)} 
Q^{-2(k_1+\cdots +k_s)-1},  
\ea
\ee
in terms of connected vevs in the matrix model (\ref{znmm}). In the case of local $\IF_0$, this matrix model was analyzed in detail in \cite{kmz}. The partition 
function (\ref{znmm}) can be written as (we set $m_{\IF_0}=1$ for simplicity)
\be
\label{zf0-bis}
Z (N)={1\over N!} 
 \int_{\IR^N}  { \rd^N u \over (2 \pi)^N}  \prod_{i=1}^N \re^{-\hbar  V(u_i, \hbar )}  \frac{\prod_{i<j} 4 \sinh \left( {u_i-u_j \over 2} \right)^2}{\prod_{i,j} 2 \cosh \left( {u_i -u_j \over 2}   \right)}, 
\ee
where the potential has an asymptotic expansion at $\hbar $ large given by 
\be
\label{vug}
V(u, \hbar)= \sum_{\ell \ge 0} \hbar^{-2\ell} V^{(\ell)} (u), 
\ee
and the leading contribution as $\hbar \rightarrow \infty$ is given by the``classical" potential,  
\be
\label{vo}
V^{(0)}(u)=-\frac{u}{2 \pi}-\frac{2}{\pi^2}  {\rm Im} \, {\rm Li}_2(-\ri \, \re^{u}). 
\ee
The variable $u$ is related to $q$ by the simple rescaling $u=q/{\sqrt{2}}$. The matrix model (\ref{zf0-bis}) is an $O(2)$ matrix model, and in particular it 
has a standard 't Hooft expansion in the 't Hooft limit (\ref{th1}), (\ref{th2}). The only difference 
with the conventional $O(2)$ matrix model analyzed e.g. in \cite{ek} is that, here, the potential itself depends on $\hbar$. This introduces additional corrections but does not 
change the structure of the $1/N$ expansion. In particular, it follows from large $N$ counting that the connected vev in the $s$-th term of (\ref{psinm}) 
goes like $\hbar^{2-s}$ as $\hbar \rightarrow \infty$. In this way, we can reorganize (\ref{psinm}) in terms of the 't Hooft expansion of the correlators of the matrix model (\ref{zf0-bis}). 

As shown in (\ref{josts}), the wavefunction $\Xi_+(q_m; \kappa)$ involves $\Psi_N(q_m)$ and an overall factor of ${\sqrt{E(q_m)}}$. Let us then 
consider the 't Hooft expansion of the function, 
\be
\label{as-thooft}
\log \left({ {\sqrt{E(q_m)}} \Psi_N(q_m) \over Z(N)} \right)  \sim \sum_{n=0}^\infty \left(-\ri \hbar_{\rm D} \right)^{n-1} \CT_n(q). 
\ee
It is easy to see that the first two terms, $\CT_0(q)$ and $\CT_1(q)$, can be written in terms of planar correlators of the 
$O(2)$ matrix model (\ref{zf0-bis}) with potential $V^{(0)}(u)$. 
Indeed, we note that 
\be
-\sum_{k=0}^\infty {1\over 2k+1}\langle  \tr M^{2k+1} \rangle Q^{-2k-1}=N \int^{Q}_{\infty} \omega_+(Q') \rd Q',
\ee
where
\be
\omega_+(Q) ={1\over 2 N} \left\langle \tr \left( {1\over Q-M} -{1\over Q+M} \right) \right\rangle
\ee
is the even part of the resolvent of the $O(2)$ matrix model studied in \cite{ek}. 
At leading order in $\hbar_{\rm D}$ we have to consider just its planar limit,  
and we conclude that
\be
\label{lead-T}
 \ri \CT_0(q)=  -2 \pi^2 V^{(0)}\left({ q \over {\sqrt{2}}} \right)+ 8 \pi^2  \lambda \int^Q_\infty \omega^0_+(Q') \rd Q'. 
\ee
Similarly, the term with $s=2$ in (\ref{cum}) is given by 
\be
\ba
&2 \sum_{k,l\ge0} {1\over (2k+1)(2l+1)} \langle\tr M^{2k+1} \tr M^{2 l +1}\rangle^{(c)} Q^{-2k-2l-2}\\
& \qquad = 2 \int^Q_{\infty} \int^Q_{\infty} W_{++}(Q_1, Q_2) \rd Q_1 \rd Q_2, 
\ea
\ee
where $W_{+ +}(Q_1 Q_2)$ is the even part of the two-point correlator, which was also considered in \cite{ek}. The subleading order in $\hbar_{\rm D}$ involves only 
the planar limit $W^0_{+ +}(P, Q)$, and we conclude that
\be
\label{s1q}
\CT_1(q)={1\over 2} \log Q+ 2 \int^Q_{\infty} \int^Q_{\infty}  W^0_{++}(Q_1, Q_2) \rd Q_1 \rd Q_2. 
\ee
Both quantities, $\CT_{0,1}(q)$, can be written as exact functions of $\lambda$ and $q$ by using the results of \cite{ek}. The planar resolvent 
$\omega^0_+(Q)$ was explicitly computed in \cite{kmz} in terms of elliptic integrals. It turns out however that the result in \cite{kmz} can be simplified to an 
expression involving essentially the spectral curve of local $\IF_0$, written in the variables $Q$, $P$. This curve reads, 
\be
\label{QPspec}
(Q+ 1/Q)(P +1/P)=\kappa. 
\ee
We can solve for $P$ as
\be
\log P(Q)=\log\left\{ {1\over 2(Q+ Q^{-1}) } \left[\kappa +{\sqrt{\kappa^2 -4 (Q+Q^{-1})^2}} \right] \right\}.
\ee
Then, one can show that 
\be
 {\rd \over \rd Q} \left\{ -{1\over 2} V^{(0)}(Q) + 2 \lambda \int^Q_\infty  \omega^0_+(Q') \rd Q' \right\}={\ri \over 2 \pi^2 Q} \log P(Q), 
\ee
and we conclude that
\be
\label{s0q}
\CT_0(q)= \int^q  p(q') \rd q'. 
\ee
This depends on the flat conifold coordinate $\lambda$ through the modulus $\kappa$. They are related by the conifold mirror map (see \cite{kmz} for details)
\be
\kappa= 4+ 8 \pi^2 \lambda+ 4 \pi^2 \lambda^2+{2 \pi^6 \lambda^3\over 3}+\cdots. 
\ee
An explicit expression for $W^0_{++}(p,q)$ is given in \cite{ek}, in terms of the endpoints of the cut $a,b$\footnote{The first minus sign of the second line is missing in \cite{ek}.}: 
\be
\ba
\label{W0plusplus}
W_{++}^0(p,q)=& \frac{1}{8\sqrt{(p^2-a^2)(p^2-b^2)}\sqrt{(q^2-a^2)(q^2-b^2)}} \left [  a^2+b^2-2b^2 \frac{E \left (1-\frac{a^2}{b^2} \right ) } {K \left (1-\frac{a^2}{b^2}  \right ) } \right . \\
&  \qquad \left.  -(p^2+q^2) \left (1-\left (\frac{\sqrt{(p^2-a^2)(p^2-b^2)} - \sqrt{(q^2-a^2)(q^2-b^2)}}{p^2-q^2}\right)^2 \right ) \right ].
\ea
\ee
Here, $p$ and $q$ are generic arguments, and $K(k^2)$ and $E(k^2)$ are elliptic functions of the first and second kind, respectively. 
In our case, the endpoints of the cut can be read from the discriminant of the curve (\ref{QPspec}):
\be
\label{disq}
\sigma_q(Q)=(Q^2-a^2)(Q^2-b^2)=(Q^2+1)^2-{\kappa^2 Q^2 \over 4}, 
\ee
and by symmetry of the potential one has 
\be
\label{apointq}
a= {1\over b}= \frac{1}{4} \left(\sqrt{\kappa ^2-16}+\kappa \right). 
\ee
It turns out that this planar two-point function can be written in terms of the annulus amplitude appearing in the standard 
topological recursion, i.e. in terms of the Bergmann kernel, as follows:
\be
\label{symannulus}
2W^0_{++}(Q_1, Q_2)={1\over 4}\left( W_0(Q_1,Q_2)+ W_0(-Q_1,Q_2)+W_0(Q_1,-Q_2)+W_0(-Q_1,-Q_2) \right), 
\ee
where $W_0(Q_1,Q_2)$ is given by Akemann's formula \cite{gernot}
\be
\label{twocutannulus}
\ba
W_0(p,q)&={1\over 4(p - q)^2} \left(   {\sqrt  {(p - x_1) (p - x_4)(q -x_2) (q - x_3) \over (p - x_2) (p - x_3)(q - x_1) (q - 
                      x_4)}} +   {\sqrt  {(p - x_2) (p - x_3)(q -x_1) (q - x_4) \over (p - x_1) (p - x_4)(q - x_2) (q - 
                      x_3)}} \right) \\ & 
                      + {(x_1-x_3)(x_2-x_4) \over 4 {\sqrt{\prod_{i=1}^4 (p-x_i)(q-x_i)}}}  {E(k^2) \over K(k^2)}  - {1\over 2
                      (p - q)^2},
                      \ea
\ee
and the modulus of the elliptic integrals is given by 
\be
\label{kpar}
k^2={(x_1-x_4)(x_2-x_3) \over (x_1-x_3)(x_2-x_4)}.
\ee
The branch points of $W_0(Q_1,Q_2)$ in (\ref{symannulus}) are given by 
\be
\{x_1, x_2, x_3, x_4\}=\{a, -a, -a^{-1}, a^{-1} \}. 
\ee
We have verified the results provided by these formulae against a perturbative computation at small $\lambda$, as in \cite{mz,kmz}. It is possible to compute higher order corrections 
in $\hbar_{\rm D}$, but in this case one has to take into account the corrections to the classical potential $V^{(0)}(u)$.

\subsection{Canonical transformation and open topological strings}

The results above for $\CT_0(q)$, $\CT_1(q)$ are very similar to the first two terms in the asymptotic expansion of a topological string wavefunction, as obtained from the curve 
(\ref{QPspec}). However, there are some differences, since for example the next-to-leading term does not involve the annulus amplitude, but the symmetrized version 
thereof (\ref{symannulus}). At the same time, the coordinates in which we are doing this calculation are not the ones in which we have an enumerative interpretation. 
For this, we have to go to the large radius coordinates, which are related to $p, q$ by (\ref{sym-trans}). In these coordinates, the curve (\ref{QPspec}) reads
  \be
  \label{ycurve}
 X+{1\over X} + Y  + {1\over Y}= \kappa, 
  \ee
  where 
  \be
  X=\re^x, \qquad  Y=\re^y.
  \ee
  The curve (\ref{ycurve}) is identical to (\ref{spec-curve}) except that $\kappa$ has the opposite sign. 
  This is equivalent to changing the sign in $X$ and $Y$. We will now give evidence that, after a canonical transformation to the 
large radius open string coordinates, we obtain precisely the topological string wavefunction (\ref{psitop}). We note that, in this parametrization, $X$ is already a flat 
coordinate for the open string modulus \cite{akv}. 

\begin{figure}[ht]
\begin{center}
\includegraphics[scale=0.55]{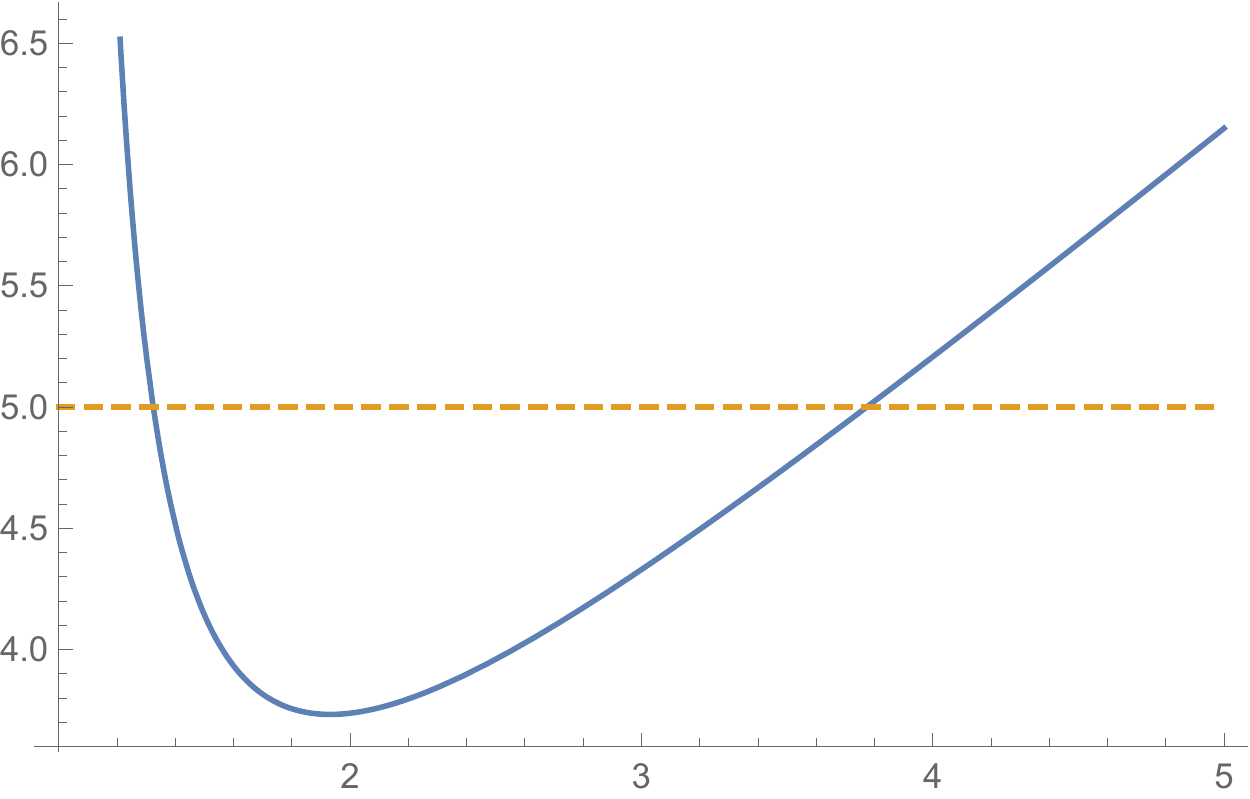}
\caption{The function in the r.h.s. of (\ref{xq}), as a function of $\zeta$, where $Q=\ri \zeta$. Given a real $X$, there are two values of $\zeta$ which lead to it, corresponding to 
two saddle points of the canonical transformation.}
\label{fsad}
\end{center}
\end{figure}

Our starting point is the asymptotic expansion (\ref{as-thooft}), 
\be
\label{wkb-str}
{ {\sqrt{E(q_m)}} \Psi_N(q_m) \over Z(N)} \sim  \exp\left[{ \ri \over \hbar_{\rm D}} \CT_0(q) + \CT_1(q)+ \CO(\hbar_{\rm D})\right]. 
\ee
The canonical transformation is given by the transform in (\ref{psin}). Since we are doing the calculation in the 't Hooft limit, we have to compute the transformed wavefunction 
in terms of the variables $x,q$ appearing in (\ref{sym-trans}), instead of the original variables $x_m,q_m$ appearing in the original spectral curve. We then find, 
\be
\label{psix}
\psi_N(x_m)= {1\over {\sqrt{2 \pi \hbar_{\rm D}}}} \int_\IR \re^{\ri  F(x, q)/ \hbar_{\rm D} } {\sqrt{E(q_m)}} \Psi_N(q_m)  \rd q, 
\ee
where
\be
F(x,q)={1\over 2} \left(x^2+ q^2 - 2 {\sqrt{2}} x q\right). 
\ee
Since the integrand in (\ref{psix}) is given as an asymptotic expansion for $\hbar_{\rm D}$ small, we can evaluate the integral transform in the saddle-point approximation. 
The saddle points are defined by 
\be
\label{saddle-gen}
\CT_0'(q)+ {\partial F \over \partial q}(x,q)=0. 
\ee
The different solutions to this equation give $q$ as a function of $x$. Let us denote by $q(x)$ the resulting function for a given saddle. By expanding around this saddle, we find
\be
 \psi_N(x_m)\sim \exp\left[ {\ri \over \hbar_{\rm D} } \CS_0(x) + \CS_1(x) + \CO(\hbar_{\rm D}) \right], 
\ee
where
\be
\ba
 \CS_0(x)&=\CT_0(q(x))+ F(x, q(x)),\\
  \CS_1(x)&= \CT_1(q(x))-{1\over 2} \log\left[-\ri  (\CT_0 +F)''(q(x))\right]. 
  \ea
  \ee
Here, we have denoted $F''= \partial^2F/\partial q^2$. The leading order term $\CT_0(q)$ has the WKB form (\ref{s0q}). This is preserved by canonical transformations: since $F(x,q)$ is the generating function of the transformation, one has
\be
\label{can-lead}
\CT_0(q(x))+ F(x, q(x))= \int^x y(x') \rd x',
\ee
where $y(x)$ is determined by (\ref{ycurve}). Since they are two different solutions for $y(x)$, this indicates that we will have at most two saddle points. 
Let us see that this is indeed the case. The saddle point equation (\ref{saddle-gen}) reads
\be
{\partial \CT_0 \over \partial q}+ q - {\sqrt{2}}x=0. 
\ee
The solution is
\be
x= {q\over {\sqrt{2}}} + {p \over  {\sqrt{2}}}= \log Q+ \log P(Q), 
\ee
or, equivalently, 
\be
\label{xq}
X= Q P(Q)= {Q\over 2(Q+ Q^{-1}) } \left[\kappa +{\sqrt{\kappa^2 -4 (Q+Q^{-1})^2}} \right].  
\ee
Given a real $X$, this equation has two different (imaginary) solutions, $Q_{1,2}(X)$ (see \figref{fsad}). The first solution is given by 
\be
\label{first-saddle}
Q_1(X)= \frac{\sqrt{-\sqrt{\left(X^2-\kappa  X+1\right)^2-4 X^2}-X^2+\kappa 
   X-1}}{\sqrt{2}}= \ri \left\{ X -{\kappa \over 2} - {\kappa^2 \over 8 X} +\cdots\right\}, 
   \ee
   where we have expanded the result for large $X$. The second solution is 
   \be
   \label{second-saddle}
   Q_2(X)= \frac{\sqrt{\sqrt{\left(X^2-\kappa  X+1\right)^2-4 X^2}-X^2+\kappa 
   X-1}}{\sqrt{2}}=\ri \left\{ 1+ {\kappa \over 2 X} + {3 \kappa^2 \over 8 X^2}+\cdots \right\}. 
   \ee
   Therefore, we have indeed {\it two} different saddle points. This is not surprising, since in the calculation of (\ref{two-psix}) at finite $\hbar$ we found that the wavefunction in the $x$ variables is a 
   sum of two pieces, coming from two different set of poles in the integral transform. The first term in the r.h.s. of (\ref{two-psix}) corresponds to the first saddle point (\ref{first-saddle}), while the second term corresponds 
   to the second saddle point (\ref{second-saddle}). These two saddles define two different functions $y_{1,2}(x)$, through 
   \be
   y_{1,2}(x)= {1\over {\sqrt{2}}} ( -q_{1,2}(x)+ p(q_{1,2}(x))).  
   \ee
They are given by 
\be
\ba
y_1(x) &=    \log \left(\frac{-1+\kappa  X-X^2+\sqrt{\left(X^2-\kappa  X+1\right)^2-4 X^2}}{2
   X}\right), \\
 y_2(x) &=    \log \left(\frac{-1+\kappa  X-X^2-\sqrt{\left(X^2-\kappa  X+1\right)^2-4 X^2}}{2
   X}\right). 
   \ea
   \ee
  These are the two different sign determinations obtained from the equation (\ref{ycurve}) (there is an additive ambiguity of an odd integer multiple $\pm \ri \pi$, since we are taking the log of a negative number.)   
We can now verify (\ref{can-lead}) explicitly. The canonical transformation of the leading order term is given by 
\be
\CT_0(q_{1,2}(x))+ {1\over 2}(x^2+ q_{1,2}^2(x)- 2 {\sqrt{2}} x q_{1,2}(x))
\ee
and it can be explicitly checked that it agrees with, 
\be
\label{lead}
\int^x y_{1,2}(X') {\rd X' \over X'}, 
\ee
up to $X$-independent terms.

Let us now consider the next-to-leading order of the canonical transformation. We first focus on the first saddle point, and we find 
\be
\label{ntl-wave}
\CT_1(q_1(x))-{1\over 2} \log\left[-\ri  (\CT_0 +F)''(q_1(x))\right] ={1\over 2} \int^X_{\infty} \int^X_{\infty} W_0(X',Y') \rd X' \rd Y' +{1\over 2} \log X, 
\ee
where $W_0(X,Y)$ is the annulus amplitude associated to the curve (\ref{ycurve}), and 
\be
\CT_0''(q)=\frac{\kappa  Q (1-Q^2)}{\left(Q^2+1\right) \sqrt{\kappa ^2 Q^2-4
   \left(Q^2+1\right)^2}}.
\ee
In $W_0(X,Y)$, the $x_i$ are the roots of the discriminant
\be
(1- \kappa X + X^2)^2 - 4 X^2,
\ee
and we have the ordering
\be
\ba
\{ x_1, x_2, x_3, x_4\}&=
\biggl\{ \frac{1}{2} \left(\kappa -\sqrt{\kappa -4} \sqrt{\kappa }-2\right),\frac{1}{2} \left(\kappa -\sqrt{\kappa +4} \sqrt{\kappa
   }+2\right), \\ 
   & \qquad \frac{1}{2} \left(\kappa +\sqrt{\kappa +4} \sqrt{\kappa }+2\right),\frac{1}{2} \left(\kappa +\sqrt{\kappa -4} \sqrt{\kappa
   }-2\right)\biggr\}.
   \ea
   \ee
   This ordering corresponds, in the topological string, to a choice of frame. In this case, this is the conifold frame which also appears in the 't Hooft limit (\ref{thooft}) of the 
   fermionic spectral traces. The result (\ref{ntl-wave}) is non-trivial, and it indicates that, after doing a canonical transformation to the large radius open coordinates, 
the first saddle point gives indeed the topological string wavefunction (\ref{psitop}), with an overall factor $X^{1/2}$. We conjecture that this is the case 
to all orders in $\hbar_{\rm D}$. 
   
What happens to the contribution of the other saddle point? The leading order, up to an $X$-independent term, is given in (\ref{lead}). The next-to-leading term can be computed by using  
the expression (\ref{pp-theta}), which involves the Abel--Jacobi map $u_q(Q)$ for the curve (\ref{QPspec}), given in (\ref{aj-q}). We can express this map 
in terms of the incomplete elliptic integral of the first kind $F(u, k)$, defined in (\ref{iei}). One finds, 
\be
\int_\infty^Q {\rd Q' \over {\sqrt{\sigma_q(Q')}}}=-{1\over a} F\left( {a \over Q},{1\over a^2}\right), 
\ee
where $a$ is given in (\ref{apointq}). We can now use the convergent expansion for $F(u, k)$ found in Theorem 4 of \cite{el-int}, to obtain the simple result
\be
\label{uqux}
u_q\left(Q_2(X)\right)= -u^{\rm or} (-X)+{\tau_q \over 4} =-u^{\rm or} (-X)+ {\tau_{\rm or} \over 2}- {1\over 4}, 
\ee
where the elliptic moduli $\tau_{\rm or}$, $\tau_q$ are defined in (\ref{tauor}), (\ref{tauq}). The minus sign in front of $X$ is due to the fact that, in the formulae of the Appendix for $u^{\rm or} (X)$, 
we consider the curve (\ref{spec-curve}), rather than the curve (\ref{ycurve}). For the curve (\ref{spec-curve}), and in the large radius frame, the transformation (\ref{uqux}) reads
\be
\label{sulr}
u(X) \rightarrow -u(X)-{\tau \over 4}-1, 
\ee
where $\tau$ is the modulus appropriate to the large radius frame, and is given in (\ref{taulr}). What 
is the interpretation of (\ref{sulr})? The shift by $-1$ is trivial in the Jacobian. One can verify, by using the results in Appendix \ref{appa}, that the points $u(X)$, $-u(X)-\tau/4$ correspond 
to the two different sheets of the Riemann surface defined by (\ref{spec-curve}). 
We will use this fact to obtain an exact result for the wavefunction when 
$\hbar=2 \pi$, in the next section. 

Let us now summarize the conclusions of the calculations in this section. The functions $\Psi_N(q)$ are the building blocks of the eigenfunctions. 
They can be computed exactly, as we showed in the last section, but they also admit a matrix integral representation which makes it possible to study their 't Hooft limit. 
The canonical transformation to the large radius open string coordinates can be computed in a saddle point approximation. 
Two different saddles contribute. The first one gives exactly the topological string wavefunction (\ref{psitop}). The second 
saddle gives a related contribution, involving the open topological string amplitude on the second sheet of the Riemann surface. 

It is possible to write down this result in a way which will be useful later on. First of all, the topological string wavefunctions 
are computed in the conifold frame. As shown in \cite{gkm}, under a change of frame, this wavefunction transforms as the topological string partition function. We can then write, 
\be
\label{asympsi}
\psi_N(x)\sim \sum_\sigma  \int \psi_{\rm top}^{(\sigma)} \left( X^{2 \pi \over \hbar}, {2 \pi \over \hbar} {\bf t}+ \pi \ri {\bf B}, \hbar_{\rm D} \right) \re^{\mJ^{\rm WS}(\mu, \hbar)-\mu N} \rd \mu. 
\ee
(In this formula we have gone back to the original notation for the open string moduli, with no subscript). The topological string wavefunction in the r.h.s. is calculated in the large radius frame for both, 
the closed and the open string moduli. The sum over $\sigma$ is over the different saddle points, which in this case correspond to the two different sheets of the Riemann surface. 
These saddle points lead to different wavefunctions through the shifts in the Abel--Jacobi map. The shift by a $B$-field $ \pi \ri {\bf B}$ is zero in this example, but we anticipate that, by analogy with what happens in the 
closed topological string, it will be nonzero for other geometries. 
 
 \subsection{The case of local $\mathbb P^2$}
 
We now make a preliminary study of the wavefunction for local $\IP^2$. In the more general case (\ref{kernelgen}), the kernel  is given in the $\sf p$ representation 
\cite{kasm}. So we adapt the notation and change the name of the $x$ variable of $\Phi_N(x)$  given in (\ref{phiNgen}) to $p_m$. To consider the 't Hooft limit of $\Phi_N(p_m)$, we need to perform a scaling similar to (\ref{tHooftscale}).  Since the parameter $\xi$ is proportional to $\hbar$ with proportionality factor depending on the case considered, we use
\be
	p={p_m \over \xi}.
\ee 
Then we have
\be
\ba
	\frac{\alpha^N \Psi_N(p_m)}{Z(N)} &= \left \langle \det  \frac{\alpha \sinh (\frac{p-\CP}{2})}{ \cosh \left(\frac{p-\CP}{2}+\ri \pi C\right) } \right \rangle
	= \exp \sum_{s=1}^{\infty} \frac{1}{s!} \left \langle \left ( \tr \log  \frac{\alpha \sinh (\frac{p-\CP}{2})}{ \cosh \left(\frac{p-\CP}{2}+\ri \pi C\right) }  \right )^s \right \rangle^{(c)},
\ea
\ee
where $\CP$ is the matrix with eigenvalues $p_1, \cdots, p_N$, and $Z(N)=\llangle 1 \rrangle$ is the partition function associated to the matrix integral (\ref{genkermm}). 
We will use exponentiated variables $P=\re^{p}$, $M=\re^{\CP}$. Using that
\be
	\label{logalphaPsi}
	\log \frac{\alpha^N \Psi_N(p_m)}{Z(N)}
	=\sum_{s=1}^{\infty}\frac{1}{s!} \sum_{k_1,...,k_s=1}^\infty \frac{(\omega^{k_1}-1)...(\omega^{k_s}-1)}{k_1 ... k_s} \langle \tr M^{k_1} ...\tr M^{k_s} \rangle^{(c)} P^{-k_1-...-k_s},
\ee
we can write
\be
\ba
	\label{leadingtHooftgen}
	\log \frac{\alpha^N \Psi_N(p_m)}{Z(N)} 
		&= \int^P \rd P'  \left \langle \tr \left ( \frac{1}{P'-M} -\frac{1}{P'-\omega M} \right ) \right \rangle+{\rm subleading}.
\ea
\ee
In this expression, we only considered the leading part at large $\hbar$, which is given by the $s=1$ contribution of (\ref{logalphaPsi}). The averaged trace can be considered as a generalization of the one-point correlator that we will call
\be
\omega(P) ={1\over N} \left\langle \tr \left( {1\over P-M} -{1\over P- \omega M} \right) \right\rangle. 
\ee
The planar limit of this average will be denoted by $\omega^0(P)$. In contrast to the case of local $\IF_0$, the matrix model underlying these more general kernels 
is a generalized $O(2)$ matrix model, and closed formulae for their planar resolvents are not known. However, from the discussion in the previous section, we 
expect $\omega^0(P)$ to be encoded in the algebraic equation for the spectral curve. 

\begin{figure}[ht]
\begin{center}
\includegraphics[scale=0.75]{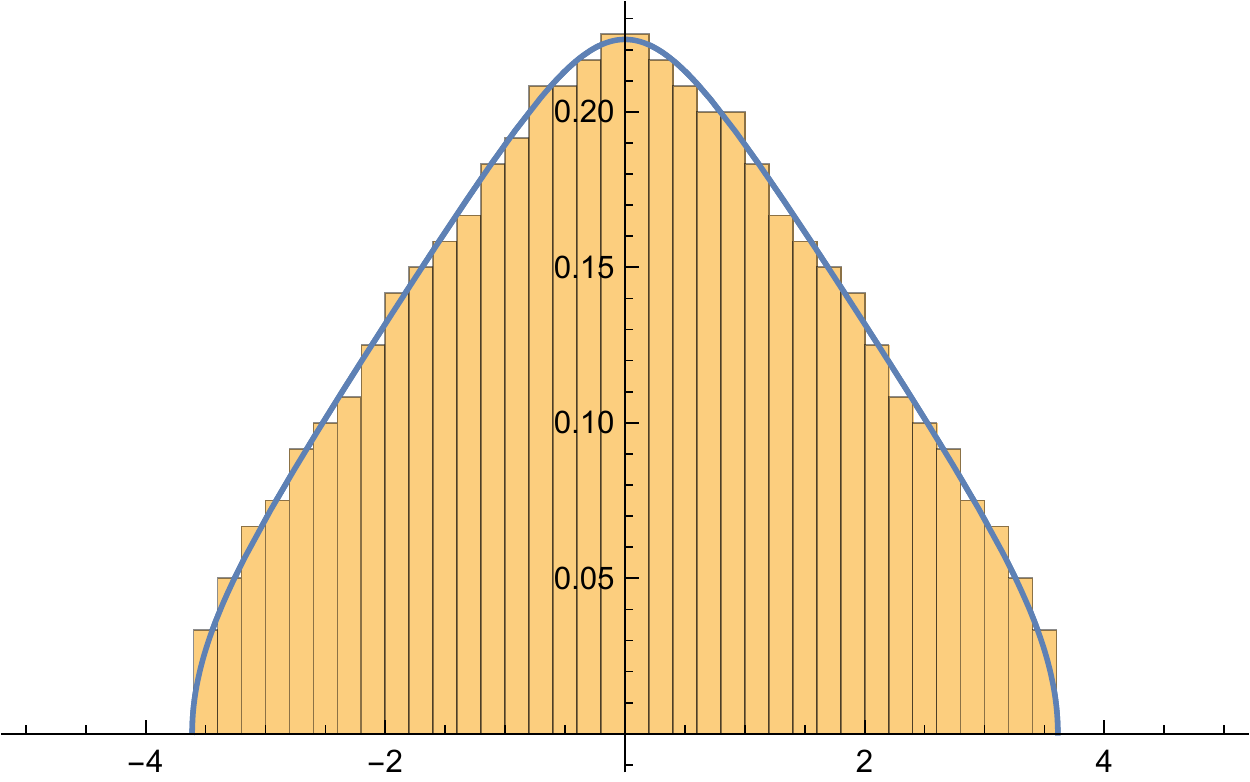}
\caption{The eigenvalue density $\rho(p)$ for local $\IP^2$ (in blue) against a numerical simulation (in orange). In this example, $\kappa=-70$, so $\lambda \approx 0.30755$. The simulation shows the probability density of $N=600$ eigenvalues of the matrix integral, relaxed to a configuration which approaches the saddle configuration.}
\label{figp2}
\end{center}
\end{figure}

We will now present evidence for this expectation in the case of local $\mathbb P^2$. 
In $Q,P$ variables (where $q$ is the conjugate of $p$ and $Q$ is the corresponding exponentiated variable), the spectral curve of local $\mathbb P^2$ is
\be
	P^2 Q+\frac{1}{PQ^2}+\frac{Q}{P}+\kappa=0.
\ee
This can be solved for the $Q$ variable as
\be
		Q(P)=\left ( \re^{\ri \pi/3} \left ( \frac{P^3+1}{2}+\frac{1}{2} \sqrt{\sigma(P)} \right )^{1/3}-\frac{P\kappa}{3}\re^{-\ri \pi/3} \left (  \frac{P^3+1}{2}+\frac{1}{2} \sqrt{\sigma(P)}  \right )^{-1/3} \right )^{-1}.
\ee
where
\be
	\sigma(Q)=(Q^3+1)^2+\frac{4}{27}\kappa^3Q^3=(Q^3-a^3)(Q^3-a^{-3}),
\ee
and the endpoint of the cut $a<1$ (when $-\infty <\kappa \leq -3 $) is given by
\be
	a=\left (-\frac{2 }{27}\kappa^3-1-\frac{2}{27}\sqrt{\kappa^6+27\kappa^3} \right )^{1/3}.
\ee
We will choose the differential on the spectral curve to be\footnote{The minus sign in the differential is convenient, 
and it can be justified by the fact that the actual canonical conjugate of $p$ is $-q$.}
\be
	-\frac{\log (Q(P)) \rd P}{P}.
\ee
The 't Hooft coupling $\lambda=N/\hbar$ is given by the period $\Pi_A$ of this differential, whose cycle encircles the cut $[a,a^{-1}]$:
\be
	\lambda=\frac{3}{8\pi^3} \Pi_A.
\ee
A perturbative calculation of (\ref{leadingtHooftgen}) from the matrix model, at small 't Hooft coupling, leads to the following conjecture, 
\be
	\label{planarspectral}
	  \omega^0(P)= \frac{3 \ri}{4\pi^2 \lambda} \frac{(-\log Q(P))}{P}+\frac{1}{2 \lambda}{V^{(0)}}'(P).
\ee
In this expression, $V^{(0)}$ is the planar potential of local $\mathbb P^2$ and it is given in \cite{mz},
\be
	V^{(0)}(P) \equiv -\frac{1}{\hbar} \log v(p)=-\frac{1}{2\pi}\log P+\frac{3\ri}{2\pi^2 } {\rm Li}_{2}(-\re^{2\pi \ri/3}P).
\ee
The conjecture (\ref{planarspectral}) for $\omega^0(P)$ also allows us to obtain the eigenvalue density $\rho(p)$ of the matrix model of local 
$\mathbb P^2$. Indeed, the usual argument in matrix model theory gives, 
\be
	\rho(P)=\frac{1}{2\pi \ri}\left ( \omega^0(P-\ri0)- \omega^0(P+\ri0) \right ), \qquad P \in (a,1/a).
\ee
We want to express it in terms of the variable $p$ (which is the one of the initial matrix integral) related to $P$ as $P=\re^p$. So we use $\rho(P)\rd P=\rho(p)\rd p$, to finally obtain
\be
	\label{p2rho}
	\rho(p)=\frac{3}{8\pi^3 \lambda} \log \left [\frac{\left ( \frac{\re^{3p}+1}{2}- \frac{\ri}{2} \sqrt{|\sigma(\re^{p})|}\right )^{1/3}+(-1)^{-2/3}\frac{\kappa \re^{p}}{3} \left (  \frac{\re^{3p}+1}{2}- \frac{\ri}{2} \sqrt{|\sigma(\re^{p})|}  \right) ^{-1/3} }{\left ( \frac{\re^{3p}+1}{2}+ \frac{\ri}{2} \sqrt{|\sigma(\re^{p})|}\right )^{1/3}+(-1)^{-2/3}\frac{\kappa \re^{p}}{3} \left (  \frac{\re^{3p}+1}{2}+ \frac{\ri}{2} \sqrt{|\sigma(\re^{p})|}  \right) ^{-1/3} }  \right ],
\ee
for $p \in (\log a, \log 1/a)$. This expression can be tested against the numerical relaxation of a large (but finite) number $N$ of eigenvalues towards a configuration which maximizes the integrand of the partition function $Z(N)$. This gives a good approximation of the saddle point configuration, which is characterised by the distribution (\ref{p2rho}). An example for $\kappa=-70$ and $N=600$ can be seen in Fig. \ref{figp2}.

 \sectiono{A conjecture for the exact eigenfunctions}

 \subsection{The general structure}
 We are now ready to extract from the above results some general statements on the exact eigenfunctions. Although we will use the example of local $\IP^1 \times \IP^1$ as our guide, 
 many of our considerations are general. 
 
 Our goal is to write an exact expression for the wavefunction $\psi(x; \kappa)$ defined in (\ref{psidef}), which is expressed in terms of coordinates appropriate to the large radius expansion. 
 This function is the analogue of the Fredholm determinant in the 
 case of closed strings. We will compute it in principle in the large radius frame for the closed string moduli as well, but since it is an entire function of $\kappa$, we expect it to be 
given by a symplectic invariant, and this will be indeed the case. 

The function $\psi(x; \kappa)$ is in the kernel of the operator $\mO +\kappa$, therefore at small $\hbar$ it can be calculated by a WKB expansion:
 \be
 \psi(x; \kappa) \sim \psi_{\rm WKB}(x; \kappa) = \exp \left[ (-\ri \hbar)^{n-1} \sum_{n=0}^\infty S^{\rm WKB}_n(x) \right]. 
 \ee
As in the closed string case, the WKB expansion in $\hbar$ in the exponent can be resummed to a function of $X$, $\hbar$ and the closed string moduli. 
When expressed in terms of flat coordinates for both the open and the closed string moduli, this resummation has the structure \cite{as,amir}
\be
\label{jopenwkb}
\mJ^{\rm WKB}_{\rm open}(\mu, \hbar, X)= \mJ^{\rm WKB}_{\rm pert}(\mu, \hbar, X)+ \sum_{{\bf d}, \ell,s} \sum_{k=1}^\infty D^{s}_{{\bf d}, \ell} 
 {q_\hbar^{k s} \over k(1-q_\hbar^k)}(-X)^{-k \ell} \re^{-k {\bf d} \cdot {\bf  t}}. 
\ee
In this equation, 
\be
q_\hbar= \re^{\ri \hbar},
\ee
$\mJ^{\rm WKB}_{\rm pert}(\mu, \hbar, X)$ is a perturbative part, which is a polynomial in $\log X=x$, and $D^{s}_{{\bf d}, \ell}$ are integer invariants 
which depend on a spin $s$, a winding number $\ell$, and the multi-degrees ${\bf d}$. The minus sign in $X$ in this equation is due to the fact that, in the WKB solution, the sign 
of $X$ is the opposite one to what is required by integrality, as one can verify in the case of local $\IF_0$. 
The WKB grand potential is obtained by adding the closed string grand potential appearing in 
(\ref{jtotal}), and the open string grand potential (\ref{jopenwkb}), i.e. 
\be
\mJ^{\rm WKB}(\mu, \hbar, X)=\mJ^{\rm WKB}(\mu, \hbar)+ \mJ^{\rm WKB}_{\rm open}(\mu, \hbar, X). 
\ee
We know from the direct calculations in the previous section that, in the 't Hooft limit, the wavefunction $\psi(x; \kappa)$ is closely related to the 
topological string wavefunction (\ref{psitop}), 
after appropriate rescaling of the variables. The standard open topological string amplitudes, in the large radius frame for both open and closed moduli, 
have an integrality structure which has been fully determined in \cite{ov,lmv}. The total open free energy of the standard topological string can be written as (see for example \cite{gkm})
\be
\label{fv}
\begin{aligned}
 F(V, {\bf t}, g_s)=& \sum_{{\bf d}} \sum_{g=0}^\infty \sum_{h=1}^\infty \sum_{{\boldsymbol \ell}} \sum_{w=1}^\infty 
 {\ri^h \over h!} n_{g, {\bf d}, {\boldsymbol \ell}} {1\over w} \left( 2\sin {w g_s \over 2} \right)^{2g-2} \\
& \qquad \qquad  \times  \prod_{i=1}^h \left(2 \sin {w \ell_i g_s \over 2}  \right) {1\over \ell_1 \cdots \ell_h}  \tr\, V^{w \ell_1} \cdots \tr \, V^{w \ell_h}{\rm e}^{-w {\bf d}\cdot {\bf   t}}.
\end{aligned}
\ee
In this expression, $n_{g, {\bf d}, {\boldsymbol \ell}}$ are integer invariants which 
generalize the Gopakumar--Vafa invariants of closed topological strings. They depend on the genus, the multi-degree ${\bf d}$ and 
the winding numbers ${\boldsymbol \ell}=(\ell_1, \cdots, \ell_h)$ 
(in fact, as shown in \cite{lmv}, the invariants $n_{g, {\bf d}, {\boldsymbol \ell}}$ can 
be written in terms of a more fundamental set of integer invariants, but we will not need them). 
The open topological string wavefunction (\ref{psitop}) is a particular case of (\ref{fv}) when we set 
\be
\tr V^n =X^{-n}, \qquad n \in \IZ. 
\ee
We then find the following integrality structure for the standard topological string wavefunction
\be
\begin{aligned}
\log \psi_{\rm top} (X, {\bf t}, g_s)= & \sum_{{\bf d}} \sum_{g=0}^\infty \sum_{h=1}^\infty \sum_{{\boldsymbol \ell}} \sum_{w=1}^\infty 
 {\ri^h \over h!} n_{g, {\bf d}, {\boldsymbol \ell}} {1\over w} \left( 2\sin {w g_s \over 2} \right)^{2g-2} \\
& \qquad \qquad \times  \prod_{i=1}^h \left(2 \sin {w \ell_i g_s \over 2}  \right) {1\over \ell_1 \cdots \ell_h} X^{-w (\ell_1+\cdots+ \ell_h)}{\rm e}^{-w {\bf d}\cdot   {\bf t}}.
\end{aligned}
\label{psi-integer}
\ee
We now introduce the worldsheet contribution for the grand potential, 
\be
\mJ^{\rm WS}(\mu, \hbar, X) =\mJ^{\rm WS}(\mu, \hbar) +\mJ^{\rm WS}_{\rm open}(\mu, \hbar,X). 
\ee
The first term in the r.h.s. is the worldsheet grand potential appearing in (\ref{jtotal}), while
\be
\label{Jpsi}
\mJ^{\rm WS}_{\rm open}(\mu, \hbar,X)=\log  \psi_{\rm top} \left( X^{2 \pi/\hbar}, {2 \pi \over \hbar} {\bf t}(\hbar) + \pi \ri {\bf B}, \hbar_{\rm D} \right). 
\ee
The total, $X$-dependent grand potential is 
\be
\mJ(\mu, \hbar,X)=\mJ^{\rm WKB}(\mu, \hbar, X)+\mJ^{\rm WS}(\mu, \hbar, X). 
\ee
The first term in the r.h.s. of this equation is a resummation of the WKB expansion, while the second term is a non-perturbative correction in $\hbar$ to the perturbative WKB result. 
Note that both terms have poles when $\hbar/2 \pi$ is a rational number. However, as in the closed string case, they cancel 
when we add both functions. Let us verify this. First, we note that the poles in the topological string contribution are only due 
to the term with $g=0$, $h=1$, and occur when $\hbar$ is of the form 
\be
\hbar = 2 \pi {w \over k}. 
\ee
For each multi-degree ${\bf d}$ and winding $\ell$, we find a simple pole with residue
\be
\label{tspole}
-\ri n_{0, {\bf d}, \ell} { (-1)^{k \ell +w {\bf B}\cdot {\bf d} }  \over k^2} \re^{- k {\bf d} \cdot {\bf t}} X^{-k \ell}. 
\ee
Let us now look at the resummed WKB eigenfunction. For the same value of $\hbar$, we find a simple pole with residue 
\be
\ri \left( \sum_{s} D^s_{ {\bf d}, \ell} \right) { (-1)^{k \ell +2 w s  }  \over k^2} \re^{- k {\bf d} \cdot {\bf t}} X^{-k \ell}. 
\ee
(\ref{tspole}) is the contribution at multi-degree ${\bf d}$ and winding $\ell$ to the disk amplitude, which is known to be equal (up to the sign in $X$ noted in 
(\ref{jopenwkb})) to the leading part of the WKB function $S_0(x)$ \cite{av,akv}. This implies in particular that
\be
 \sum_{s} D^s_{ {\bf d}, \ell}=n_{0, {\bf d}, \ell}. 
 \ee
 We now see that poles cancel, provided that 
 \be
 \label{pole-cancel}
 (-1)^{{\bf B}\cdot {\bf d}}=(-1)^{2s}
 \ee
 for all ${\bf d}$ and $s$ such that $D^s_{ {\bf d}, \ell}\not=0$. This condition on the B-field has also been found in \cite{amir}, in the framework of a different proposal 
 for the eigenfunctions. The B-field has to be chosen in such a way that poles cancel in the closed string sector, and it is natural to conjecture that the same choice will 
 satisfy (\ref{pole-cancel}). This is the case in local $\IF_0$, where ${\bf B}=0$ \cite{ghm} and the spins $s$ of the non-vanishing invariants are all integers \cite{amir}.

 We now conjecture that the wavefunction $\psi(x; \kappa)$ is obtained from the total grand potential by
\be
\label{psiJ}
\psi(x; \kappa) =\sum_{\sigma} \sum_{n \in \IZ}  \exp\left[ \mJ_\sigma (\mu+ 2 \pi \ri n , \hbar,X) \right]. 
\ee
Here, $\sigma$ labels the saddle points which contribute in the topological string sector, as in (\ref{asympsi}). In the example of local $\IF_0$ analyzed in detail in the previous section, 
the disk contribution of the different saddle points is the same up to an overall sign, and by changing the corresponding sign in the WKB contribution, pole cancellation is again achieved. 
This mechanism is likely to be present in other examples as well. Note that (\ref{psiJ}) is a natural generalization of the conjecture (\ref{spec-det}) for the 
Fredholm determinant. In (\ref{psiJ}), the contribution of the standard open topological string gives a non-perturbative correction to the resummed WKB wavefunction. The sums over $n \in \IZ$ and 
over $\sigma$ can be regarded as sums over the different sheets for the closed and the open string moduli, respectively. 

Let us verify that the conjecture (\ref{psiJ}) incorporates the results of the previous section. From the expansion (\ref{psixn}), we have
\be
\psi_N(x)=  \oint_{0} {\rd \kappa \over 2 \pi \ri} \kappa^{-N-1} \psi(x; \kappa). 
\ee
As in the derivation of (\ref{int-Z}), the contour integral in the r.h.s. can be combined with the sum over $n$ in (\ref{psiJ}) into an Airy type of integral
\be
\psi_N(x) =\sum_{\sigma} \int_{{\cal C}} \re^{\mJ_\sigma (\mu,\hbar,X)- \mu N} { \rd \mu \over 2 \pi \ri}. 
\ee
In the 't Hooft limit, the only contribution to $\mJ_\sigma(\mu, \hbar, X)$ comes from the topological open string wavefunction in (\ref{Jpsi}), and one recovers the asymptotic result (\ref{asympsi}). 
%
%
%

 \subsection{The maximally supersymmetric case}
 
 In the previous subsection we have generalized the arguments in \cite{ghm} and we have obtained a conjectural expression for the eigenfunctions. As pointed out in \cite{cgm8,ghm}, the simplest
 situation occurs in the so-called maximally supersymmetric case $\hbar=2 \pi$. This is the fixed point of the transformation that maps $\hbar$ into the dual Planck constant 
 $\hbar_{\rm D}$, hence $\hbar =2 \pi$ is also called the self-dual case. As in the closed string case, many of the contributions to $\mJ(\mu, \hbar, X)$ vanish for this value of $\hbar$. 

Let us then calculate the functions $\mJ_\sigma (\mu, 2 \pi, X)$ for the different saddle points. We will first consider the contribution from the first saddle point (\ref{first-saddle}), which we will denote as $\sigma=-$, since it corresponds to 
the first piece in (\ref{two-psix}). We start by looking at the topological string part (\ref{psi-integer}). When $\hbar=2 \pi$, all terms vanish except the one with $g=0, \, h=1$, and the one with $g=0, \, h=2$. 
The term $g=0$, $h=1$ gives the following contribution
 \be
{\ri \over 2 \pi} \sum_{d, \ell} n_{0, d, \ell} \sum_{w=1}^\infty {1\over w^2} \left(-1- \ell w \log(X) - d w t \right) \re^{-d w t } (-X)^{- w \ell}. 
\ee
Let us define the function 
\be
D(X,t)= \sum_{d, \ell} n_{0, d, \ell} \sum_{w=1}^\infty {1\over w^2} \re^{-d w t } (-X)^{- w \ell}. 
\ee
Then, the finite part of the term with $g=0$, $h=1$ at $\hbar=2 \pi$ is given by\footnote{A similar calculation gives the simplified form of Faddeev's quantum dilogarithm for $\mathsf{b}=1$ obtained e.g. in \cite{gk}.} 
\be
\label{disk-finite}
{\ri \over 2 \pi} \left( x {\partial D(X, t) \over \partial x} + t {\partial D(X,t) \over \partial t} - D(X,t) \right). 
\ee
 The function $D(X,t)$ is essentially the disk amplitude, up to a change of sign $X \rightarrow -X$, therefore it agrees precisely with the leading WKB amplitude when $\hbar=2 \pi$. 
 It is convenient to introduce the function 
\be
\label{yminus}
y_-(X)=   \log \left(\frac{-1-\kappa  X-X^2+\sqrt{\left(X^2+\kappa  X+1\right)^2-4 X^2}}{-2
   X}\right)= y(X) -\ri \pi. 
   \ee
Note that
\be
y_-(X) \approx -x, \qquad x\gg 1,
\ee
so we can write
\be
D(X, t)= \int_\infty^X \left( y_-(X') +x'\right) {\rd X' \over X'}. 
\ee
To evaluate the derivatives appearing in (\ref{disk-finite}), we take into account that 
\be
\label{dyk}
{\partial \over \partial \kappa}\left(  {y(X) \over X} \right)= -{1\over {\sqrt{\sigma(X)}}}, 
\ee
where 
\be
\label{sig-x}
\sigma(X)=  \left(X^2+\kappa  X+1\right)^2-4X^2.  
 \ee
We then find, 
\be
{\partial D(X,t) \over \partial t} =-\ri \pi u(X), 
\ee
where $u(X)$ is the Abel--Jacobi map in the large radius frame, given explicitly in (\ref{uxlr}). We also have 
\be
{\partial D(X, t) \over \partial x} =y_-(X)+x. 
\ee
The disk amplitude has a perturbative or ``classical" part given by 
\be
\mJ^{\rm WKB}_{\rm pert}(\mu, 2 \pi, X)= -{\ri  x^2\over 4 \pi}.
\ee
From the topological string, we also have the contribution of the annulus at $\hbar=2 \pi$. This gives the contribution 
\be
{1\over 2}  \int_{\infty}^X \int_{\infty}^X W_0(X_1,X_2 ) \rd X_1 \rd X_2 +{1\over 2} \log X, 
 \ee
 where $W_0(X_1,X_2 )$ is computed for the curve (\ref{spec-curve}), due to a an additional minus sign for $X_1$, $X_2$ which is obtained when (\ref{fv}) is evaluated at $\hbar=2\pi$. 
 
 Let us now consider the resummed WKB expansion at $\hbar=2 \pi$. The poles in this part cancel the poles coming from the disk amplitude, 
 as we saw in the previous subsection. There is in addition 
 a finite contribution which turns out to be exactly the next-to-leading WKB piece 
 \be
 \label{ntlwkb}
 S^{\rm WKB}_1(x)=-{1\over 4} \log \left( {\sigma(X) \over 4 X^2} \right).
\ee
Putting everything together, we find:
\be
\label{expdisk}
\ba
\mJ_-(\mu, 2 \pi, X)&= {\ri \over 2 \pi} \left\{ x y_-(X) +{1\over 2} x^2 - \int_\infty^X \left( y_-(X') +x'\right) {\rd X' \over X'} -\ri \pi t u(X) \right\}\\
&+{1\over 2}  \int_{\infty}^X \int_{\infty}^X W_0(X_1,X_2 ) \rd X_1 \rd X_2 +{1\over 2} \log X+ S^{\rm WKB}_1(x)+\mJ(\mu, 2\pi). 
\ea
\ee
This is very similar to what is found in the closed string case, i.e. in the structure of $\mJ(\mu, 2\pi)$: 
we have a contribution from the disk amplitude and its derivatives, similar to the contribution 
from the prepotential $F_0$ and its derivatives in $\mJ(\mu, 2\pi)$. In addition, we have the contribution from the next-to-leading terms in the topological string 
wavefunction and in the WKB wavefunction. This is similar 
to the contribution from $F_1$ and $F_1^{\rm NS}$ in $\mJ(\mu, 2\pi)$ \cite{cgm8,ghm}.
 
 %
 %
 %
 %
The next step is to sum over all the shifts of $\mu$ by $2 \pi \ri n$. This is again very similar to what is done in \cite{cgm8,ghm}. 
The dependence on $\mu$ in (\ref{expdisk}) is through $t=2 \mu+\cdots$. There is a contribution from 
the closed string grand potential $\mJ(\mu, 2\pi)$. In the open string sector, the only contribution comes from the term involving the Abel--Jacobi map. 
We then obtain the following modification of the theta function involved in the spectral determinant of 
local $\IF_0$ computed in \cite{ghm}:
\be
\vartheta_3\left( \xi -{1\over 3} + u(X) \biggl|\tau \right). 
\ee
Here, $\xi$ and $\tau$ are given by 
\be
\xi= {1\over 2 \pi^2} \left( t \partial_t^2 F_0(t) - \partial_t F_0(t)\right), \qquad \tau= {2 \ri \over \pi} \partial_t^2 F_0(t). 
\ee
The elliptic modulus $\tau$ can be written explicitly as a function of $\kappa$, see (\ref{taulr}). Note that both $\xi$ and the Abel--Jacobi map transform as modular functions of weight $-1$.  
We then find, 
\be
\ba
\psi_-(x; \kappa)& ={\sqrt{2}} X \re^{J(\mu, 2 \pi)}   \vartheta_3\left( \xi -{1\over 3} +u(X) \biggl|\tau\right)  {1\over \left(\sigma(X)\right)^{1/4}}  \exp\left[ {1\over 2}  \int_{\infty}^X \int_{\infty}^X W_0(X_1,X_2) \rd X_1 \rd X_2 \right]
\\ &\, \, \times 
\exp\left[ {\ri \over 2 \pi} \left\{  x y_-(X) +{1\over 2} x^2 - \int_\infty^X \left( y_-(X') +x'\right) {\rd X' \over X'} -\ri \pi t u(X) \right\} \right]. 
\ea
\ee
This can be written as 
\be
\ba
\label{psim-x}
\psi_-(x; \kappa)& =  -{\re^{-\pi \ri/4} \over 2{\sqrt{\pi}}} \re^{J(\mu, 2 \pi)} {\sqrt {4 X^2 \over \sigma(X)}} {  \kappa \vartheta'_1(0|\tau) \over 4 K(16/\kappa^2)} {\vartheta_3\left(\xi -{1\over 3}+u(X) \biggl|\tau\right)  
\over \vartheta_1 \left(u(X) | \tau \right)} \re^{{\ri \over 2\pi} \Sigma(x; \kappa)}. 
\ea
\ee
In writing this expression we have denoted
\be
\Sigma(x;\kappa)=  x y_-(X) +{1\over 2} x^2 - \int_\infty^X \left( y_-(X') +x'\right) {\rd X' \over X'} -\ri \pi t u(X),
\ee
we have used the explicit formula for the annulus amplitude given in (\ref{int-annulus}), 
and we have included the appropriate multiplicative constant to compare to the results in section \ref{sect-2}. 
The wavefunction (\ref{psim-x}) is very similar to a classical Baker--Akhiezer function on the mirror curve (see for example \cite{dubrovin,bbt}), 
although there are some important differences. For example, the first two terms in the exponent, 
 in the second line of (\ref{wave-or}), are not part of the standard Baker--Akhiezer function, nor is the next-to-leading WKB contribution in (\ref{ntlwkb}). 
 
 In order to study the wavefunction (\ref{psim-x}) on-shell, we have to consider the regime in which $\kappa=-\re^{E}$ is negative, and $E$ is interpreted as an energy. 
 As in \cite{ghm}, going to negative $\kappa$ leads to a non-trivial 
 monodromy in the periods. We have for example, 
 \be
 t \rightarrow t+ 2 \pi \ri, \qquad \xi \rightarrow \xi+{\tau \over 2}-1. 
 \ee
The periods $t$, $\partial_t F_0$ are computed with the expressions written down in (\ref{lr-periods}), which depend on $\kappa^2=\re^{2E}>0$, and are real. 
One obtains, up to an overall function of the modulus $\kappa$, 
\be
\label{psime}
 \psi_-(x; E) \propto {\sqrt {4 X^2 \over \sigma(X)}}  {\vartheta_2\left(\xi -{1\over 3}+u(X) \biggl|\tau\right)  
\over \vartheta_1 \left(u(X) | \tau \right)} \re^{{\ri \over 2\pi} \Sigma(x; \kappa)}, 
\ee
where we set $\kappa=-\re^{E}$. 

It is instructive to check directly that the wavefunction (\ref{psime}) has the property of independence w.r.t. the integration path typical of 
Baker--Akhiezer functions. Let us suppose that the integration path defining the Abel--Jacobi map encircles $n$ times an A-cycle and $m$ times a B-cycle. We then have
\be
\label{nmshift}
u(X) \rightarrow u(X) + n +m \tau, \qquad  n, m \in \IZ, 
\ee
where $\tau$ is given in (\ref{taulr}). 
The integral of $y_-(X)$ picks a linear combination of periods, 
\be
\int^X_{\infty} y_-(X') {\rd X' \over X'} \rightarrow \int^X_{\infty} y_-(X') {\rd X' \over X'}- \pi \ri n t + 2 m \left( \partial_t F_0 -{ \pi^2 \over 3} \right). 
\ee
Using standard properties of the $\vartheta$ functions under shifts of the arguments, we find that $\psi_-(x;E)$ remains invariant.

 The similarity between (\ref{psim-x}), (\ref{psime}) and a Baker--Akhiezer function could have been anticipated by using the results of \cite{em,be}. In those papers, it was shown that the 
 topological string wavefunction (\ref{psitop}), after summing over all filling fractions, leads to a ``quantum" Baker--Akhiezer function. 
 As $g_{\rm st}$ goes to zero, it becomes a classical Baker--Akhiezer function. In our case, 
 the grand potential includes the topological string wavefunction, and the sum over $n \in \IZ$ in (\ref{psiJ}) roughly corresponds to the sum over filling fractions in \cite{em,be}. However, 
 in contrast to what happens in \cite{em,be}, the classical limit of the ``quantum" Baker--Akhiezer function is obtained here in the self-dual or maximally supersymmetric case $\hbar=2\pi$.  
 
 When $\hbar$ is arbitrary, we expect our proposal (\ref{psiJ}) 
 to be closely related to the ``quantum" Baker--Akhiezer function studied in \cite{em,be}. There are however a number of important differences with the 
 results in \cite{em,be}, which also appear in the closed string case (see \cite{grassi} for a 
 detailed discussion). First of all, the wavefunction in \cite{em,be} is given as a formal expansion in $1/N$. In contrast, 
 the wavefunction (\ref{psiJ}) can be effectively computed as an expansion at large radius or as 
 an expansion around $\hbar=2\pi$ \cite{grassi}. Second, our wavefunction 
 requires information from both the WKB expansion and the topological string wavefunction, while the quantum Baker--Akhiezer function in \cite{em,be} is 
 constructed solely with topological string information. Finally, in (\ref{psiJ}) there are various contributions coming from different saddle points, which are not present in \cite{em,be}. 
 
It is easy to see that the wavefunction (\ref{psim-x}) has the same properties under 
symplectic transformation as the spectral determinant, i.e. 
it is invariant up to a change in the characteristics of the theta functions (see \cite{em,be,grassi}). 
Therefore, we can easily go to the orbifold frame, which is more convenient in order to compare with the results in section 2. 
We find,
\be
\label{wave-or}
\ba
\psi_-(x; \kappa)& ={\re^{-\pi \ri /4} \over 2 {\sqrt{\pi}}} \re^{J(\mu, 2 \pi)} { \vartheta'_1(0|\tau_{\rm or}) \over 2 \CK}  {\vartheta_1 \left( - \xi_{\rm or}+ u^{\rm or}(X) |\tau_{\rm or} \right) \over \vartheta_1(u^{\rm or}(X)|\tau_{\rm or})}
{\sqrt{ 4 X^2 \over \sigma(X)}} \re^{{\ri \over 2 \pi} \Sigma_{\rm or}(x; \kappa)}.  
\ea
\ee
In this equation, $\CK$, $\tau_{\rm or}$, $t_{\rm or}$ and $u_{\rm or}(X)$ are given in (\ref{grandk}), (\ref{tauor}), (\ref{orb-pers}) and (\ref{ajor}), respectively. 
In addition, $\xi_{\rm or}$ is essentially as in \cite{ghm}, namely
\be 
\xi_{\rm or}={1\over 4}- {1 \over 2 \pi^2} \left(t_{\rm or} \partial_{t_{\rm or}} ^2 F_0-\partial_{t_{\rm or}}  F_0 \right) ,
\ee
where
\be
   \partial_{t_{\rm or}} ^2 F_0= -{\pi \ri \over 2} \tau_{\rm or}.
   \ee
Finally, we have denoted
\be
\label{sigx}
 \Sigma_{\rm or}(x; \kappa)=x y_{-}(X) +{1\over 2} x^2 -\int_\infty^X \left( y_-(X') +x'\right) {\rd X' \over X'} -\ri \pi t_{\rm or} u^{\rm or}(X). 
 \ee
The expression (\ref{wave-or}) is completely explicit. It is written down in the orbifold frame, so we can expand it around $\kappa=0$ 
and compare the resulting coefficients with the functions $\psi_N^{(-)}(x)$ 
appearing in (\ref{two-psix}). It is quite satisfying to verify that the intricate details of these functions, 
as it is apparent in the explicit examples in (\ref{psix-ex}), are precisely reproduced by (\ref{wave-or}) (we have 
checked this up to order 7 in the $\kappa$ expansion). 

\begin{figure}[tb]
\begin{center}
\resizebox{150mm}{!}{\includegraphics{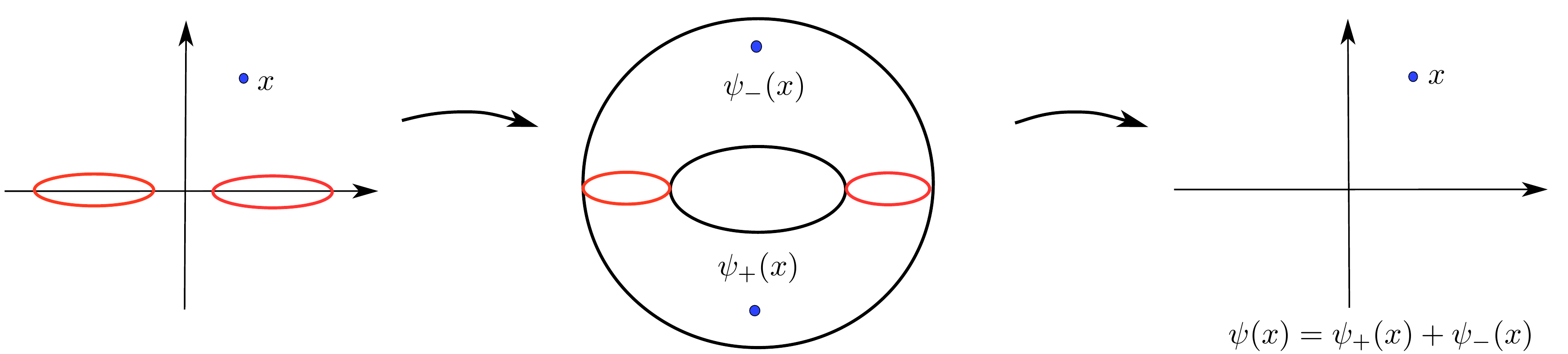}}
\end{center}
  \caption{The wavefunctions $\psi_\pm (x;\kappa)$ obtained with the topological string data are defined on the cut $x$-plane, and in particular they have 
  singularities at the branchpoints. After adding 
  the contribution of the two Riemann sheets, we obtain an entire function defined on the $x$ plane with no cuts.}
\label{cuts-nocuts}
\end{figure}

The wavefunctions (\ref{psim-x}), (\ref{wave-or}) are singular 
at the branch points of the curve (i.e. at the zeroes of (\ref{sig-x})). This is the counterpart in our framework of the well-known 
singular behavior of WKB wavefunctions at turning points. However, (\ref{psim-x}) is not the whole story, since 
we have to add the contribution of the other saddle point. This contribution can be easily computed by taking into account the transformation (\ref{sulr}). 
This is similar to (\ref{nmshift}), but now the value of $m$ is fractional. By using (\ref{dyk}), it follows that the integral of $y_-(X)$ changes as, 
\be
\int^X_{\infty} y_-(X') {\rd X' \over X'} \rightarrow -\int^X_{\infty} y_-(X') {\rd X' \over X'}- {1\over 2} \left( \partial_t F_0 -{ \pi^2 \over 3} \right). 
\ee
We change accordingly the arguments of the theta functions, and we obtain
\be
\ba
\label{psi-p}
\psi_+(x;\kappa)& =  -{\re^{-{\ri \pi /4}} \over 2 {\sqrt{\pi}}} \, \re^{J(\mu, 2 \pi) -\pi \ri (\xi-1/3) /2} {\sqrt {4 X^2 \over \sigma(X)}} {  \kappa \vartheta'_1(0|\tau) \over 4 K(16/\kappa^2)} {\vartheta_3\left(u(X)- \xi +{1\over 3}+ {\tau \over 4}  \biggl|\tau\right)  
\over \vartheta_1 \left(u(X) +{\tau \over 4} \Bigl| \tau \right)}\re^{-{\ri \over 2\pi} \Sigma(x; \kappa)}
\ea
\ee
It can be verified that the total wavefunction 
\be
\label{tpsi}
\psi(x;\kappa)=\psi_-(x; \kappa) + \psi_+(x; \kappa) 
\ee
has no singularities at the zeroes of $\sigma(X)$ (the calculation involves the values of the Abel--Jacobi map at the branch points listed in (\ref{aj-branch})). In addition, this function is 
entire on the $x$ plane. As illustrated in \figref{cuts-nocuts}, the wavefunctions $\psi_{\pm}(x; \kappa)$ are defined on the $x$ plane with cuts, corresponding to the Riemann surface described by (\ref{spec-curve}). After 
adding the contribution of the two sheets of the Rieman surface, we obtain a function (\ref{tpsi}) 
defined on the $x$ plane with no cuts (since it is entire). As we mentioned in the 
introduction, this gives a more complex implementation of the situation found in \cite{mmss} in the context of 
non-critical strings. 

When $\kappa=-\re^{E}$, the wavefunction (\ref{psi-p}) reads 
\be
\label{psipe}
\psi_+(x;E)\propto -\re^{-\pi \ri (\xi-1/3) /2} {\sqrt {4 X^2 \over \sigma(X)}} {\vartheta_2\left(u(X)- \xi +{1\over 3}+ {\tau \over 4}  \biggl|\tau\right)  
\over \vartheta_1 \left(u(X) +{\tau \over 4} \Bigl| \tau \right)} \re^{-{\ri \over 2 \pi} \Sigma(x; \kappa)}, 
\ee
up to an overall function of $\kappa$ which is the same as the one in (\ref{psime}). We can also 
perform a symplectic transformation to obtain the wavefunction (\ref{psi-p}) in the orbifold frame. One finds, 
\be
\label{wave-dos}
\ba
\psi_+(x;\kappa)& =-{1 \over 2 {\sqrt{\pi}}} \re^{J(\mu, 2 \pi)} { \vartheta'_1(0|\tau_{\rm or}) \over 2 \CK} 
 {\vartheta_4\left( \xi_{\rm or} + u^{\rm or}(X) +{1\over 4} |\tau_{\rm or} \right) \over \vartheta_4\left(u^{\rm or}(X) +{1\over 4} |\tau_{\rm or}\right)}
{\sqrt{ 4 X^2 \over \sigma(X)}}  \re^{ -{\ri \over 2 \pi} \Sigma_{\rm or} (x; \kappa) }.
\ea
\ee
We have checked that, when (\ref{wave-dos}) is expanded around $\kappa=0$, we recover the 
functions $\psi^{(+)}_N(x)$ obtained in (\ref{two-psix}), (\ref{psix-ex}) up to $N=7$. 
\begin{figure}[tb]
\begin{center}
\begin{tabular}{cc}
\resizebox{70mm}{!}{\includegraphics{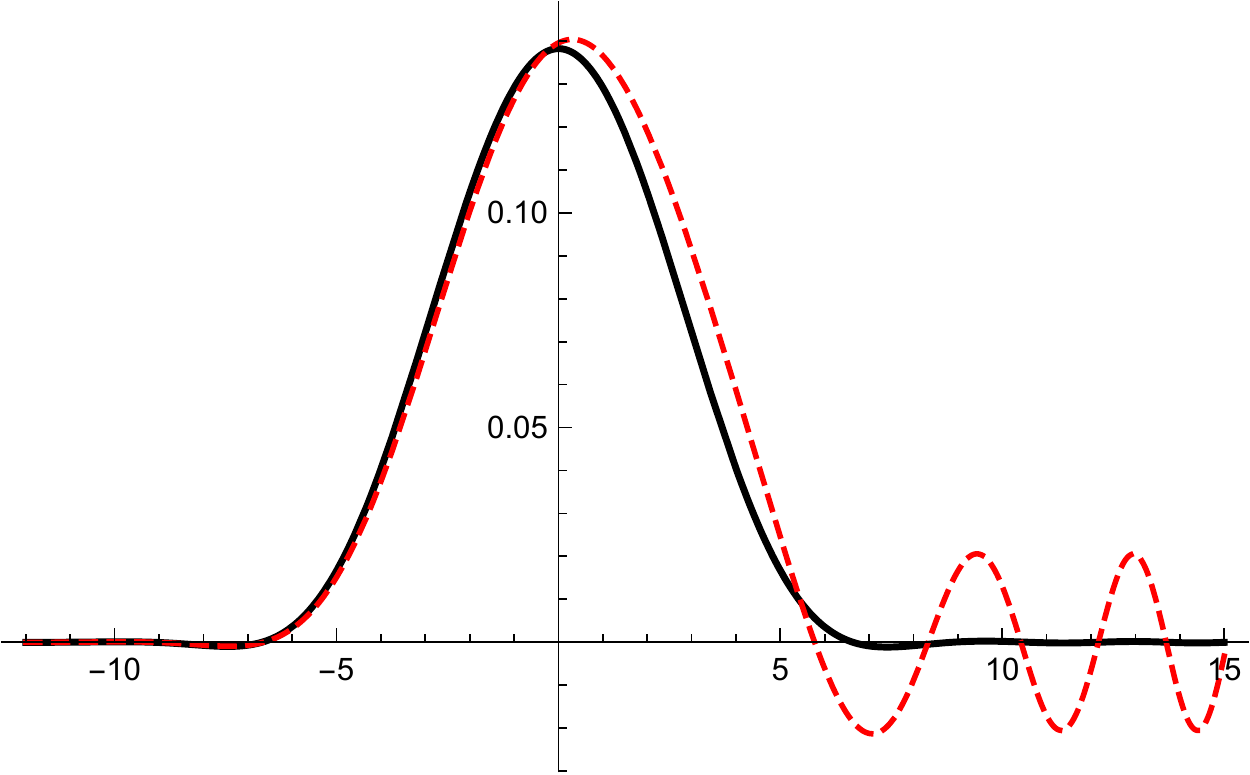}}
\hspace{10mm}
&
\resizebox{70mm}{!}{\includegraphics{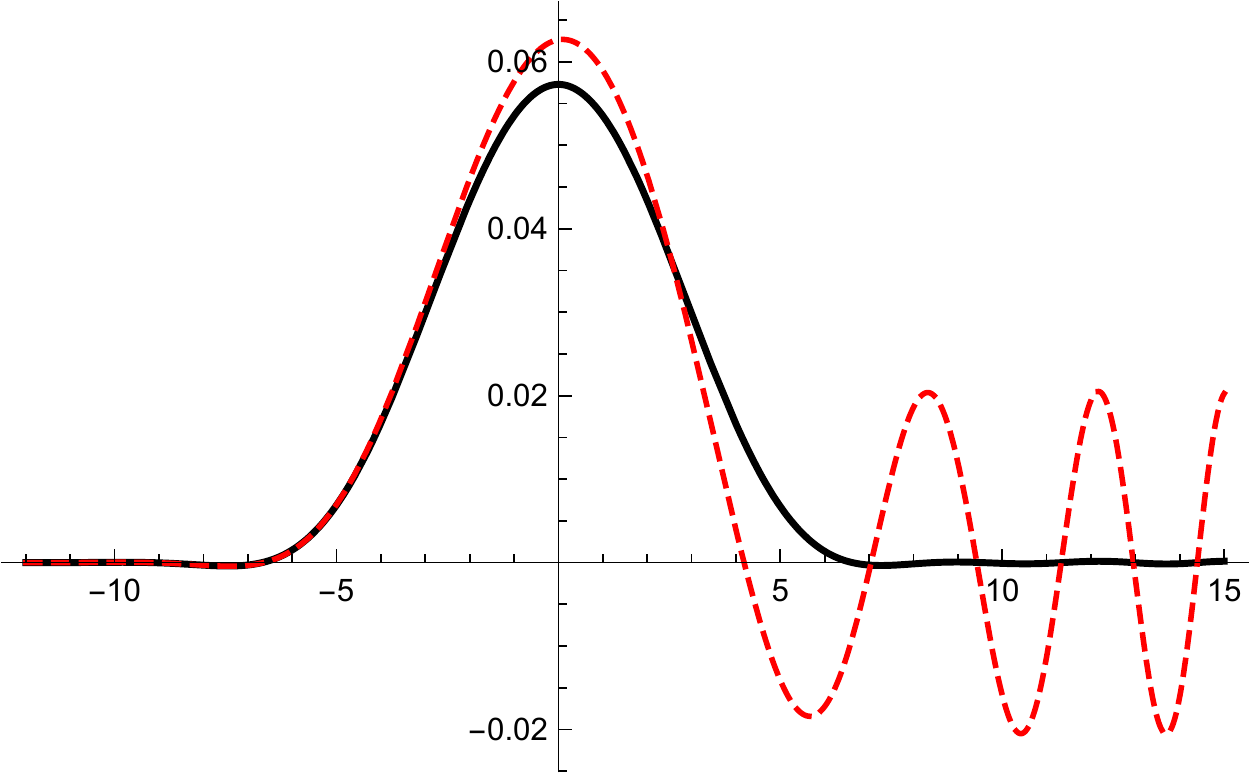}}
\end{tabular}
\end{center}
  \caption{The real (left) and the imaginary (right) parts of $\psi(x; \kappa)$, plotted as a function of $x$, for two values of $\kappa$. The black line corresponds to 
  $\kappa=-\re^{E_0}$, where $E_0=2.8818154...$ is the energy of the ground state, while the red, 
dashed line corresponds to the nearby value $\kappa=-\re^{28/10}$. For this value, $\psi(x; \kappa)$ is not square integrable due to the oscillatory behavior as $x \rightarrow \infty$. }
\label{refundf}
\end{figure}

The final expression for the wavefunction in the orbifold frame is the sum of (\ref{wave-or}) and (\ref{wave-dos}), 
\be
\label{totalf}
\ba
\psi(x; \kappa)&= {\re^{J(\mu, 2 \pi)} \over 2  {\sqrt{\pi} } } { \vartheta'_1(0|\tau_{\rm or}) \over 2 \CK} {\sqrt{ 4 X^2 \over \sigma(X)}}
 \biggl\{ \re^{ {\ri \over 2 \pi} \Sigma_{\rm or}(x; \kappa) -{\pi \ri /4}}  {\vartheta_1 \left( - \xi_{\rm or}+ u^{\rm or}(X) |\tau_{\rm or} \right) \over
  \vartheta_1(u^{\rm or}(X)|\tau_{\rm or})} \\  & \qquad \qquad \qquad - \re^{ -{\ri \over 2 \pi} \Sigma_{\rm or} (x; \kappa) } 
  {\vartheta_4\left( \xi_{\rm or} + u^{\rm or}(X) +{1\over 4} |\tau_{\rm or} \right) \over \vartheta_4(u^{\rm or}(X) +{1\over 4} |\tau_{\rm or})} \biggr\}. 
\ea
\ee

\begin{figure}[tb]
\begin{center}
\begin{tabular}{cc}
\resizebox{70mm}{!}{\includegraphics{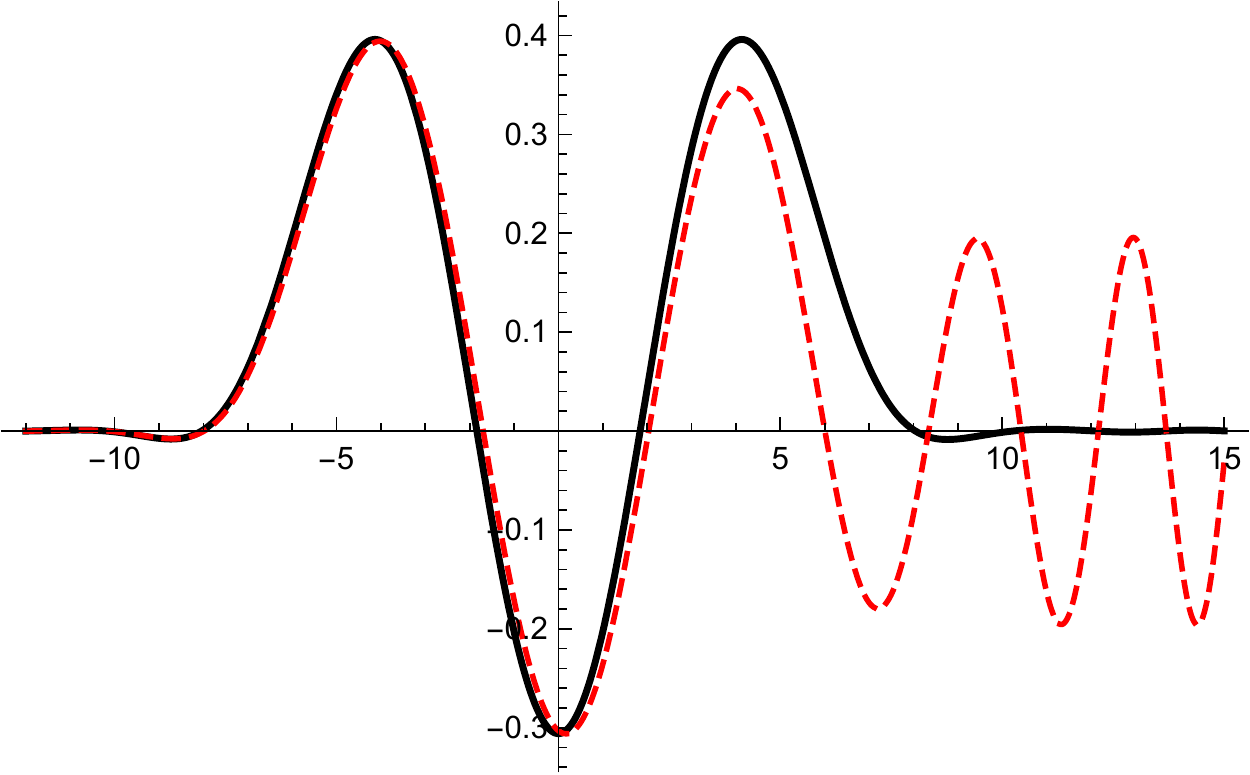}}
\hspace{10mm}
&
\resizebox{70mm}{!}{\includegraphics{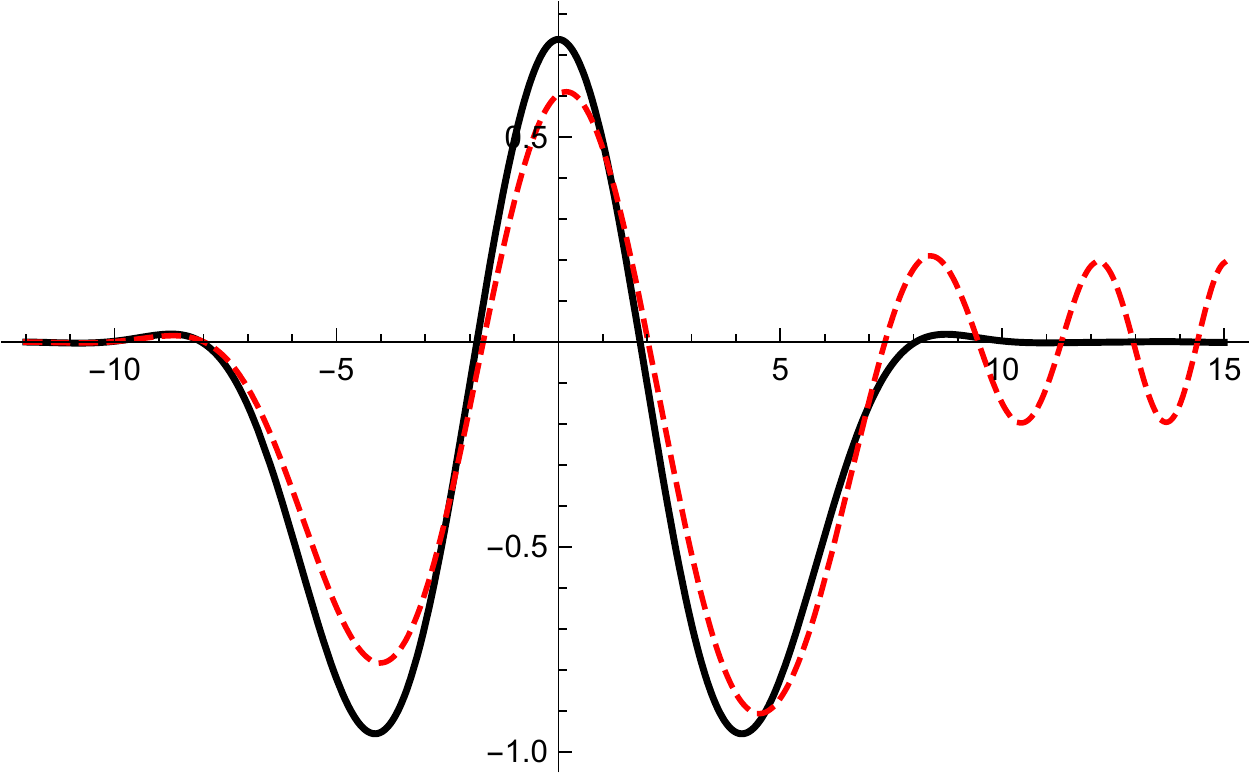}}
\end{tabular}
\end{center}
  \caption{The real (left) and the imaginary (right) parts of $\psi(x; \kappa)$, plotted as a function of $x$, for two values of $\kappa$. The black line corresponds to 
  $\kappa=-\re^{E_2}$, where $E_2=5.2881953...$ is the energy of the second excited state, while the red, 
dashed line corresponds to the nearby value $\kappa=-\re^{52/10}$. }
\label{rexcitedf}
\end{figure}

 As we have already emphasized, all these wavefunctions are meaningful ``off-shell," for arbitrary values of $\kappa$. In particular, (\ref{totalf}) can be computed explicitly in 
spectral theory, as we did in section 2. However, it will not 
be in general square integrable: as $x\rightarrow -\infty$, the wavefunction decays exponentially, but as $x \rightarrow \infty$ it goes to a function of $\kappa$, 
times the oscillatory function $\exp(-\ri x^2/(4 \pi))$. This behavior 
comes uniquely from $\psi_-(x; \kappa)$, since $\psi_+(x; \kappa)$ decays in both directions. This function of $\kappa$ is given by
\be
{\re^{\pi \ri/4}  \over \sqrt{\pi}}\,  \re^{J(\mu, 2 \pi)} \vartheta_1 \left(  \xi_{\rm or} |\tau_{\rm or} \right) ={\re^{\pi \ri/4} \over \sqrt{2 \pi}} \, \Xi(\kappa), 
\ee
i.e. it is proportional to the Fredholm determinant. The last equality follows from the conjecture in \cite{ghm}. Therefore, in order to find an appropriate eigenfunction, 
which decreases as $x\rightarrow \infty$, one has to 
set $\kappa$ to be a zero of the Fredholm determinant, in agreement with our general discussion in section 2. In other words, the quantization condition appears when 
we require the wavefunction to be square integrable along the 
real axis\footnote{In some situations, the quantization condition can be obtained by requiring 
monodromy invariance of the wavefunction, but this is not the case for (\ref{totalf}). Monodromy invariance means here 
independence w.r.t. the integration path as we circle closed loops in the Riemann surface. Since 
(\ref{totalf}) behaves in this respect like a Baker--Akhiezer function, it is monodromy invariant for {\it any} $\kappa$.}. 
This is illustrated in detail in \figref{refundf}, which shows the real and the imaginary parts of the wavefunction 
for two values of $\kappa$. The black line is the wavefunction plotted for $\kappa=-\re^{E_0}$, where $E_0=2.8818154...$ is the energy of the ground state, while the red, 
dashed line is the wavefunction for a nearby value $\kappa=-\re^{28/10}$. The oscillatory part 
which dominates the asymptotics as $x\rightarrow \infty$ disappears when we tune the value of $\kappa$ to be a 
zero of the Fredholm determinant. Of course, the function $\psi(x; \kappa)$ contains 
information on all the different eigenfunctions, as we change the value of $\kappa$. In \figref{rexcitedf} we show how to 
construct the eigenfunction for the second excited state, as well as a nearby 
oscillating function. We should note that the real and the imaginary parts of the on-shell wavefunctions are proportional to each other, as required by the reality of the integral kernel 
(\ref{p1k}).  

An equivalent analysis can be performed by studying the wavefunction in the large radius frame, by using the expressions (\ref{psime}) and (\ref{psipe}). 
For (\ref{psime}) to decay at infinity, we have to tune 
$\xi$ such as to obtain a zero of the theta function in the numerator. This leads to the exact quantization condition 
\be
\xi-{1\over 3}=n+{1\over 2} , \qquad n =0,1,2, \cdots,  
\ee
which was first written down in the context of ABJ(M) theory in \cite{cgm8}. This if of course equivalent, in the case $\hbar=2 \pi$, to having a zero of the 
Fredholm determinant, as explained in \cite{cgm8,ghm}. For these ``quantized" values of the moduli, leading to quantized energies $E_n$, $n=0,1,2, \cdots$, 
the quotient of theta functions in (\ref{psime}) and (\ref{psipe}) simplifies, and one obtains\footnote{After this paper was posted to the arXiv, this on-shell version of the wavefunction 
was obtained by R. Kashaev and S. Sergeev \cite{ks}  by studying the difference equation (\ref{or-eq}).}
\be
\psi(x; E_n)= \psi_-(x; E_n)+\psi_+(x; E_n) \propto {\sqrt{ 4 X^2 \over \sigma(X)} }\left\{ \re^{ {\ri \over2 \pi} \Sigma(x; -\re^{E_n})} - \ri^{-n-1/2} \re^{- {\ri \over2 \pi} \Sigma(x; -\re^{E_n})} \right\}. 
\ee

There are a number of additional remarks that can be made about our results. It is easy to check that both (\ref{wave-or}) and (\ref{wave-dos}) satisfy the 
difference equation (\ref{or-eq}), where $\mO$ is given in (\ref{f0op}) and $m_{\IF_0}=1$. This is guaranteed by the 
the first two terms in (\ref{sigx}), since everything else depends on $X=\re^x$ and is invariant under shifts by $\pm 2 \pi \ri$. 
A second remark is that the operator $\mO$ for local $\IF_0$ is the Hamiltonian of the 
$SU(2)$ relativistic Toda lattice (see for example \cite{hm}), therefore our result (\ref{totalf}) gives the exact eigenfunctions 
for this problem, in the self-dual case $\hbar=2\pi$. 

%
 %

\sectiono{Conclusions and open problems}

In this paper we have taken the first steps in the computation of exact eigenfunctions for the spectral problem introduced in \cite{ghm,cgm}. 
Clearly, there are many things that should be extended and clarified. To begin with, it would be desirable to reformulate some of the 
ingredients in a more intrinsic and geometric way. For example, the functions $\Psi_N(x)$ are our basic objects in the study of the open string sector, but we have introduced them in a way 
which depends on the details of the kernel. Is there a more intrinsic 
formulation, valid for any trace class operator coming from a mirror curve? In the same vein, we should find a general 
(and, if possible, more geometric) characterization of the saddle points which appear in the final answer. In our example, 
the second saddle point has in the end a simple form (it suffices to compare (\ref{wave-dos}) with (\ref{wave-or})), but this is the result 
of a relatively involved calculation. There is for sure a simpler logic behind. There are also many concrete issues which deserve a more careful 
study. For example, the pole cancellation in the open string sector requires an additional condition on the B-field (already noted in \cite{amir}), 
which should be explored in more detail. 

Generalizations come to life when one considers more examples. In this paper we have given some of the required ingredients to obtain the eigenfunctions for local $\IP^2$, but a more thorough 
study of this model is needed. In addition, one should calculate the eigenfunctions beyond the maximally supersymmetric case. This might be involved in practice, 
since calculations based on BPS invariants and the topological string provide 
expansions of the wavefunctions at large $x$, which will have to be compared to the results of Fredholm theory. 

One possible application of the exact results for the eigenfunctions is the relationship between the single quantization condition of \cite{cgm}, appropriate for quantum curves, and the 
$g_\Sigma$ quantization conditions of \cite{hm,fhm}, appropriate for cluster integrable systems. We expect that, as more quantization conditions are imposed, the 
wavefunctions solving the spectral problem for the quantum curve acquire 
additional analyticity properties that make them suitable solutions of the Baxter equation. 

One intriguing aspect of our calculation is that it involves the topological string wavefunction evaluated at the second sheet of the Riemann surface. This gives rise to different integrated 
open string amplitudes, and their computation requires exact expressions as functions of $X$. It does not seem to be feasible, for example, to compute these amplitudes 
by using the topological vertex. It would be interesting to have a deeper understanding of these amplitudes, both conceptually and technically. 

Finally, the results in this paper might provide a first step in defining topological open string amplitudes beyond perturbation theory. 
We have shown that the wavefunction $\Psi_N(x)$ gives a non-perturbative 
meaning to certain open string amplitudes. However, in order to reconstruct non-perturbatively 
the full set of $n$-point correlation functions appearing in \cite{mmopen,bkmp}, we probably need additional quantities in the spectral theory of the quantum mirror curve. 

\section*{Acknowledgements}
We would like to thank Vincent Bouchard, Santiago Codesido, Alba Grassi, Jie Gu, Yasuyuki Hatsuda, Amir Kashani--Poor, Rinat Kashaev, 
Ricardo Schiappa and Antonio Sciarappa for useful discussions, correspondence, and comments on a preliminary version of this paper.
This work is supported in part by the Fonds National Suisse, 
subsidies 200021-156995 and 200020-141329, and by the NCCR 51NF40-141869 ``The Mathematics of Physics" (SwissMAP). 

\appendix

\sectiono{Bergmann kernel and the Abel--Jacobi map}

\label{appa}

It is well-known that the Bergmann kernel can be expressed in terms of an odd theta function, see for 
example \cite{be}. This gives a useful formula for the annulus ampitude $W_0(p,q)$. 
We will now present some detailed results for this amplitude in the case of curves of genus one involving the quartic polynomial
\be
\sigma(x) =(x-x_1)(x-x_2)(x-x_3)(x-x_4). 
\ee
The Abel--Jacobi map is given by 
\be
\label{udef}
u(x)=c  \int_{\infty}^ {x} {\rd z \over {\sqrt{\sigma(z)}}}, 
\end{equation}
where we have chosen the base-point at infinity and $c$ is an appropriate function of the moduli of the curve. 
The annulus amplitude is built upon the Bergmann kernel, which can be written in terms of the theta function 
of odd characteristic, 
\be
B(u,u')\rd u \rd u'= \rd u \rd u' \partial_u \partial_{u'} \log \vartheta_1 (u-u'|\tau). 
\ee
In this expression, $u$ is a coordinate in the Jacobian, and it is related to a coordinate in the curve $p$ by the Abel--Jacobi map. Since 
\be
W_0(p,q) = B(p,q) -{1\over (p-q)^2}, 
\ee
we can write
\be
W_0(p,q)= \partial_p \partial_q \log \left( {\vartheta_1 (u(p)-u(q)|\tau) \over p-q} \right).
\ee
This expression is particularly useful in order to compute the integral appearing in (\ref{psitop}). We obtain 
\be
\ba
\label{double-int}
\int_{p_r}^p \int_{q_r}^q W_0(p,q) \rd p \rd q&= \log \left( {\vartheta_1 (u(p)-u(q)|\tau) \over p-q} \right)+  \log\left( {\vartheta_1 (u(p_r)-u(q_r)|\tau) \over p_r-q_r} \right)
\\&  -  \log \left( {\vartheta_1 (u(p_r)-u(q)|\tau) \over p_r-q} \right)- \log\left( {\vartheta_1 (u(p)-u(q_r)|\tau) \over p-q_r} \right). 
\ea
 \ee
 Here, $p_r$ and $q_r$ are reference points that we will eventually take to infinity. We first calculate the limit of the r.h.s. of (\ref{double-int}) when 
 $q$ goes to $p$, and when $q_r$ goes to $p_r$. We find, 
 \be
 \log \left( u'(p) \vartheta_1'(0|\tau) \right) +\log \left( u'(p_r) (p_r-p)^2 \vartheta_1'(0) \right) -  \log \left( \vartheta_1 (u(p)-u(p_r)|\tau) ^2 \right). 
 \ee
 Here, we have used that $\vartheta_1(u|\tau)$ is an odd function of $u$. We can now take the limit $p_r \rightarrow \infty$. In this limit, $u(p_r) \rightarrow 0$, and 
 \be
 \lim_{p_r\to \infty} u'(p_r) (p_r-p)^2= c. 
 \ee
 We conclude that:
 \be
 \label{int-annulus}
 {1\over 2}  \int_{\infty}^p \int_{\infty}^p W_0(p',q') \rd p' \rd q'=- \log\left[{( \sigma(p))^{1/4}  \over c \vartheta_1'(0|\tau)}  \vartheta_1 (u(p)|\tau)  \right]. 
 \ee
 A similar formula can be obtained for the symmetrized annulus amplitude (\ref{symannulus}). We find, 
 \be
 2 W^0_{++}(p,q)= {1\over 2} \partial_p \partial_q \log {\vartheta_1(u(p)-u(q)) \over \vartheta_1(u(p)+u(q))}. 
\ee
 We can also perform the integration to obtain 
 \be
 \label{pp-theta}
2 \int_{\infty}^p \int_{\infty}^p W^0_{++}(p',q') \rd p' \rd q'= -{1\over 2} \log \left[ -{( \sigma(p))^{1/2} \over  2 p   c \vartheta_1'(0|\tau)}   \vartheta_1 (2u(p)|\tau)  \right].
\ee

In the paper we are interested in computing the Abel--Jacobi map $u(X)$, the elliptic modulus $\tau$ and the annulus amplitude for the curves (\ref{spec-curve}) (when $m_{\IF_0}=1$), 
and for the curve (\ref{QPspec}). The calculation depends on the choice of symplectic frame. In the case of the curve (\ref{spec-curve}), 
there are two closed string frames which will be particularly useful: the large radius frame, appropriate for the expansion around $\kappa=\infty$, 
and the so-called {\it orbifold} frame, which is 
appropriate for the expansion around $\kappa=0$, as explained in \cite{cgm8,ghm}.

Let us summarize some important features of the large radius frame. In this frame, the periods of local $\IF_0$ with $m_{\IF_0}=1$ are given by 
\be
\label{lr-periods}
\ba
\omega_1(z)&= \log (z) +4 z \, _4F_3\left(1,1,\frac{3}{2},\frac{3}{2};2,2,2;16 z\right), \\
\omega_2(z)&= {1\over \pi} G_{3,3}^{3,2}\left(16 z\left|
\begin{array}{c}
 \frac{1}{2},\frac{1}{2},1 \\
 0,0,0
\end{array}
\right.\right)-{2 \pi^2 \over 3},
\ea
\ee
where
\be
z={1\over \kappa^2}. 
\ee
They determine the genus zero free energy $F_0(t)$ as 
\be
t= -\omega_1(z), \qquad {\partial F_0 \over \partial t}= \omega_2(z). 
\ee
The periods defined above are period integrals of the one-form $y_-(X) \rd X/X$, where $y_-(X)$ is defined in (\ref{yminus}). 
We will denote the zeroes of the quartic polynomial $\sigma(X)$ in (\ref{sig-x}) by 
\be
X_{\pm 1}, \quad X_{\pm 1/k}. 
\ee
They are the branch points of the curve (\ref{spec-curve}). Their explicit expression is 
\be
\ba
X_1(\kappa)&=\frac{1}{2} \left(2-\kappa -\sqrt{\kappa -4} \sqrt{\kappa }\right),\\
X_{1/k} (\kappa)&=\frac{1}{2} \left(-\kappa -\sqrt{\kappa +4} \sqrt{\kappa }-2\right),\\
X_{-1}(\kappa)&=1/X_1(\kappa)=\frac{1}{2} \left(-\kappa +\sqrt{\kappa -4} \sqrt{\kappa }+2\right),\\
X_{-1/k}(\kappa)&=1/X_k(\kappa)=\frac{1}{2} \left(-\kappa +\sqrt{\kappa +4} \sqrt{\kappa }-2\right). 
\ea
\ee
The A-cycle goes around the interval 
\be
A: \qquad [X_{-1}(\kappa), X_{-1/k} (\kappa)], 
\ee
while the B-cycle is around the interval
\be
B: \qquad [X_{1}(\kappa), X_{-1} (\kappa)].
\ee
One then has, 
\be
\ba 
 \oint_A y_-(X) {\rd X \over X}=&-\pi \ri t, \\
   \oint_B y_-(X) {\rd X \over X}= &2 \partial_t F_0 - {2 \pi^2\over 3}. 
   \ea
   \ee

The Abel--Jacobi map in the large radius frame is given by 
\be
\label{uxlr}
u(X)= -{\ri \kappa \over 4 K(16 z)}  \int_\infty^X {\rd X' \over {\sqrt{ \sigma(X')}}}.  
\ee
Let us introduce the modulus
\be
k^2= 1-{16 \over \kappa}. 
\ee
Then, the elliptic modulus of the curve is 
\be
\label{taulr}
\tau=2 \ri {K(k)\over K'(k)}, 
\ee
where as usual $K'=K(k')$, and $(k')^2+k^2=1$. The Abel--Jacobi map satisfies
\be
\oint_A \rd u(X)= 1, \qquad \oint_B \rd u(X) = 
\tau. 
\ee
It is very useful to express this map in terms of the incomplete elliptic integral of the first kind: 
\be
\label{iei}
F(u,k)=\int_0^u {\rd x \over {\sqrt{(1-x^2) (1-k^2 x^2)}}}. 
\ee
Let us define
  \be
  B^2={\kappa-4 \over \kappa}.
  \ee
Then, one finds, 
\be\label{uci}
u(X)= -{\ri \over 2 K'} \left\{  F\left({1\over B} {X+1\over X-1}, k\right)- F\left( {1\over B}, k\right)  \right\}.   
\ee

From the explicit expression (\ref{uci}), and standard properties of elliptic functions (see for example \cite{akhiezer}), 
we can deduce the values of the Abel--Jacobi map at the branch points, 
\be
\label{aj-branch}
\ba
u(X_{-1/k})&= {3 \tau \over 8}, \qquad u(X_{-1})= {3 \tau \over 8}-{1\over  2},\\
u(X_{1})&= -{\tau \over 8}-{1\over 2}, \qquad  u(X_{1/k})= -{\tau \over 8}, 
\ea
\ee
modulo $m+ n \tau$.

Let us now consider the orbifold frame. The periods can be taken to be, 
\be
\label{orb-pers}
\ba
t_{\rm or}=& \kappa  \, _3F_2\left(\frac{1}{2},\frac{1}{2},\frac{1}{2};1,\frac{3}{2};\frac{\kappa ^2}{16}\right),\\
   \partial_{t_{\rm or}} F_0&=
 { \kappa  \over 8 \pi} G^{2,3}_{3,3} \left( \begin{array}{ccc} {1\over 2}, & {1\over 2},& {1\over 2} \\ 0, & 0,&-{1\over 2} \end{array} \biggl| {\kappa^2\over 16}\right)
 +{ \ri \pi  \kappa  \over 4}   {~}_3F_2\left(\frac{1}{2},\frac{1}{2},\frac{1}{2};1,\frac{3}{2};\frac{\kappa^2
   }{16}\right). 
   \ea
\ee
The Abel--Jacobi map reads in this case
 \be
 \label{ajor}
 u^{\rm or} (X)= -{\ri  \over 2 \CK}  \int_\infty^X {\rd X' \over {\sqrt{ \sigma(X')}}}, 
 \ee
 where
 \be
 \label{grandk}
 \CK= K(\kappa^2/16). 
 \ee
The elliptic modulus appearing in the theta function is
 \be
 \label{tauor}
 \tau_{\rm or}= {\ri \over 2} {K(1-\kappa^2/16) \over K(\kappa^2/16)} -{1\over 2}, 
 \ee
 as in section 4.3 of \cite{cgm}. We note that the large radius frame and the orbifold frame are related by the modular transformation 
 implemented by the ${\rm SL}(2, \IZ)$ matrix
 \be
\label{eps-matrix}
\begin{pmatrix} 0& -1 \\
1 & 2 \end{pmatrix}, 
\ee
so that
\be
\tau_{\rm or}= -{1\over \tau+2}. 
\ee

 In the case of the curve (\ref{QPspec}), the Abel--Jacobi map is given by 
\be
\label{aj-q}
u_q(Q)={1  \over 2 \CK}  \int_\infty^Q {\rd Q' \over {\sqrt{ \sigma_q(Q')}}}, 
\ee
where $\sigma_q(Q)$ is defined in (\ref{disq}). The elliptic modulus is given by 
\be
\label{tauq}
\tau_q= 2 \tau_{\rm or}-1. 
\ee
%

 %
 %
 %
   %
 %
 %
\sectiono{Numerical calculation of the eigenfunctions}
\label{app-deux}

Here we describe how to get numerical approximations of the wavefunctions of the kernel (\ref{p1k}) in an orthogonal basis. We rely on the method described in 
\cite{hmo}, but instead of building the basis using Chebyshev polynomials of the second kind (which are Gegenbauer polynomials with $\alpha=1$), we use Gegenbauer polynomials 
$C_n^{\left(\alpha \right)}(y)$ with $\alpha=3/2$ which are better suited to our kernel. To be concrete, let us define in the $\mq$-representation the following basis functions:
\be
	c_n(q)= \langle q | c_n \rangle = \sqrt{\frac{1}{2\sqrt{2}} } \sqrt{\frac{n+\frac{3}{2}}{(n+1) (n+2)}} \frac{
   \, C_n^{\left(3/2\right)}\left(\tanh \left(\frac{ 
   q}{2\sqrt{2}}\right)\right)}{\cosh^2\left(\frac{  q}{2\sqrt{2}}\right)}.
\ee
\begin{figure}[ht]
\begin{center}
\includegraphics[scale=0.63]{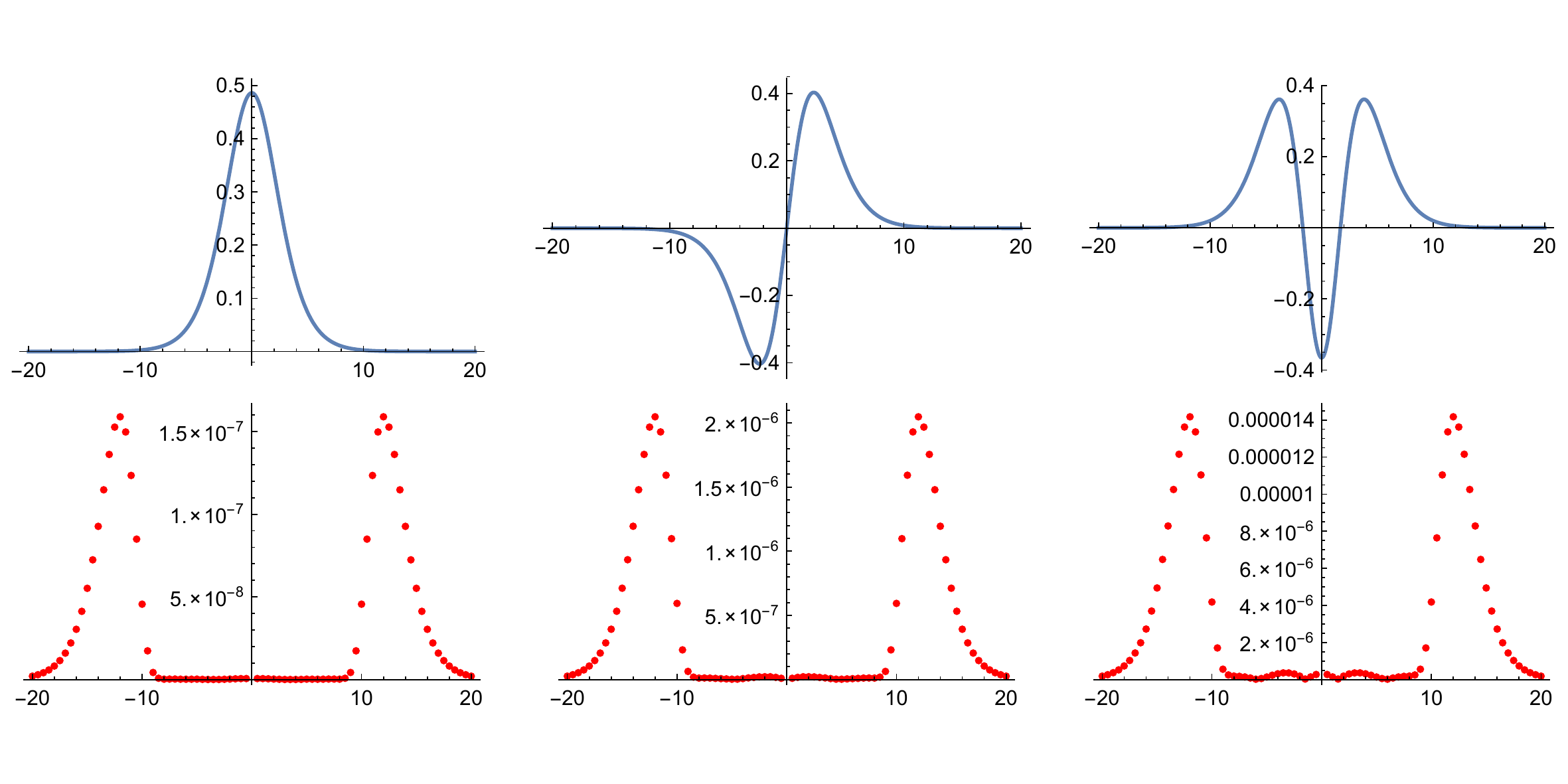}
\caption{The blue plots are the numerical wavefunctions of the ground state and the first two excited states, obtained by diagonalizing a $150 \times 150 $ matrix. They are undistinguishable
 from the ones obtained with Fredholm theory, which are shown in Fig. \ref{figwf}. The red plots give the absolute difference between the numerical and the exact 
 wavefunctions at some sample points. At this level of precision, the maximal difference is of the order $10^{-7}$ for the ground state, $10^{-6}$ for the first excited state, and $10^{-5}$ for the second excited state. }
\label{numgraph}
\end{center}
\end{figure}
They satisfy
\be
	\langle m | n\rangle = \int_{-\infty}^{\infty} \rd q \, c_m(q) c_n(q)=\delta_{mn},
\ee
and they have the following exponential decay at infinity:
\be 
\ba
	c_n(x) &\sim \re^{-q/\sqrt{2}} \qquad \qquad \qquad \quad q \rightarrow +\infty, \\
		& \sim (-1)^n\re^{+q/\sqrt{2}} \qquad \qquad \,\,q \rightarrow -\infty.
\ea
\ee
The eigenvalue equation is 
\be
	\rho | \psi \rangle = \lambda | \psi \rangle,
\ee
which in our basis becomes
\be
	\sum_{n=0}^\infty M_{mn} v_n =\lambda v_m,
\ee
where
\be
	M_{mn}= \langle c_m | \rho | c_n \rangle, \qquad \qquad v_m = \langle c_m | \psi \rangle.
\ee
After solving numerically  the (truncated) matrix eigenvalue equation, the wave function can be rebuilt as 
\be
	| \psi \rangle =\sum_{n =0}^{\infty} v_n | c_n \rangle.
\ee
In order to compute the matrix elements $M_{mn}$, there is a more efficient way than computing numerically the double integral, as explained in \cite{hmo}. It relies on the decomposition of the kernel
\be
	\rho(q_1,q_2)=\sum_{k=0}^\infty \rho_k(q_1) \rho_k(q_2),
\ee
with
\be
	\rho_k(q)=\sqrt{\frac{1}{16 \sqrt{2}\pi}} \, \frac{\tanh^k \left ( \frac{q}{2\sqrt{2}}\right )}{\cosh^2 \left ( \frac{q}{2\sqrt{2}}\right )}.
\ee
We thus have
\be
	M_{mn}=\sum_{k=0}^{\infty} R_{mk} R_{nk},
\ee
where
\be
\ba
	R_{mk} &= \langle c_m | \rho_k \rangle \\
	& = \frac{2^{m+1/2} (1+(-1)^{m+k})}{\pi}\sqrt{\frac{(m+3/2)}{(m+1)(m+2)}} \\
	& \qquad \qquad  \times \sum_{\ell=0}^{\lfloor m/2 \rfloor}(-1)^\ell 2^{-2\ell} \frac{\Gamma \left ( m-\ell +\frac{3}{2} \right )}{\Gamma \left ( \ell+1 \right )\Gamma \left ( m-2\ell \right )} \frac{1}{(m+k+1-2\ell)(m+k+2-2\ell)}.
\ea
\ee
The numerical results for the eigenfunctions obtained with this method agree with those in the section 2, as pictured in Fig. \ref{numgraph}.


\begin{thebibliography}{99}
\bibliographystyle{plain}

\bibitem{adkmv}
 M.~Aganagic, R.~Dijkgraaf, A.~Klemm, M.~Mari\~no and C.~Vafa, ``Topological strings and integrable hierarchies,''
  Commun.\ Math.\ Phys.\  {\bf 261}, 451 (2006)
  [hep-th/0312085].

\bibitem{ghm}
 A.~Grassi, Y.~Hatsuda and M.~Mari\~no, ``Topological Strings from Quantum Mechanics,''  Annales Henri Poincare {\bf 17}, no. 11, 3177 (2016) 
 [arXiv:1410.3382 [hep-th]].

\bibitem{cgm} S.~Codesido, A.~Grassi and M.~Mari\~no, ``Spectral Theory and Mirror Curves of Higher Genus,''
  arXiv:1507.02096 [hep-th].
  
  \bibitem{mmrev}
   M.~Mari\~no, ``Spectral Theory and Mirror Symmetry,''
  arXiv:1506.07757 [math-ph].

\bibitem{mirmor}  
 A.~Mironov and A.~Morozov, ``Nekrasov Functions and Exact Bohr-Sommerfeld Integrals,''
  JHEP {\bf 1004}, 040 (2010) [arXiv:0910.5670 [hep-th]].
  
\bibitem{acdkv}
M.~Aganagic, M.~C.~N.~Cheng, R.~Dijkgraaf, D.~Krefl and C.~Vafa, ``Quantum Geometry of Refined Topological Strings,''
  JHEP {\bf 1211}, 019 (2012)
  [arXiv:1105.0630 [hep-th]].
  
       \bibitem{ns}
   N.~A.~Nekrasov and S.~L.~Shatashvili, ``Quantization of Integrable Systems and Four Dimensional Gauge Theories,''
  arXiv:0908.4052 [hep-th].

\bibitem{km}
J.~Kallen and M.~Mari\~no, ``Instanton effects and quantum spectral curves,''
  Annales Henri Poincare {\bf 17}, no. 5, 1037 (2016) [arXiv:1308.6485 [hep-th]].

\bibitem{hw}
  M.~x.~Huang and X.~f.~Wang, ``Topological Strings and Quantum Spectral Problems,''
  JHEP {\bf 1409}, 150 (2014) [arXiv:1406.6178 [hep-th]].


 \bibitem{mp}
 M.~Mari\~no and P.~Putrov, ``ABJM theory as a Fermi gas,''
  J.\ Stat.\ Mech.\  {\bf 1203}, P03001 (2012)
  [arXiv:1110.4066 [hep-th]].

\bibitem{hmo}
 Y.~Hatsuda, S.~Moriyama and K.~Okuyama, ``Exact Results on the ABJM Fermi Gas,''
  JHEP {\bf 1210}, 020 (2012) [arXiv:1207.4283 [hep-th]].
  
  \bibitem{hmo2}
 Y.~Hatsuda, S.~Moriyama and K.~Okuyama, ``Instanton Effects in ABJM Theory from Fermi Gas Approach,''
  JHEP {\bf 1301}, 158 (2013) [arXiv:1211.1251 [hep-th]].
  
  \bibitem{calvo-m}
   F.~Calvo and M.~Mari\~no, ``Membrane instantons from a semiclassical TBA,''
  JHEP {\bf 1305}, 006 (2013) [arXiv:1212.5118 [hep-th]].
  
  \bibitem{hmo3}   Y.~Hatsuda, S.~Moriyama and K.~Okuyama, ``Instanton Bound States in ABJM Theory,''
  JHEP {\bf 1305}, 054 (2013) [arXiv:1301.5184 [hep-th]].
  
 \bibitem{hmmo}
   Y.~Hatsuda, M.~Mari\~no, S.~Moriyama and K.~Okuyama, ``Non-perturbative effects and the refined topological string,''
  JHEP {\bf 1409}, 168 (2014) [arXiv:1306.1734 [hep-th]].
  
\bibitem{cgm8}
 S.~Codesido, A.~Grassi and M.~Mari\~no,``Exact results in $ \mathcal{N}=8 $ Chern-Simons-matter theories and quantum geometry,''
  JHEP {\bf 1507}, 011 (2015) [arXiv:1409.1799 [hep-th]].
  
\bibitem{kasm}
  R.~Kashaev and M.~Mari\~no, ``Operators from mirror curves and the quantum dilogarithm,'' 
  Commun.\ Math.\ Phys.\  {\bf 346}, no. 3, 967 (2016) 
  [arXiv:1501.01014 [hep-th]].
  
 \bibitem{kmz}
 R.~Kashaev, M.~Mari\~no and S.~Zakany, ``Matrix models from operators and topological strings, 2,''
  Annales Henri Poincare {\bf 17}, no. 10, 2741 (2016) [arXiv:1505.02243 [hep-th]].
  
\bibitem{lst}
 A.~Laptev, L.~Schimmer and L.~A.~Takhtajan, ``Weyl type asymptotics and bounds for the eigenvalues of functional-difference operators for mirror curves,'' 
 Geom. Funct. Anal. {\bf 26}, 288 (2016) [arXiv:1510.00045 [math.SP]].  
 
     \bibitem{cgum}
 S.~Codesido, J. Gu and M.~Mari\~no, 
 ``Operators and higher genus mirror curves,'' arXiv:1609.00708 [hep-th].
 
  

\bibitem{py}
 P.~Putrov and M.~Yamazaki, ``Exact ABJM Partition Function from TBA,''
  Mod.\ Phys.\ Lett.\ A {\bf 27}, 1250200 (2012) [arXiv:1207.5066 [hep-th]].

\bibitem{oz} 
 K.~Okuyama and S.~Zakany, ``TBA-like integral equations from quantized mirror curves,''
  JHEP {\bf 1603}, 101 (2016) [arXiv:1512.06904 [hep-th]].

   \bibitem{gkmr}
 J.~Gu, A.~Klemm, M.~Mari\~no and J.~Reuter, ``Exact solutions to quantum spectral curves by topological string theory,''
  JHEP {\bf 1510}, 025 (2015) [arXiv:1506.09176 [hep-th]].

 \bibitem{bgt} G.~Bonelli, A.~Grassi and A.~Tanzini, ``Seiberg-Witten theory as a Fermi gas,'' Lett.\ Math.\ Phys.\  {\bf 107}, no. 1, 1 (2017) [arXiv:1603.01174 [hep-th]].


\bibitem{mz}
 M.~Mari\~no and S.~Zakany, ``Matrix models from operators and topological strings,''
  Annales Henri Poincare {\bf 17}, no. 5, 1075 (2016) [arXiv:1502.02958 [hep-th]].
  
     
  \bibitem{eynard}
  B.~Eynard, ``Large N expansion of convergent matrix integrals, holomorphic anomalies, and background independence,''
  JHEP {\bf 0903}, 003 (2009) [arXiv:0802.1788 [math-ph]].
  
  \bibitem{em}
  B.~Eynard and M.~Mari\~no, ``A holomorphic and background independent partition function for matrix models and topological strings,''
  J.\ Geom.\ Phys.\  {\bf 61}, 1181 (2011) [arXiv:0810.4273 [hep-th]].
  
  \bibitem{grassi}
 A.~Grassi, ``Spectral determinants and quantum theta functions,''
  arXiv:1604.06786 [hep-th].

\bibitem{wzh}
 X.~Wang, G.~Zhang and M.~x.~Huang, ``New Exact Quantization Condition for Toric Calabi-Yau Geometries,''
  Phys.\ Rev.\ Lett.\  {\bf 115}, 121601 (2015) [arXiv:1505.05360 [hep-th]].
  
  \bibitem{hm}
 Y.~Hatsuda and M.~Mari\~no, ``Exact quantization conditions for the relativistic Toda lattice,''
  JHEP {\bf 1605}, 133 (2016) [arXiv:1511.02860 [hep-th]].
  
  
  \bibitem{fhm}
  S.~Franco, Y.~Hatsuda and M.~Mari\~no, ``Exact quantization conditions for cluster integrable systems,"  J.\ Stat.\ Mech.\  {\bf 1606}, no. 6, 
  063107 (2016) [arXiv:1512.03061 [hep-th]].

 
 \bibitem{butterfly}
  Y.~Hatsuda, H.~Katsura and Y.~Tachikawa, ``Hofstadter's Butterfly in Quantum Geometry,'' New J.\ Phys.\  {\bf 18}, no. 10, 103023 (2016) [arXiv:1606.01894 [hep-th]].
    
  \bibitem{tw}
 C.~A.~Tracy, H.~Widom, ``Proofs of two conjectures related to the thermodynamic Bethe ansatz,''
  Commun.\ Math.\ Phys.\  {\bf 179}, 667-680 (1996).
  [solv-int/9509003].
  
     
     
 \bibitem{be}
  G. Borot and B. Eynard, ``Geometry of spectral curves and all order dispersive integrable system," SIGMA {\bf 8}, 100 (2012) [arXiv:1110.4936 [math-ph]].



\bibitem{mmss}
J.~M.~Maldacena, G.~W.~Moore, N.~Seiberg and D.~Shih, ``Exact vs. semiclassical target space of the minimal string,''
  JHEP {\bf 0410}, 020 (2004)  [hep-th/0408039].


  \bibitem{berry}
  M. Berry, ``Two-State Quantum Asymptotics," Annals of the New York Academy of Sciences {\bf 755}, 303 (1995).
  
  \bibitem{amir}
 A.~K.~Kashani-Poor, ``Quantization condition from exact WKB for difference equations,''
  JHEP {\bf 1606}, 180 (2016) [arXiv:1604.01690 [hep-th]].
  
  \bibitem{sciarappa}
  A.~Sciarappa, ``Bethe/Gauge correspondence in odd dimension: modular double, non-perturbative corrections and open topological strings,''
  JHEP {\bf 1610}, 014 (2016) [arXiv:1606.01000 [hep-th]].
 
   
\bibitem{eo}
 B.~Eynard and N.~Orantin, ``Invariants of algebraic curves and topological expansion,''
  Commun.\ Num.\ Theor.\ Phys.\  {\bf 1}, 347 (2007)
  [math-ph/0702045].

 \bibitem{gs}
   S.~Gukov and P.~Sulkowski, ``A-polynomial, B-model, and Quantization,''
  JHEP {\bf 1202}, 070 (2012) [arXiv:1108.0002 [hep-th]].
  
  
  \bibitem{mulase}
 O.~Dumitrescu and M.~Mulase, ``Lectures on the topological recursion for Higgs bundles and quantum curves,''
  arXiv:1509.09007 [math.AG].
  
  \bibitem{norbury}
   P.~Norbury, ``Quantum curves and topological recursion,''
  arXiv:1502.04394.
  
 \bibitem{boucharde}
 V.~Bouchard and B.~Eynard, ``Reconstructing WKB from topological recursion,''
  arXiv:1606.04498 [math-ph].
 
\bibitem{hkp}
 M.~X.~Huang, A.~Klemm and M.~Poretschkin, ``Refined stable pair invariants for E-, M- and $[p, q]$-strings,'' JHEP {\bf 1311}, 112 (2013)
  [arXiv:1308.0619 [hep-th]]. 

  
  \bibitem{hkrs}
 M.~x.~Huang, A.~Klemm, J.~Reuter and M.~Schiereck, ``Quantum geometry of del Pezzo surfaces in the Nekrasov-Shatashvili limit,''
  JHEP {\bf 1502}, 031 (2015) [arXiv:1401.4723 [hep-th]].
  
\bibitem{integral}
S. M. Zemyan, {\it The classical theory of integral equations}, Springer--Verlag, 2012. 
  
  \bibitem{zamo}
  A.~B.~Zamolodchikov, ``Painlev\'e III and 2-d polymers,''
  Nucl.\ Phys.\ B {\bf 432}, 427 (1994) [hep-th/9409108].

  
\bibitem{kwy}
A.~Kapustin, B.~Willett and I.~Yaakov, ``Nonperturbative Tests of Three-Dimensional Dualities,''
  JHEP {\bf 1010}, 013 (2010)
  [arXiv:1003.5694 [hep-th]].

\bibitem{voros-zeta}
 A.~Voros, ``The Zeta Function Of The Quartic Oscillator,''
  Nucl.\ Phys.\ B {\bf 165}, 209 (1980).

\bibitem{voros-complex}
A. Voros, ``The return of the quartic oscillator. The complex WKB method," Annales de l'I.H.P. {\bf 39}, 211 (1983).

\bibitem{leon}
L. Takhtajan, {\it Quantum Mechanics for mathematicians}, American Mathematical Society, 2008. 


  
\bibitem{moshinsky}
 M.~Moshinsky and C.~Quesne, ``Linear canonical transformations and their unitary representations,''
  J.\ Math.\ Phys.\  {\bf 12}, 1772 (1971).


\bibitem{gk}
 S.~Garoufalidis and R.~Kashaev, ``Evaluation of state integrals at rational points,''
  Commun.\ Num.\ Theor.\ Phys.\  {\bf 09}, no. 3, 549 (2015) [arXiv:1411.6062 [math.GT]].


\bibitem{gv}
 R.~Gopakumar and C.~Vafa, ``M theory and topological strings. 2.,''
  hep-th/9812127.

\bibitem{abk}
 M.~Aganagic, V.~Bouchard and A.~Klemm, ``Topological Strings and (Almost) Modular Forms,''
  Commun.\ Math.\ Phys.\  {\bf 277}, 771 (2008) [hep-th/0607100].

  
  
  \bibitem{cpw}
   C.~Kozcaz, S.~Pasquetti and N.~Wyllard, ``A and B model approaches to surface operators and Toda theories,''
  JHEP {\bf 1008}, 042 (2010) [arXiv:1004.2025 [hep-th]].

\bibitem{awata}
H.~Awata, H.~Fuji, H.~Kanno, M.~Manabe and Y.~Yamada, ``Localization with a Surface Operator, Irregular Conformal Blocks and Open Topological String,''
  Adv.\ Theor.\ Math.\ Phys.\  {\bf 16}, no. 3, 725 (2012) [arXiv:1008.0574 [hep-th]].


\bibitem{eynard-co}
B.~Eynard, ``Topological expansion for the 1-Hermitian matrix model correlation functions,''
  JHEP {\bf 0411}, 031 (2004) [hep-th/0407261].
    
 
 \bibitem{mmopen} 
  M.~Mari\~no, ``Open string amplitudes and large order behavior in topological string theory,''
  JHEP {\bf 0803}, 060 (2008) [hep-th/0612127].
   
  \bibitem{bkmp}
V.~Bouchard, A.~Klemm, M.~Mari\~no and S.~Pasquetti, ``Remodeling the B-model,'' Commun.\ Math.\ Phys.\  {\bf 287}, 117 (2009) [arXiv:0709.1453 [hep-th]].
  
   
 \bibitem{av}
 M.~Aganagic and C.~Vafa, ``Mirror symmetry, D-branes and counting holomorphic discs,''
  hep-th/0012041.

  
\bibitem{ek}
B.~Eynard and C.~Kristjansen, ``Exact solution of the O(n) model on a random lattice,''
  Nucl.\ Phys.\ B {\bf 455}, 577 (1995) [hep-th/9506193].

\bibitem{gernot}
 G.~Akemann, ``Higher genus correlators for the Hermitian matrix model with multiple cuts,''
  Nucl.\ Phys.\ B {\bf 482}, 403 (1996) [hep-th/9606004].

  
 \bibitem{akv}
  M.~Aganagic, A.~Klemm and C.~Vafa, ``Disk instantons, mirror symmetry and the duality web,''
  Z.\ Naturforsch.\ A {\bf 57}, 1 (2002) [hep-th/0105045].


 \bibitem{el-int} 
D. Karp and S. M. Sitnik, ``Asymptotic approximations for the first incomplete elliptic integral near logarithmic singularity," Journal of computational and applied mathematics {\bf 205} 186 (2007) 
[arXiv:math/0604026 [math.CA]].

\bibitem{gkm}
A.~Grassi, J.~Kallen and M.~Mari\~no, ``The topological open string wavefunction,''
  Commun.\ Math.\ Phys.\  {\bf 338}, no. 2, 533 (2015) [arXiv:1304.6097 [hep-th]].

\bibitem{as}
M.~Aganagic and S.~Shakirov, ``Knot Homology and Refined Chern-Simons Index,''
  Commun.\ Math.\ Phys.\  {\bf 333}, no. 1, 187 (2015) [arXiv:1105.5117 [hep-th]].
  
\bibitem{ov}
H.~Ooguri and C.~Vafa, ``Knot invariants and topological strings,''
  Nucl.\ Phys.\ B {\bf 577}, 419 (2000) [hep-th/9912123].
  
\bibitem{lmv}
J.~M.~F.~Labastida, M.~Mari\~no and C.~Vafa, ``Knots, links and branes at large N,''
  JHEP {\bf 0011}, 007 (2000) [hep-th/0010102].

\bibitem{dubrovin}
B. Dubrovin, ``Theta functions and non-linear equations," Russian mathematical surveys {\bf 36}, 11-92 (1981).

\bibitem{bbt}
O. Babelon, D. Bernard and M. Talon, {\it Introduction to classical integrable systems}, Cambridge University Press, 2003. 


\bibitem{ks}
R. Kashaev and S. Sergeev, work in progress. 

\bibitem{akhiezer}
N. I. Akhiezer, {\it Elements of the theory of elliptic functions}, American Mathematical Society, 1990. 
  
\end{thebibliography}
\end{document}